\documentclass[10pt,preprint]{aastex}

\shorttitle{FEPS Cold Debris Disks}
\shortauthors{Hillenbrand et al.}

\begin{document}

\title{ 
The Complete Census of 70-$\mu$m-Bright Debris Disks within the $FEPS$ 
(Formation and Evolution of Planetary Systems) Spitzer Legacy Survey
of Sun-like Stars 
}

\author{
Lynne A. Hillenbrand\altaffilmark{1}, John M. Carpenter \altaffilmark{1},
Jinyoung Serena Kim\altaffilmark{2}, Michael R. Meyer \altaffilmark{2},
Dana E. Backman\altaffilmark{3}, 
Amaya Moro-Mart{\'{\i}}n\altaffilmark{4},
David J. Hollenbach\altaffilmark{5},
Dean C. Hines\altaffilmark{6}, 
Ilaria Pascucci\altaffilmark{2},
Jeroen Bouwman\altaffilmark{7}. 
}
\email{lah@astro.caltech.edu}
\altaffiltext{1}{Astronomy/Astrophysics, California Institute of Technology,
Pasadena, CA 91125.} 

\altaffiltext{2}{Steward Observatory, The
University of Arizona, 933 N. Cherry Ave., Tucson, AZ 85721}

\altaffiltext{3}{SOFIA / SETI Institute, Mountain View, CA 94043}

\altaffiltext{4}{Astrophysics, 
Princeton University, Peyton Hall, Princeton, NJ, 08540}

\altaffiltext{5}{NASA-Ames, Moffet Field, CA 94035-1000}

\altaffiltext{6}{Space Science Institute, 4750 Walnut Street,
Suite 205 Boulder, CO 80301}

\altaffiltext{7}{Max-Planck-Institut fu¬ r Astronomie, D-69117
Heidelberg, Germany.}


\begin{abstract}

We report detection with the Spitzer Space Telescope of cool dust
surrounding solar type stars.  The observations were performed  
as part of the Legacy Science Program, ``Formation and Evolution 
of Planetary Systems'' ($FEPS$). From the overall $FEPS$ sample 
(Meyer et al. 2006) of 328 stars having ages $\sim$0.003-3 Gyr 
we have selected sources with 70 $\mu$m flux densities indicating excess 
in their spectral energy distributions above expected 
photospheric emission.  Six strong excess sources are likely
primordial circumstellar disks, remnants of the star formation process.
Another 25 sources having $\ge3\sigma$ excesses 
are associated with dusty debris disks, generated by collisions within 
planetesimal belts that are possibly stirred by existing planets. 
We draw attention to six additional sources with $\ge 2\sigma$ excesses
which require confirmation as debris disks. In our analysis, most ($>$80\%) 
of the debris disks identified via 70 $\mu$m excesses 
have $\ge3\sigma$ excesses at 33 $\mu$m as well, while only a minority 
($<$40\%) have $\ge3\sigma$ excesses at 24 $\mu$m.  

The rising spectral energy distributions towards 
- and perhaps beyond - 70 $\mu$m imply dust temperatures $T_{dust} <$45-85 K  
for debris in equilibrium with the stellar radiation field.
We infer bulk properties such as characteristic
temperature, location, fractional luminosity, and mass of the dust  
from fitted single temperature  blackbody models. 
For $>$1/3 of the debris sources we find that multiple temperature 
components are suggested, implying a spatial distribution of dust 
extending over many tens of AU.  Because the disks are dominated by
collisional processes, the parent body (planetesimal) belts 
may be extended as well.
Preliminary assessment of the statistics of cold debris around sun-like
stars shows that $\sim$10\% of $FEPS$ targets with masses 
between 0.6 and 1.8 $M_\odot$ and ages between 30 Myr and 3 Gyr 
exhibit 70 $\mu$m
emission in excess of the expected photospheric flux density. 
We find that fractional excess amplitudes appear higher for younger stars and
that there may be a trend in 70 $\mu$m excess frequency with stellar mass. 

\end{abstract}

\keywords{
circumstellar matter --- infrared --- Kuiper Belt: 
debris disks: 
stars --- planetary systems  --- stars : individual 
(\objectname
[HD 104860, HD 105, HD 107146, HD 122652, HD 141943, HD
145229, HD 150706, HD 17925, HD 187897, HD 191089, HD 201219, HD 202917, HD
204277, HD 206374, HD 209253, HD 219498, HD 22179, HD 25457, HD 31392, HD
37484, HD 377, HD 38207, HD 38529, HD 61005, HD 6963, HD 70573, HD 72905, HD
85301, HD 8907 MML 17, HD 35850, PDS 66, HD
143006, RX J1111.7-7620, RX J1842.9-3532, RX J1852.3-3700, \[PZ99\] J161411.0-230536]
{HD 104860, HD 105, HD 107146, HD 122652, HD 141943, HD
145229, HD 150706, HD 17925, HD 187897, HD 191089, HD 201219, HD 202917, HD
204277, HD 206374, HD 209253, HD 219498, HD 22179, HD 25457, HD 31392, HD
37484, HD 377, HD 38207, HD 38529, HD 61005, HD 6963, HD 70573, HD 72905, HD
85301, HD 8907 MML 17, HD 35850, PDS 66, HD
143006, RX J1111.7-7620, RX J1842.9-3532, RX J1852.3-3700, [PZ99] J161411.0-230536}
)}

\section{Introduction}

Revolutionary improvements in astronomical observing capability 
are directed not only towards the distant reaches of the universe, 
but also to our nearest neighbors beyond, and even within, the Solar System.
In exploiting these new capabilities we scrutinize old paradigms in new detail. 
The Spitzer Space Telescope is no exception and
a major science area for Spitzer has been the investigation of dusty 
circumstellar disks -- both young primordial and older debris systems. 
Spitzer has unprecedented ability to detect the Rayleigh-Jeans tail of 
stellar photospheres in the 3-70 $\mu$m mid-infrared regime -- and hence 
small excesses above those photospheres due to circumstellar dust -- 
since for sizable samples of nearby stars, the photometric accuracy 
is dominated by calibration uncertainty rather than by 
signal-to-noise considerations. 
Spitzer has thus extended our knowledge of disks beyond 
the brightest/nearest objects of various classes, 
to previously unexplored realms of completeness in e.g. volume, 
spectral type, and age.  In particular, Spitzer has
enabled statistically significant surveys for warm (T$_{\rm dust} \sim 125-300$K) 
dust in the outer terrestrial zone (R$_{\rm dust}\sim 1-5$ AU) and
cold (T$_{\rm dust} <$ 40-120K) dust in the  
Jovian/Kuiper zone (R$_{\rm dust} > 5-50$ AU) of potential
Solar System analogs having a wide range of ages.

The current Kuiper Belt dust mass is estimated at $\approx 1\times 10^{-5} M_\earth$ 
in sub-cm-sized particles, based on several different measures such as: 
IRAS and COBE upper limits to cold emission in the ecliptic plane 
and associated modeling (e.g. Backman et al. 1995; Teplitz et al. 1999)
of assumed 2-5000 $\mu$m grains,
or detections of outer Solar System dust thought to originate 
from the Kuiper Belt (Landgraf et al. 2002) and the dynamical model 
of Moro-Martin \& Malhotra (2003) for $\sim$1-150 $\mu$m grains.
There is currently an additional $M\approx 0.1 M_\earth$ in large 
-- more than cm-sized --  bodies (Gladman et al. 2001).  
Although the numbers are uncertain by 
probably 1-1.5 orders of magnitude, the above may be compared to 
an inferred $M\approx 1\times 10^{-3} M_\earth$ in dust 
and $10-50 M_\earth$ for the large bodies during the early debris stages 
at a few tens of Myr (Stern \& Colwell 1997; Stern, 1996b). This dust level is
easily detected by Spitzer for nearby solar-type stars.  In contrast, 
the low dust mass observed at the current solar age is not detectable.

The Spitzer Legacy program $FEPS$ (Formation and Evolution of Planetary Systems)
was designed to study the final stages of primordial disk dissipation and the 
development and evolution of debris disks 
around solar-mass stars over a range of ages. 
The Spitzer data include IRAC and MIPS photometry and
IRS spectrophotometry for 328 sources.  
Approximately 55 stars in each of 6 logarithmic age
bins between 3 Myr and 3 Gyr of age
were observed for the $FEPS$ program.  The targets span a narrow mass range 
(95\% are within 0.8-1.5~M$_\odot$) in order to focus on Sun-like stars.  
The targets are proximate enough (d $\approx$ 10-200 pc) 
to enable a complete census for circumstellar dust comparable 
in quantity to predictions from simple models of our Solar System's 
collisional evolution as a function 
of stellar age (such as that discussed in Meyer et al. 2007).  Sensitivity 
was further maximized by choosing targets in regions of lower infrared 
background over those of the same age in regions of higher background/cirrus.
Signal-to-noise (SNR) $>$30 is obtained on the underlying stellar photosphere 
at 3.6, 4.5, 8, and 24 $\mu$m with IRAC and MIPS photometric observations 
while SNR $>$ 4 is achieved out to 35 $\mu$m from IRS spectophotometric 
observations for $>$90\% of the objects. 


The main aim of $FEPS$ is to trace dust evolution via 
spectral energy distribution interpretation and thereby to 
probe the detritus
indicative of planet formation and evolution. 
In this contribution we focus on sources that are detected 
at 70 $\mu$m with flux densities in excess 
of those expected from the stellar photosphere.  Six such $FEPS$ objects
(RX J1852.3-3700, HD 143006, RX J1842.9-3532,
1RXS J132207.2-693812 a.k.a. PDS 66, RX J1111.7-7620, and 
1RXS J161410.6-230542 a.k.a. $[$PZ99$]$ J161411.0-230536)
are considered ``primordial disks" and have been presented 
also by Silverstone et al. (2006) and Bouwman et al. (2008).
These young disks have strong excess emisson not only at 70 $\mu$m 
but also shortward, down to at least 3-8 $\mu$m. 
We provide their data again here, for completeness and for context. 
However, our main focus is on
the larger sample of ``debris disks" which
generally have weaker excess emission at 70 $\mu$m, 24-33 $\mu$m
flux densities consistent with, or only moderately in excess 
of, expected photospheric values, and $<$3-13 $\mu$m flux densities
which are purely photospheric.  Our debris sample includes
25 sources with $>3\sigma$ significant and 6 possible sources with
$2-3\sigma$ significant cold disk systems detected at 70 $\mu$m. 
Of the total, 14 are newly appreciated debris disk systems announced here
while the remainder have been reported previously including
in $FEPS$ contributions by Meyer et al. (2004); Kim et al. (2005); and
Pascucci et al. (2006); see Table 1 for details. 

We begin with a description of the Spitzer observations 
and data handling (\S2). We then present our methods for distinguishing
detections from noise at 70 $\mu$m and our results 
in the form of: color-color diagrams, excess signal-to-noise histograms,
and spectral energy distributions
which demonstrate the existence of 70 $\mu$m excesses indicative of cool
circumstellar material (\S3). We proceed 
to analyze the spectral energy distributions in terms of single-temperature blackbody
models and argue that in $>$1/3 of the cases multi-temperature models 
indicative of a range of dust radii, are a better match to the data
than are the narrow ring implied by single-temperature models (\S4). 
Our modeling results indicate dust temperatures typically $<$85 K which imply, 
depending on the stellar parameters, corresponding dust inner radii 
typically 5-50 AU and (poorly constrained) dust outer radii 
typically several hundred AU.  Comparison (\S5) with inferred parent star ages 
of the theoretical time scales for dust depletion mechanisms 
such as: inward drag due to Poynting-Robertson (P-R) or corpuscular effects,
outward push due to stellar radiative or mechanical effects, 
and in situ collisional destruction, suggests that 
the dust is continuously generated debris resulting from 
collisions among an unseen population of planetesimals.
Planetesimal orbits can be perturbed either by the largest embryos 
in the planetesimal population or by planetary mass bodies, generating 
in both cases a steady state collisional cascade. Alternately, debris dust
may be the result of individual, large catastrophic collisions
that artifically raise the mass in small dust particles over steady state 
evolution values. 
Trends in debris disk detections with stellar age and mass are investigated
(\S6).  Finally,
we place our results into a larger context in \S7 and then conclude in \S8.

\section{Sample, Observations, and Data Processing}

The $FEPS$ program utilized all three Spitzer science instruments -- IRAC, IRS, and MIPS -- 
to observe 328 solar-type stars.  Meyer et al. (2006) provides 
a description of the $FEPS$ observing strategy. 
Among the $FEPS$ sample are 15 
previously suspected (based on IRAS or ISO literature)
debris or long-lived primordial disk systems, 
only 11 of which are in fact confirmed by Spitzer.  Ten of these 15 
were observed by $FEPS$ for the purpose of probing primordial gas disk dissipation 
(e.g. Pascucci et al 2006, 2007) while the others were either serendipitously
on our lists or discovered as excess sources after the $FEPS$ program 
was submitted.
The sources selected {\it ab initio} because they were claimed to
exhibit infrared excess emission 
can not be included in statistical analyses of $FEPS$
Spitzer data for debris characteristics as a function of 
e.g. stellar age, stellar mass, stellar metallicity, stellar rotation, 
etc.  However, we do include them in this paper which presents disk detections
and simple dust models.

Carpenter et al. (2008b) provides a detailed discussion of 
realized $FEPS$ observation, data reduction, and data validation procedures.
We review here only those details of particular relevance to the present 
discussion of 70 $\mu$m excess.
Four sources -- HD 17925, HD 72905, HD 202917, HD 216803 --
were observed at 70 $\mu$m not by $FEPS$ but instead by the GTO program 
described by Bryden et al. (2006) who employ the same observing strategy 
as is standard for $FEPS$; the data were obtained from the Spitzer archive 
and processed using standard $FEPS$ techniques.  These particular
sources were included 
in $FEPS$ for the purpose of the gas disk dissipation experiment discussed above.
MIPS 160 $\mu$m photometry was obtained by the $FEPS$ project for 
a sub-sample of the full target list. It derives for most  of the sources 
discussed here from follow--up Spitzer GO-2 and GO-3 programs in which
additional 160 $\mu$m data 
\footnote{
MIPS 160\micron\ data from GO-2 and GO-3 followed the standard $FEPS$ observations,
using 10 sec exposure time per data collection event with 2--4 cycles (typically 4).
Raw data were processed with SSC pipeline S14.4.0 and
the MIPS DAT pipeline (Gordon et al. 2005) version 3.02. The final mosaic image 
has 8\arcsec/pixel. The flux conversion factor from instrumental units 
to MJy/Sr) is 44.7 and the absolute calibration uncertainty is 12\%
 (http://ssc.spitzer.caltech.edu/mips/calib/conversion.html).
 For aperture photometry we used an aperture radius of 24\arcsec, 
 a sky annulus spanning 64-128\arcsec, and  an aperture correction of 2.380.
Uncertainty was calculated by propogating the measured
root-mean-square deviation in the sky area over the source aperture.
}
was obtained for $FEPS$ sources with detected 70 (and/or 33) $\mu$m excesses.  

%
%

Exposure times at 70 $\mu$m were 10 seconds per data collection event 
or image, with 8 images taken per cycle.  The number of cycles varied 
between (1, for the four GTO targets)
2 and 14 in order to reach the desired depth.  
Our original intent to detect photospheres at 70 $\mu$m was predicated upon
pre-launch sensitivity estimates. However, the higher than expected rate of
large cosmic-ray hits reduced the on-orbit sensitivity by about a factor of
three (Rieke et al. 2004), so we attempted instead for each of our targets
to reach a common sensitivity relative to an estimate for the outer Solar
System dust level at the age of the star (see \S3.1).

Spitzer data were processed initially by the Spitzer Science Center pipeline 
S13. 
Post-pipeline processing of IRAC, IRS, and MIPS data, including further 
reduction details, photometry/spectral extraction, error derivation, 
and flux density calibration discussions are all given 
in Carpenter et al. (2008b); 
see also Kim et al. (2005) for 160 $\mu$m procedures. 
The photometric uncertainties are also discussed in Carpenter et al. (2008b). 
They were assessed for the IRAC and MIPS 24 $\mu$m data 
by computing the error in the mean of the flux densities derived 
from individual frames, with an adopted floor. These (presumed) 
photometric uncertainties are then validated by examination of flux density
histograms and of source colors. 
For MIPS 70 $\mu$m data, the photometry was performed not on individual images 
as for data at all other wavelengths, but on the final stacked/mosaicked image 
only, with the error calculated by propogating the measured
root-mean-square deviation in the sky area over the source aperture.
Thus the only validation
of the internal uncertainty comes from the Kurucz model comparison (illustrated
below).  
The random errors in the photometry at 70 
and 160 $\mu$m were estimated from the square root of the variance observed
in the sky annuli of the final resampled mosaics for these background--limited observations. 
After careful analysis we find that the MIPS 70 $\mu$m internal 
uncertainties must be inflated by a factor of 1.5 to account for the scatter 
in the quantity (data - model),
as discussed in detail in Carpenter et al. (2008a).  
Calibration uncertainties were taken from the Spitzer Observers' Manual version 7.0 
($<$2\% for the IRAC bands, 4\% for MIPS 24 $\mu$m and 7\% for MIPS 70 $\mu$m )
and dominate the error in our absolute photometry for sources 
with SNR $>>$ 1/$\sigma_{calib}$ (see Table 2 for relevant details).
We make use of internal and internal $+$ calibration uncertainties
at separate points in our analysis.
For comparison between simple models and the data, 
we also use synthetic photometry points 
constructed from the IRS spectrophotometric data 
(S14 processing) with square bandpasses of 12-15\% width
centered at 13 and 33 $\mu$m
(flux-weighted average wavelengths 13.17 and 32.36 $\mu$m, assuming 
a Rayleigh-Jeans spectral energy distribution).

\section{Identifying 70 $\mu$m Detections and Excess Sources}

In this section, we first establish the reliability of the 70 $\mu$m source
detections and their association with the intended $FEPS$ target.
We consider sensitivity, cirrus, and confusion as limitations.
The possible detections at 70 $\mu$m  are defined by:
1) photometric measurements with SNR $>$2 at 70$\mu$m using internal errors,
2) visibility to the human eye and point source morphology
on the 70$\mu$m images, and
3) positional alignment with the corresponding 24 $\mu$m point sources.
From the detections at 70 $\mu$m, we then define
via color-color diagrams and excess signal-to-noise histograms
the sub-sample with SNR $>$2 excesses at 70 $\mu$m,
using both internal and calibration errors.  Finally in this section,
we present spectral energy distributions for these sources.
Table 1 contains our 70 $\mu$m excess candidates and
includes notes on several sources for which the determination of
excess detection in 70 $\mu$m is not straightforward.

\subsection{Sensitivity Considerations}

$FEPS$ achieves {\it photospheric sensitivity}
with SNR $>$30 for 100\% of the program objects at all Spitzer broad band
wavelengths $\leq$24 $\mu$m, and with SNR $>$3 for 90\% of the program
objects in IRS spectrophotometry out to 35 $\mu$m. At 70 $\mu$m, however,
detecting photospheres of solar type stars at distances greater than about 12 pc
is not feasible in the launched version of Spitzer
in less than several hours of integration.
As our targets range from tens to hundreds of parsecs, a very small
fraction of our sample was proximate enough for detection at 70 $\mu$m
in the absence of excess emission.
A few such photospheres are indeed detected:
HD 13974 (11 pc) and HD 216803 (7.6 pc),
the latter observed as part of GTO time (Rieke), and
also potentially HD 17925 (10 pc) which is noted below as only a
low significance excess object.

Our integration times (\S2) were chosen to be sensitive
to a minimum dust level relative to that inferred for dust in our own
Solar System  (e.g. Landgraf et al. 2002) as it appeared earlier
in its evolutionary history.
Such evolution has been described as having a power-law behavior in
certain regimes -- roughly $\tau^{0}$ until collisional equilibrium is reached, 
transitioning to $\tau^{-1}$ by several hundred million years, 
then to $\tau^{-2}$ beyond a few billion years
(cf. Dominik and Decin, 2003; Wyatt, 2005).  These canonical regimes 
are well sampled by the $FEPS$ age distribution.  A realistic model
has more structure than the simple power-law estimates above, which are just
guides to the behavior.  The simulations 
that we used (Backman et al., private communication; see also Meyer et al. 2007)
assume an initial planetesimal belt of 30 M$_\earth$
distributed between 30 and 50 AU that undergoes collisional evolution;
material is subsequently parsed according to a Dohnanyi (1969)
fragment mass distribution down to small sizes.  For a fiducial source 
at distance 30 pc
and luminosity 1 L$_\odot$, the dust evolution predicts a change
in 70 $\mu$m flux density from 180 to 50 mJy for source ages between
150 and 1500 Myr.  Given the actual age and distance/luminosity distribution
of our sample, approximately 1/3 of our targets are younger than 150 Myr and
almost all $FEPS$ targets 150-1500 Myr have 3-sigma sensitivity
at 70 $\mu$m exceeding this dust model.  Our survey is sensitive to dust
emission $\times$5 greater than that estimated from the projected  
young Solar System model for most of the remaining $\sim$2/3 of the targets,
and sensitive to $\times$10 greater emission for all but a few
(with the limitations primarily driven by
the increased distance range required to find young targets).
For older (nearby) stars, our survey was sensitive to dust emission roughly
5-20 times the current Solar System level
(or 9-36 mJy in the excess, for the fiducial source above).

\begin{figure}[t]
  \begin{center}
     \epsscale{.9}
     \vspace*{-10mm}
     \plotone{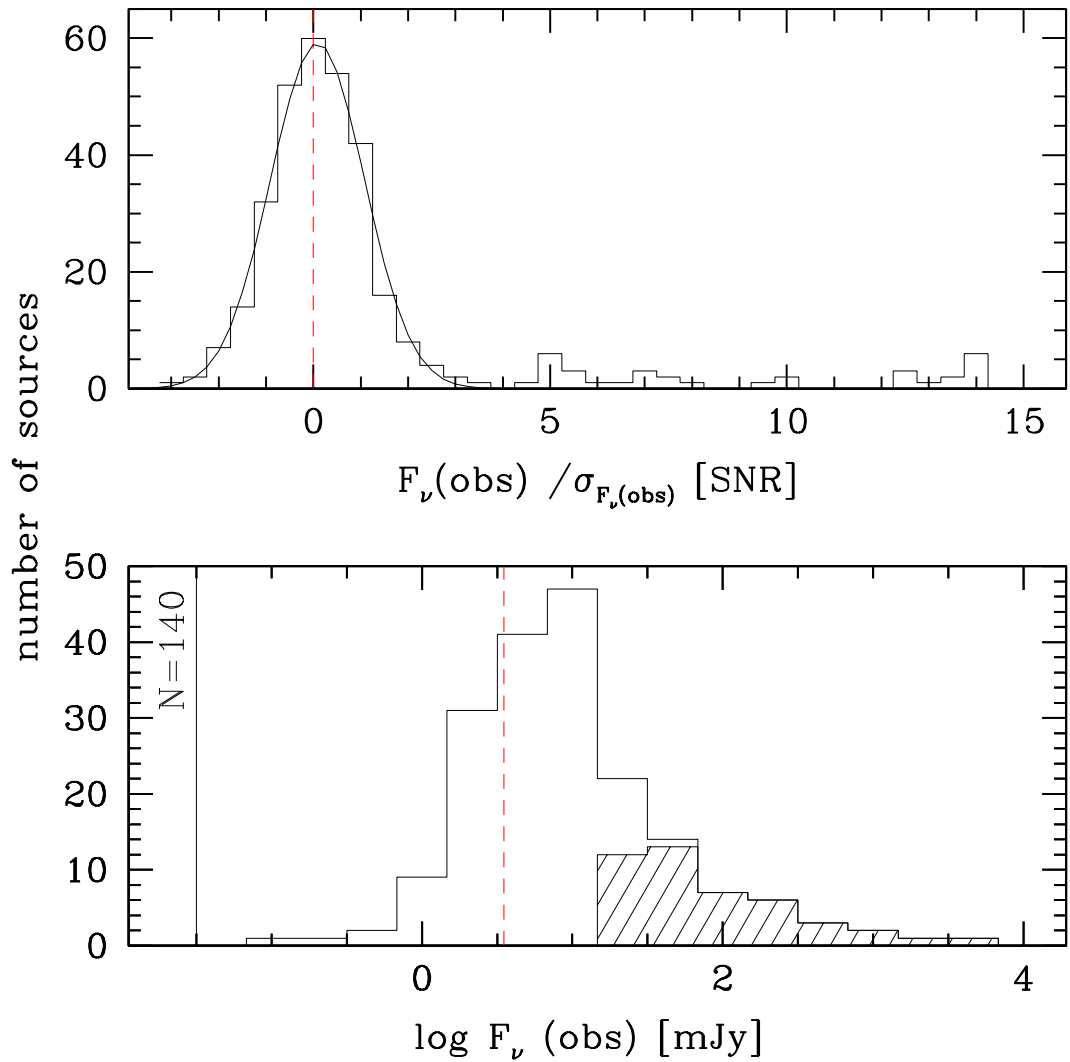}
  \end{center}
  \caption{ 
  Histograms of 70 $\mu$m signal-to-noise and measured flux density.
  Top panel: Measured flux density divided by uncertainty.
  A gaussian fit at $|SNR|<3$ has a mean of 0.09 and a dispersion of 0.99,
  and is shown as the solid curve; vertical dashed line 
  (colored red in the electronic edition) indicates zero, 
  for reference.  Based on this distribution, which validates the arbitrarily 
  inflated (by a factor of 1.5) noise estimates, 
  secure 70 $\mu$m detections are those
  sources with SNR$>3$ where each SNR bin contains less than 
  1 spurious noise source. 
  Bottom panel: Logarithm of the 70 $\mu$m flux densities for candidate  
  detections having flux density larger than twice the error 
  (hatched histogram) compared to the measured 70 $\mu$m flux densities 
  for all sources (open histogram).  The unclosed bin to the left
  represents objects with formally negative flux densities 
  (left side of the gaussian in top panel). 
  For comparison, the typical 3-$\sigma$ detection limits 
  from IRAS and ISO at 60 $\mu$m were 500 and 100 mJy, respectively.
  The vertical dashed line (colored red in the electronic edition)
  indicates the estimated ``5$\sigma$" confusion limit 
  of 3.2 mJy based on the MIPS 70um Source Density Criterion for confusion
  (Dole et al. 2003, 2004b).
  When the non-detections (open histogram) are plotted as 1-$\sigma$ 
  rather than measured values, the histogram indeed piles 
  up at this limit (see e.g., lower bounds in Figure 2).  
  \label{fig:fluxhist} 
  }
\end{figure}

Figure~\ref{fig:fluxhist} shows 
for the full sample of stars observed under the auspicies of $FEPS$,
the distribution of signal-to-noise at 70 $\mu$m 
and the distribution of measured
70 $\mu$m flux density, separately for detections and non-detections.
There is significant overlap among the detected and non-detected
flux densities due to source-to-source variation in astrophysical
background, the main sensitivity limitation.  
The typical $<$2-sigma non-detection has
measured flux density about 5-10 mJy (median noise = 9 mJy)
while the typical $>$2-sigma detection has measured flux density
$>$30 mJy (median = 60 and mean = 80 mJy).
The 2-sigma level is used rather than a more stringent 3-  
or 5-sigma threshold in order to identify all reasonable candidate 
70 $\mu$m sources, including those that require confirmation.

Assuming our (estimated) uncertainties are accurate,
for our sample of 328 sources we expect $<$1 to fall above
$+$3-sigma and $\sim$8 to fall above $+$2-sigma if the data
follow a gaussian distribution.
Therefore, $\sim$7 should fall between 2- and 3-sigma. We observe 33  
sources above 3-sigma and 11 sources between 2- and 3-sigma.
Accordingly, possibly 1 of the $>$3-sigma detections
and probably most of the 2-3 sigma detections are noise, and should be
treated with caution. Only a portion of the latter survive our other
cuts for source detection (image visibility and positional alignment),
and are subsequently identified as objects with excess emission. They are  
noted in Table 1 as those also having {\it excess} SNR between 2- and 3-sigma
(except for HD 17925 which is a 4.6-sigma detection
in observed flux density but only a 2.9-sigma significant excess).

\begin{figure}[t]
    \epsscale{1.2}
    \plottwo{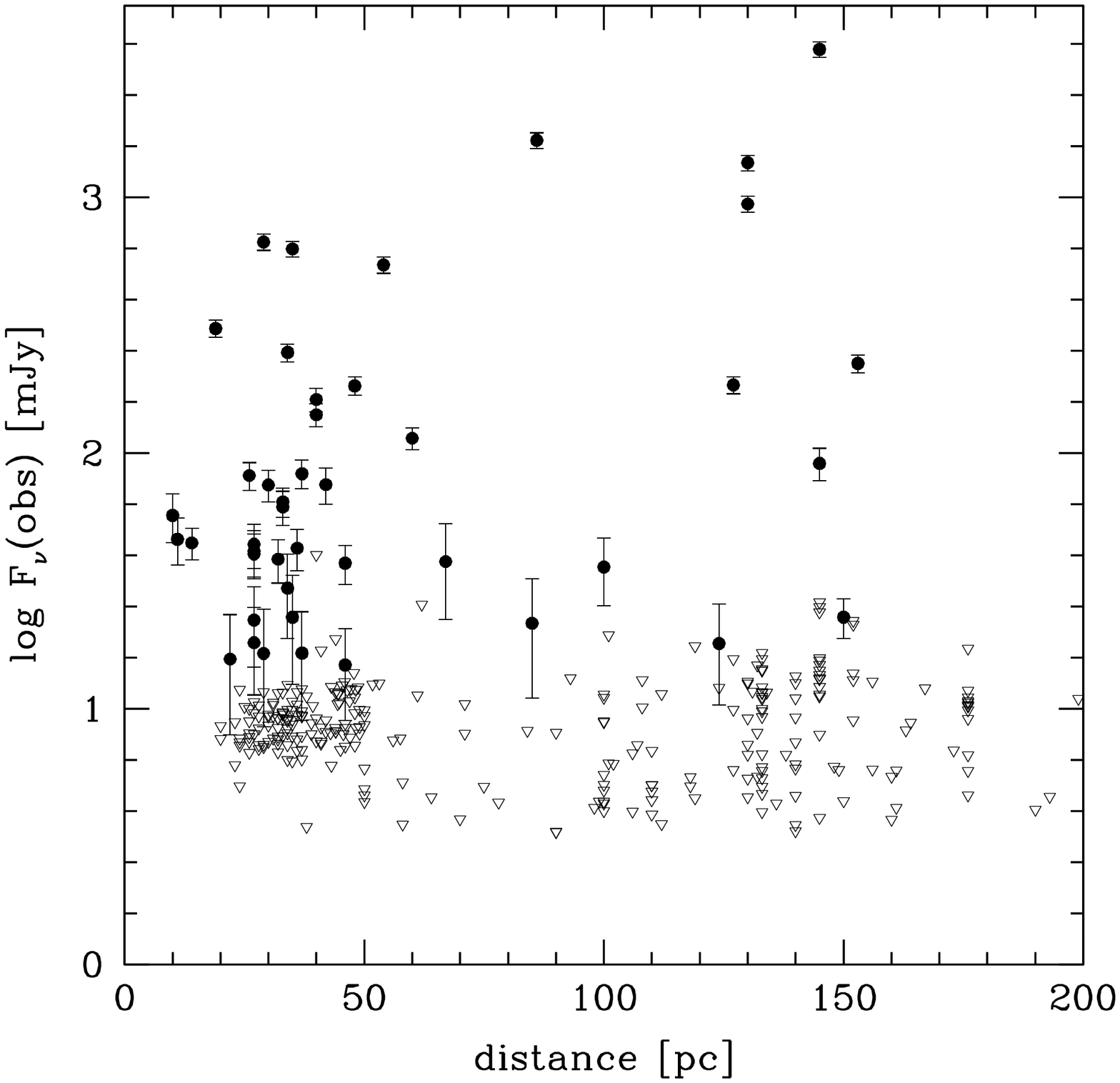}{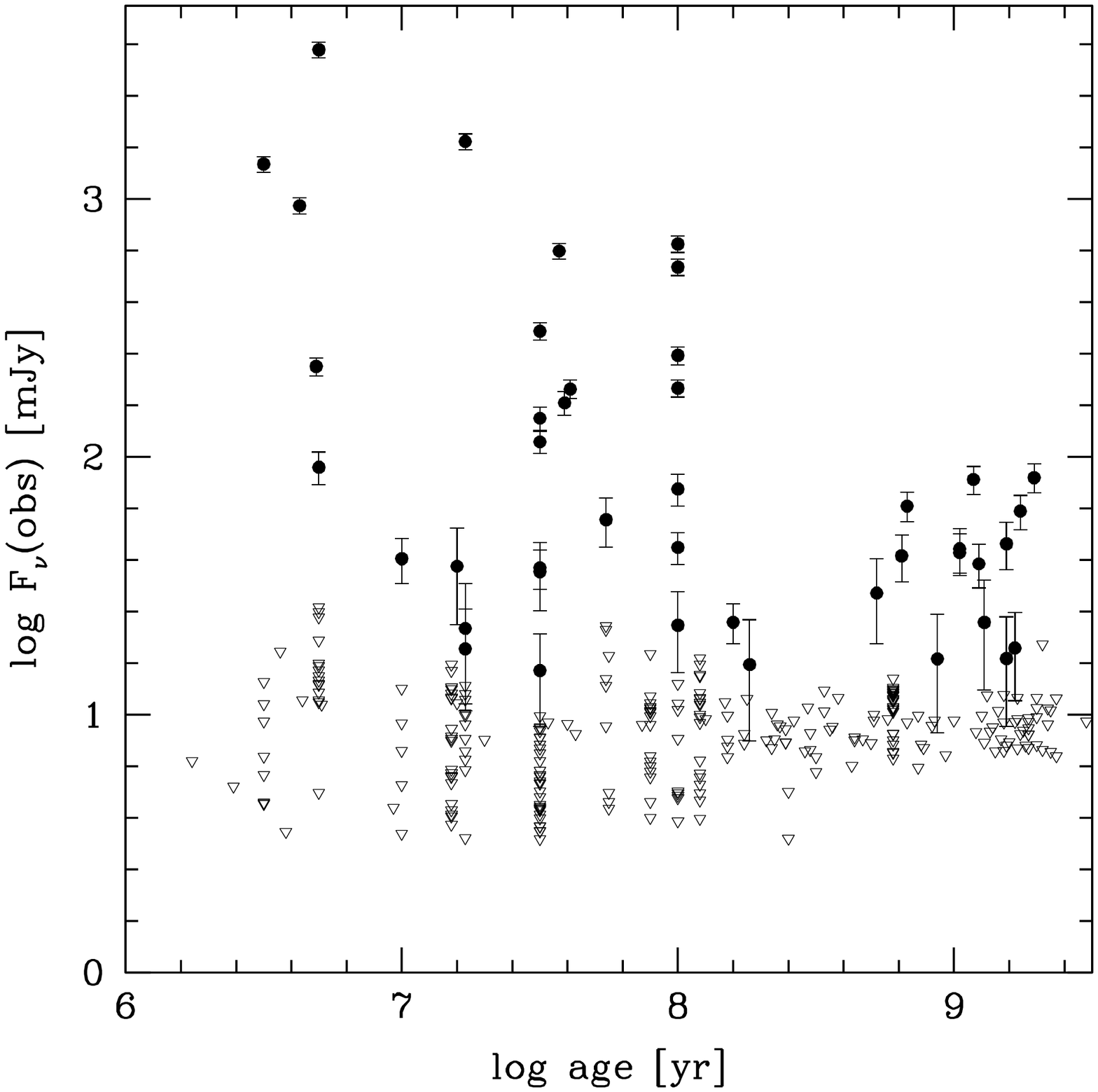}
    \epsscale{1.0}
  \caption{
  Sensitivity of the $FEPS$ 70 $\mu$m data as a function 
  of source distance and age.  Filled symbols with error bars are candidate
  detections (flux density larger than twice the error) while open triangles 
  are upper limits 
  (plotted now at their 1-sigma values rather than the ``measured" values
  illustrated in Figure 1).  Although there is wider scatter in the upper
  limits for distances $>$50 pc and ages $<$300 Myr with approximately
  1/2 of such cases having more sensitive limits than closer and older
  stars, any systematic trends with distance or age in the relative 
  distribution among the upper limits are weak. This is consistent with the
  interpretation that 70 $\mu$m sensitivity is dominated by infrared 
  background and cirrus as intended with our integration time strategy. 
  \label{fig:sens}
  }
\end{figure}

Approximately 10\% of $FEPS$ targets are detected at 70 $\mu$m and the
remainder are undetected, having flux density upper limits.
\footnote{We note that some of the sources represented here as upper
limits at 70 $\mu$m have been identified as having dust excesses from  
IRS
data and subsequently were detected at 70 $\mu$m in deeper follow-up
observations conducted by the $FEPS$ team through GO programs
(see Kim et al. 2008).} In Figure~\ref{fig:sens} we illustrate
the measured flux densities and 1$\sigma$ noise values,
versus source distance and stellar age.  No trends are apparent
in the relative distribution of the upper limits with these variables.
This is consistent with the interpretation that our sensitivity
at 70 $\mu$m is dominated by infrared background and cirrus effects,
as expected.
A K-S test comparing the distributions of
photometric background for the detected and non-detected sources
indicates that they might not be consistent with having been
drawn from the same parent population (P(d) $<$ a few percent).  However,
the distance and age distributions of detected and nondetected sources 
are both consistent with having
been drawn from the same parent based on the K-S test 
($P(d) >$ a few percent), the distances being potentially more
distinguishable than the ages.  The only obvious trend in
Figure~\ref{fig:sens} is that younger sources (which typically are  
more distant than the older sources in our sample) tend to have a 
larger upper range to their 70 $\mu$m flux densities, 
(indicating higher values of $L_{dust}/L_\ast$).  We return
to this point in \S6.1.

\subsection{Cirrus and Confusion Considerations}

We take care to ensure not only the detection of signal above the noise
at 70 $\mu$m, but that the signal is from the intended $FEPS$ target.  
Peaks in the galactic cirrus structure
or confusion with extragalactic contaminants are considerations at 70 $\mu$m.

Each candidate 70 $\mu$m source was thus inspected by-eye to check 
for point-like appearance.
No obvious examples of resolved extended emission, which could
indicate contamination from cirrus, were identified.
Such emission can also be identified via relative photometry in
larger vs smaller apertures.  As reported in Carpenter et al. (2008b),  
two sources are identified with larger than expected flux ratios
in photometry derived from bigger versus smaller apertures.
Both are confused by nearby 70 $\mu$m bright objects which  
were removed by PSF fitting before the final photometry 
of the $FEPS$ target was measured and reported. 
Based on this analysis, cirrus contamination is an
unlikely explanation for the 70 $\mu$m point source detections  
reported here.

The MIPS 70 $\mu$m FWHM is 16" (9.8" pixels) compared to 5".4 FWHM
(2".5 pixels) at 24 $\mu$m. Because of the dependence of diffraction 
on wavelength, a 70 $\mu$m source near the center of the MIPS field 
may not be spatially coincident with the targeted source 
that is detected with a higher accuracy centroid and at higher SNR
at shorter wavelengths.  Unassociated  
contaminants such as AGN and ULIRGs have 70/24 flux density ratios 
of 0.5-3 (e.g. Frayer et al. 2006), similar to those observed 
for our sources (e.g. Figure~\ref{fig:cc} below)
which we interpret as due to circumstellar dust disks.  Thus
comparing the centroids of detections at 24 and 70 $\mu$m 
is particularly important.
The absolute pointing of Spitzer's focal plane array pixel centers is
assessed by the Pointing Control System, which is astrometrically tied
to the 2MASS survey. The 1$\sigma$ uncertainties on the absolute  
pointing
reconstruction are better than
1".4 at 24 $\mu$m ($\sim$1/4 beam) and 1".7 at 70 $\mu$m ($\sim$1/10  
beam).
Therefore, 24 $\mu$m coordinates should be within 1".4 of  
corresponding 2MASS
sources and the difference between 24 and 70 $\mu$m positions should be
$<2".2 = \sqrt{1.4^2+1.7^2}$ in the high SNR limit.
Measured positional offsets thus provide a good, though not robust,
discriminant between associated and unassociated sources.
We use the 24 $\mu$m and 70 $\mu$m images for this comparison
in order to keep the relative investigation to within the Spitzer  
focal plane and free of absolute positional calibration.

\begin{figure}[t]
  \begin{center}
   \plotone{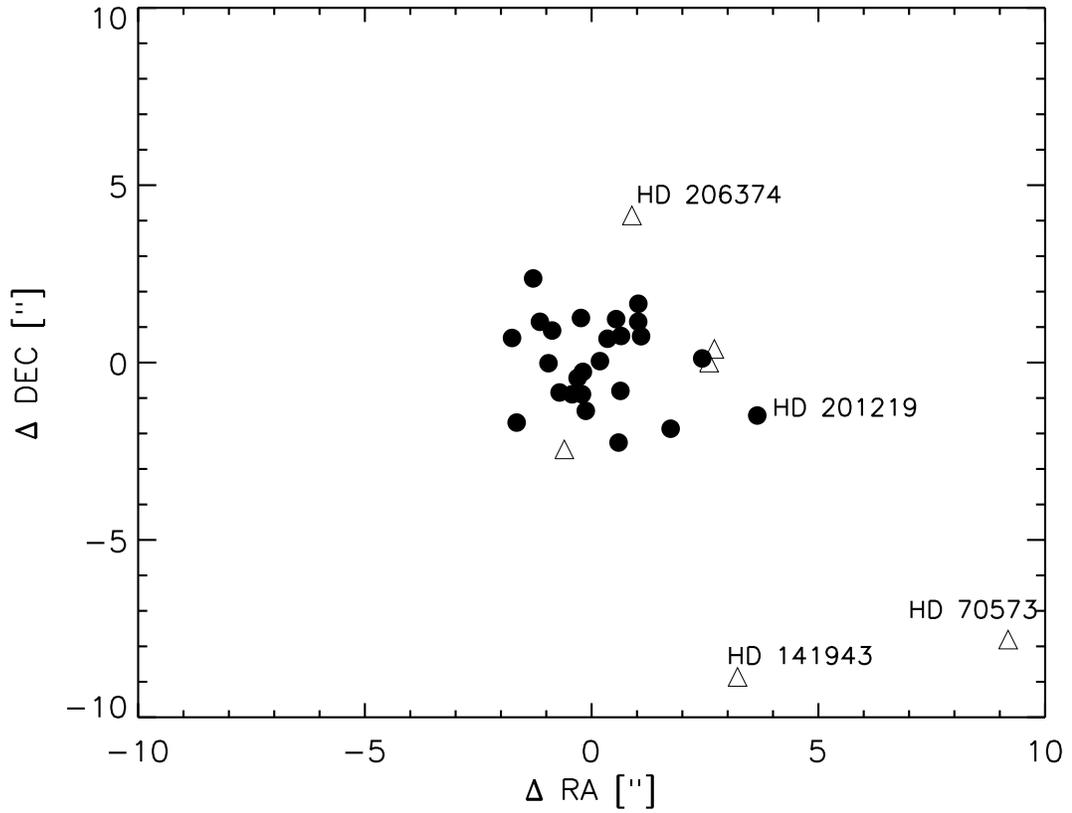}
  \end{center}
  \caption{
  Positional offsets between the centroids of 24 $\mu$m and 70 $\mu$m 
  point sources. 
  Centroiding errors depend on signal-to-noise, which has a large range for
  our sample as illustrated for the 70 $\mu$m data in Figure 1.  
  Filled symbols indicate sources with 70 $\mu$m detection SNR$>4$ 
  while open triangles are lower SNR detections.
  The empirical scatter (1$\sigma$) in the position differences is 
  2".10 in right ascension, 2".53 in declination, and 2".47 in total separation.
  \label{fig:pos}
   }
\end{figure}

Figure~\ref{fig:pos} shows the right ascension and declination
offsets between the 24 $\mu$m and 70 $\mu$m point source positions.
At 24 $\mu$m, pixel positions were determined from Gaussian centroiding and
the corresponding RA and DEC derived from the distortion-corrected
image headers.  At 70 $\mu$m, Gaussian centroiding was applied to a 44"
squared region centered at the expected source position.
In Figure~\ref{fig:pos}
two sources (HD 141943 and HD 70573) are rather large outliers,
$>$1/2 of the 70 $\mu$m beam size, while two
others (HD 206374 and HD 201219) are offset by $\sim$1/4 beam.
The empirical 1$\sigma$ scatter in  Figure~\ref{fig:pos} is 2".47,
roughly 10\% higher than the minimum 2".2 from above. We implement
a cutoff of 2".75 ($<$2 sigma)
to consider a 70 $\mu$m detection as being
coincident with the source seen at 24 $\mu$m.
We retain HD 201219, however, as its known companion,
which has been subtracted for photometry purposes,
still influences the 70 $\mu$m image centroid.
We also retain HD 206374 which is in the low signal-to-noise regime
and thus a large offset is possible.

We also consider the probability of false association of the 70 $\mu $m point
source with the $FEPS$ target even when there is apparent spatial coincidence
with a 24 $\mu$m point source that can be robustly associated 
itself with the intended target -- both positionally and, in many cases,
by having the expected photospheric flux density.
We estimate the probability of a chance superposition 
with a background galaxy that dominates the flux at 70 $\mu$m, adopting
the methodology of Downes et al.  (1986).  For a surface density
of objects $\Sigma(F>B)$ having flux densities, $F$, brighter than $B$,
the probability of finding one of these galaxies
within radius $r$ of the $FEPS$ target is given by the Poisson  
distribution $P = 1 - e^{-\Sigma(F>B)\pi r^2}$.
From Dole et al. (2004a), we expect $\Sigma$
$<1.1\times10^{-1}$, $<1.1\times10^{-2}$, and $<3\times10^{-3}$
galaxies per sq. arcmin at 70 $\mu$m flux density levels
of $>$10, $>$50, and $>$100 mJy, respectively,
or $<$2.1, $<$0.2, and $<$0.06 galaxies per MIPS 70 $\mu$m mosaic
($\sim$20 sq. arcmin after combining several individual 2'.6 x 5'.25 
raster images).  Considering the $FEPS$ data set as a whole, 
approximately 1/3 of all (324 visually examined)
$FEPS$ 70 $\mu$m mosaiced images have an obvious source
somewhere in the field, with $>$1/3 of these 1/3 (or 40 sources)
within 1/2-beam width of the image center, the expected position 
of the $FEPS$ target.  Among the near-coincident sources we observe
0, 18, and 26 sources at flux density levels $<$10, $<$50, and $<$100 mJy
with 14 sources $>$100 mJy.
(including those with the relatively large offsets noted above).
Thus, the probability of a chance superposition of a galaxy within
our search radius of 2".75 emitting $F_{70 \mu m} >= 10, 50, 100$ mJy is
$P = 7.26\times10^{-4}$, $7.25\times10^{-5}$, $1.98\times10^{-5}$ per object.
This corresponds to a probability of $<$24\%, $<$2\% and $<$1\%
that one target is  contaminated at the $>$10, $>$50, and $>$100 mJy 
levels respectively, for the whole sample of 328.

At 24 $\mu$m, the faintest detection among our 328 $FEPS$ targets is $F_{24 \mu m}
\sim$ 1 mJy.  The 24 $\mu$m surface density due to extragalactic objects
at 1 mJy is $\Sigma \sim 0.15$ per sq. arcmin (Papovich et al. 2004),
implying a maximum probability per star of $P = 9.9\times10^{-4}$ for
a faint galaxy within 2".75 contributing to the measured 24 $\mu$m 
flux density (using the radius appropriate to 
the agreement between the 24 $\mu$m and 70 $\mu$m positions; Figure ~\ref{fig:pos}).
Despite having a higher potential than at 70 $\mu$m for extragalactic
contamination, our 24 $\mu$m sources in most cases have
flux densities consistent with expected photospheric emission; this
argues that they indeed emanate from the intended $FEPS$ target.
For those with measured excesses at the mJy level, there is a
$<$ 33 \% chance that one source suffers extragalactic contamination
among the sample as a whole.

We conclude that the 70 $\mu$m emission,
distributed in flux as we have shown in Figure~\ref{fig:fluxhist},
and emanating from the same source as the 24 $\mu$m emission 
as we have shown in Figure~\ref{fig:pos},
is likely associated directly with the targeted $FEPS$ stars.

\subsection{Color-Color Diagrams}

\begin{figure}[t]
  \begin{center}
   \plotone{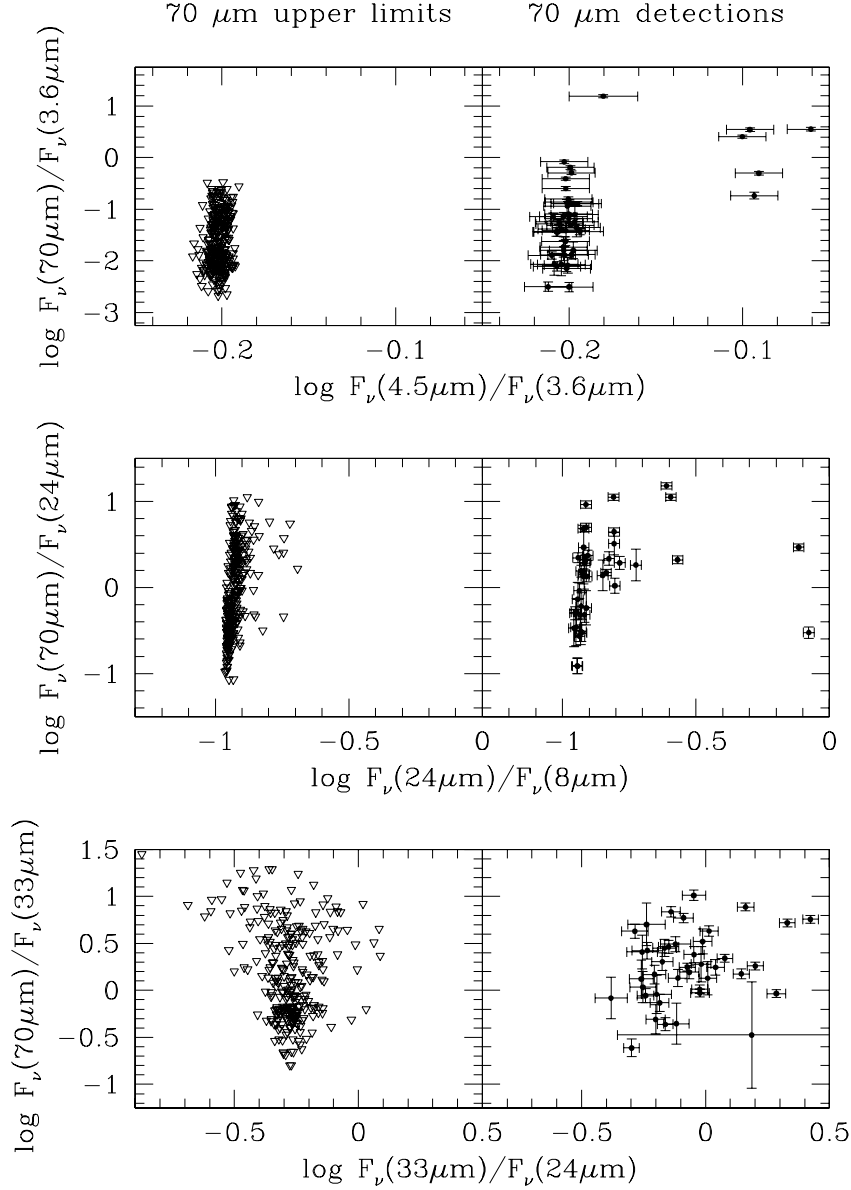}
  \end{center}
  \caption{Color-color diagrams highlighting 70 $\mu$m photometry.
Sources detected at SNR(70 $\mu$m) $>$2 are distinguished
in the right panels from the non-detections in the left panels;
the latter are plotted as 1-sigma upper limits to illustrate the admixture
in color among sources in the $FEPS$ sample between 70 $\mu$m signal and 
70 $\mu$m noise. Note the horizontal broadening of the data distribution 
in the 24-33 $\mu$m color (bottom) panels relative to the upper panels. 
The source with large error bar in lower right panel
is due to a particularly noisy long wavelength IRS spectrum; see Table 2. 
  \label{fig:cc}
  }
\end{figure}

Color-color diagrams are an efficacious way to identify objects with
unusually red colors due to circumstellar dust.  In Figure~\ref{fig:cc}
we show several different flux ratios
involving the 70 $\mu$m band observed with Spitzer.  As mentioned above, 
the $FEPS$ 70 $\mu$m data are dominated by upper limits.  For clarity,
we therefore indicate separately the maxima and the measured
colors involving 70 $\mu$m photometry.  
Although we can not use exclusively these color-color 
diagrams to identify 70 $\mu$m excess sources, we can employ them in a  
rudimentary assessment of the hot, warm, and cold dust components in the
circumstellar environments of $FEPS$ sources.

The top panels of Figure~\ref{fig:cc}
show 4.5/3.6 $\mu$m and 70/3.6 $\mu$m flux density ratios;
the abscissa is approximately photospheric for the great majority of $FEPS$ stars
while the ordinate is sensitive to cool dust.  The few red outliers in the
4.5/3.6 $\mu$m flux density ratio are also amongst
the reddest objects in the 70/3.6 $\mu$m flux density ratio, as expected
if they have both hot inner and cool outer dust.  These sources exhibit
evidence for primordial (gas--rich) disks.
In contrast to the narrow 4.5/3.6 $\mu$m flux density ratio,
there is a large range in the 70/3.6 $\mu$m flux density ratio for
those stars detected at 70 $\mu$m (top right panel), but an
equally large range in the distribution of color limits (top left panel).
Notable is the admixture along the ordinate of the detections and upper limits.
Even accounting for the fact that the limits are plotted at
1$\sigma$ levels (consistent with Figure 2) while the detections
are all $>$2-3$\sigma$,  the most stringent upper limits in the top left panel
appear a factor of several lower compared to the detections reported
in the top right panel.  Such variation along the ordinate among the
detected sources likely reflects  real differences in debris disk properties.
Recall, however, as argued above based on K-S statistics, that
variation in source background may play a significant role in 70 $\mu$m
detection despite our attempts to observe the lowest background sources
of given age and distance.

The middle and lower panels of Figure~\ref{fig:cc} illustrate
70/24 vs 24/8 $\mu$m and 70/33 vs 33/24 $\mu$m flux density ratios.
Again, by comparing the left and right panels it can been seen 
that 70 $\mu$m detections are interspersed in color
with 70 $\mu$m upper limits.  Further, a subset of the stars
is redder in the 24/8 $\mu$m and/or 33/24 $\mu$m flux density ratios 
compared to the bulk of the sample.  These are ``warm" excess sources.
Some but not all such objects are also detected at 70 $\mu$m, 
which enables better constraints on the bulk dust characteristics 
than in cases in which the excess is detected in only a single band.

Our focus in this paper is on the sub-set of objects with excesses
detected at 70 $\mu$m. Typically these sources are blue 
along the abscissae of Figure~\ref{fig:cc}, implying that they are
close to photospheric at wavelengths shorter than 24-33 $\mu$m.
Several of the brightest debris disks in our sample
(specifically, HD 61005, HD 107146, HD 38207, HD 191089, HD 104860)
can be distinguished in the color-color plots;
however, additional analysis is needed to identify most debris disk candidates.

\subsection{Excess Signal-to-Noise Histograms}

The majority of sources detected at 70 $\mu$m are dominated
by the circumstellar contribution to the flux density.
However, the photospheric contribution at 70 $\mu$m is not
negligible for all sources, and
must be modeled accurately in order to characterize the excess.
We employ a Kurucz model of the underlying stellar photosphere
in order to more robustly identify individual objects with 70 $\mu$m excess 
than is possible from color-color diagrams and to
analyze the signal-to-noise in the excess.

As described in more detail by Carpenter et al. (2008b),
available BV (Johnson, Tycho),
$vby$ (Stromgren), H$_p$ (Hipparcos), RI (Cousins), and JHK$_s$ (2MASS)
photometry data were used in combination with initial estimates of temperature,
surface gravity, and metallicity based on spectroscopic data
from the literature, to find a best-fit Kurucz model. 
Kurucz model flux densities were converted
to magnitudes in each of the available optical/near-infrared filters
via multiplication with the combined filter, atmospheric transmission, and
detector response curves as in Cohen et al. (2003a,b and references therein).
In general, surface gravity and metallicity were fixed at log g = 4.5 cm/s$^2$
and [Fe/H]=0.0, and the effective
temperature and normalization constant were the fitted parameters.
The line-of-sight extinction was fixed to A$_V=0.0$ for stars within 75 pc;
beyond this distance A$_V$ was initially estimated from the literature
but then varied as a free parameter in the fits for all stars not in clusters
or with estimated ages younger than 30 Myr (which may suffer some
obscuration). Best-fit was defined in a least-squared sense.  
The formal uncertainty in the resulting
photospheric projection to the Spitzer bands is typically 2-3\%.

\begin{figure}[t]
  \begin{center}
   \epsscale{0.8}
   \plotone{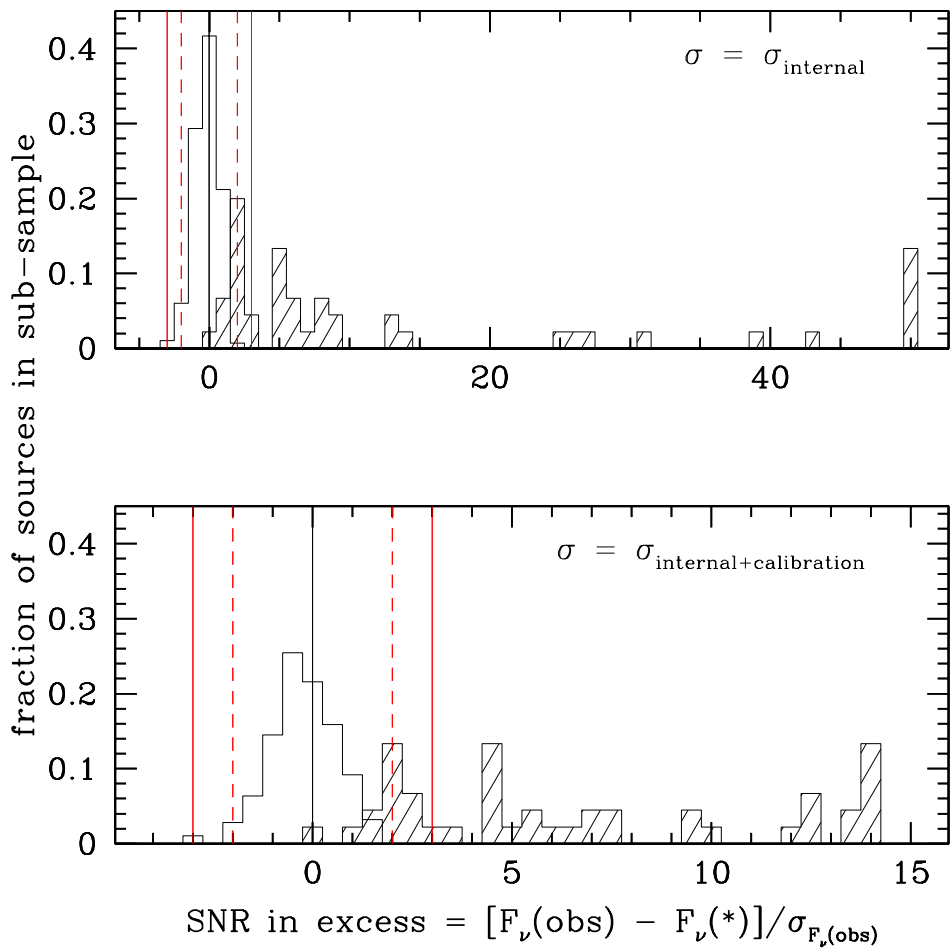}
   \epsscale{1.0}
  \end{center}
  \caption{ 
  Signal-to-noise in the 70 $\mu$m excess following subtraction of a
  photospheric model from the observed photometry. Top panel
  considers only internal errors in computing excess SNR, 
  while bottom panel considers both internal (random) and calibration 
  (systematic) errors.
  Vertical lines indicate excess SNR=0 (black) and $\pm$2,3 (gray, 
  or colored red in the electronic edition).  
  Note that plots are fractional for each of the two subsamples, which
  separately add across the bins to unity.
  Open histogram represents sources not detected
  at greater than 2-sigma at 70 $\mu$m. The excess SNR of this sample 
  is centered near zero excess (though formally negative, with 
  mean/median about -0.2, perhaps suggesting we have over-estimated
  the contribution from the stellar photosphere) and has a roughly Gaussian
  distribution as expected from pure noise; no sources have
  excess SNR $<$ -3 and seven have excess SNR between -3 and -2.
  Hatched histogram includes only those $FEPS$ objects 
  convincingly detected at 70 $\mu$m, as detailed in the text,
  and is biased towards significant positive excess.  HD 13974
  is the only statistically significant photospheric detection 
  at 70 $\mu$m among $FEPS$ objects (flux density SNR = 5 and
  excess SNR = 0.88/0.82 from top/bottom panels).  HD 216803,
  observed as part of a GTO program, is a also a detected
  photosphere at flux density SNR=5 and excess SNR = -0.16/-0.15 
  in the top/bottom panels. Thirty-one $FEPS$   
  sources have excess SNR $>$ 3 and six have excess SNR between 2 and 3.
  \label{fig:excesshist}
  }
\end{figure}

With a model of the expected photospheric flux, the excess above the
photosphere is computed as the difference between the observed and
Kurucz in-band flux densities. The signal-to-noise in the excess is defined
as this difference divided by the root-sum-squared error in the
observed flux densities and the photospheric projection.
Figure~\ref{fig:excesshist} shows histograms of the resulting 
signal-to-noise {\it in the excess} at 70 $\mu$m.
The excess SNR distribution 
for the nondetections and indeed for the full $FEPS$ sample 
is peaked near zero, suggesting the expected dominance by photometric noise
at this wavelength. The median, mean and dispersion of the distribution
are: -0.25, -0.22 and 0.84, in units of SNR.  
The $>2\sigma$ 70 $\mu$m excess sources have
median, mean and dispersion of 6, 15, and 17 in units of SNR.
Using only the internal uncertainty, the significance of the detected
excesses can be as high as SNR=50, while using the total uncertainty
(root-sum-squared of common calibration and individual internal
uncertainty terms), no source has excess SNR$>$10

We have defined a sample of 70 $\mu$m excess sources as follows.
The excess signal-to-noise distribution of Figure ~\ref{fig:excesshist}
is centered near (but not exactly at) zero with dispersion that is close to
(but not exactly) the value of unity that would be expected from Gaussian noise
(including the imposition of an additional 50\% scale factor in
the 70 $\mu$m flux density uncertainties as discussed in \S2).
We therefore consider most robust those sources which have $>3\sigma$
(formally 99.6\% confidence) excesses at 70 $\mu$m when they are apparent from
{\it both} the internal--only and total uncertainty assessments.  These are
our ``Tier 1" sources.  Our ``Tier 2" sources are those with $2-3\sigma$
excesses at 70 $\mu$m. That the mean in Figure ~\ref{fig:excesshist}
is significantly (with respect to the error in the mean) negative 
suggests a systematic offset with respect to the Kurucz models,
in the sense that we are somehow over-correcting for the photosphere.
This may indicate that some of the 2-3$\sigma$ excess sources
we have designated in Table 1 are in fact slightly more significant,
by 0.22 sigma, than our estimates.

Note in Figure ~\ref{fig:excesshist}
that no sources have excess SNR $<$ -3 and seven have excess SNR
between -3 and -2, while thirty-one sources
have excess SNR $>$ $+$3 and six have excess SNR between $+$2 and $+$3.
\footnote{
We can compare these {\it excess detection} numbers (31 and 6) to
the {\it source detection} numbers at these same significance
levels (33 and 11, as reported in \S3.1).
}
The number of 2-3$\sigma$ significant excess
sources is seemingly consistent with random noise,
both empirically and from gaussian statistics.
We note these sources with caution, and distinguish them
clearly as ``tier 2" objects in the remainder of this paper.

\subsection{Summary and Spectral Energy Distributions}

In summary, we find 31 primordial and debris disk
targets with excess $SNR_{70~ \mu{\rm m}} > 3$.
Excess $SNR_{70~ \mu{\rm m}}$ between 2 and 3 is measured
for an additional 6 candidate debris excess objects.  These sources
meet, in addition to the flux density criteria, the point-like
appearance and positional coincidence requirements stated earlier.
In Table 1 we present the 70 $\mu$m excess sources selected as described
above, along with the stellar parameters
(distance, spectral type, luminosity, and age as adopted by $FEPS$).
In Table 2 we present corresponding
Spitzer photometry (measured flux densities, uncertainties) and in Table 3
the calculated excesses and significances above
the adopted model stellar photospheres at 13, 24, 33, 70, and 160 $\mu$m.
Of the sources selected to have 70 $\mu$m excess attributed to debris dust,
more than half, less than half, and a single source
also have significantly measured excesses at 
$>$33, 24, and 13 $\mu$m respectively; none of the debris disk candidates
has excess detected at 8 $\mu$m.  
\footnote{Carpenter et al. (2008a) discuss additional sources 
within our 70 $\mu$m excess sample with low amplitude excesses 
at wavelengths $<$35 $\mu$m which were not apparent from our 
analysis comparing to Kurucz models. 
}
The excess amplitudes at 70 $\mu$m
range from 1.6 to over 300 times the photosphere
(median is 20 times photosphere), while at 33 $\mu$m the median 
excess amplitude is equal to (100\% of) the photosphere,
at 24 $\mu$m it is 40\% of the photosphere, and
at 13 $\mu$m 17\% of the photosphere.

Several sources deserve specific comment.  First,
some objects selected for the $FEPS$ probe of disk gas evolution 
based on claimed IRAS- or ISO-based 60 and/or 90$\mu$m excesses
are not confirmed from this analysis with Spitzer.
These include ScoPMS 214
\footnote
{Carpenter et al. (2008a) find that this source has weak MIPS/24 $\mu$m and IRS excess}, HD 41700, HD 216803, and HD 134319, which
were discussed in Pascucci et al. (2006).
We include these 4 objects in Table 1 for
completeness, but they do not appear in subsequent Tables or Figures.
Second, there are additional sources selected for the gas
experiment for which $FEPS$ is not obtaining 70 $\mu$m observations because
the objects are part of GTO programs with MIPS
(HD 216803, HD 202917, HD 17925, HD 72905).
These objects exhibit excess emission based on
Spitzer data and are included in Table 1 and our subsequent analysis.

\begin{figure}[t]
  \begin{center}
    \epsscale{0.9}
    \plotone{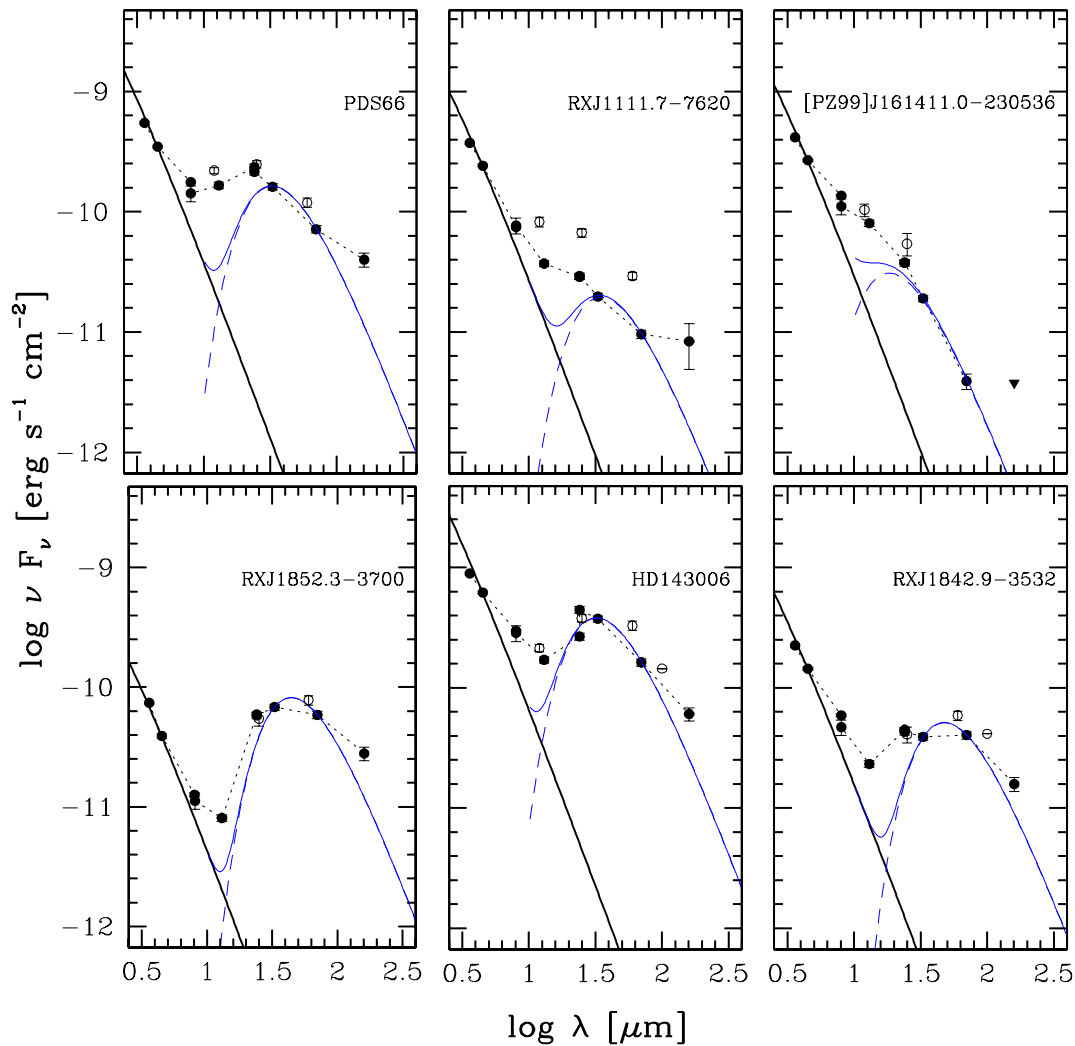}
    \epsscale{1.0}
  \end{center}
  \caption{
  Spectral energy distributions for the 70 $\mu$m excess sources 
  in the $FEPS$ ``primordial" disk sample.
  We use the ``average weighted" wavelengths 
  for the Spitzer 3.6, 4.5, 8.0, 24, 70, and 160 
  $\mu$m bands and include synthetic photometry points at 
  13, 24, and 33 $\mu$m created from the IRS spectra.  
  No color corrections have been applied; see text.  
  Error bars are indicated but generally 
  are smaller than the plotted points.  Open symbols are IRAS measurements.
  Light solid curves (colored blue in the electronic edition)
  are single temperature blackbody fits to the 33-70 $\mu$m color 
  excess (dashed curve) summed with the photosphere (heavy solid curve).  
  In all cases, the observed flux densities (connected by dotted lines) 
  are broader than a single temperature blackbody
  and indicate cooler outer disk material as well as 
  warmer inner disk material in addition to material at the distance of
  the plotted fiducial single temperature.  At the same time, the shortest
  wavelength near-infrared bands are photospheric, indicating that the disks 
  do not extend inward of about 0.2-0.5 AU,
  with the excesses beginning by $\sim 8 \mu$m. Detailed modeling of these
  sources will be presented elsewhere (e.g. Cortes et al. 2008 for PDS 66).
  \label{fig:sedsprim}
  }
\end{figure}

\begin{figure}[t]
  \begin{center}
    \epsscale{1.2}
    \plottwo{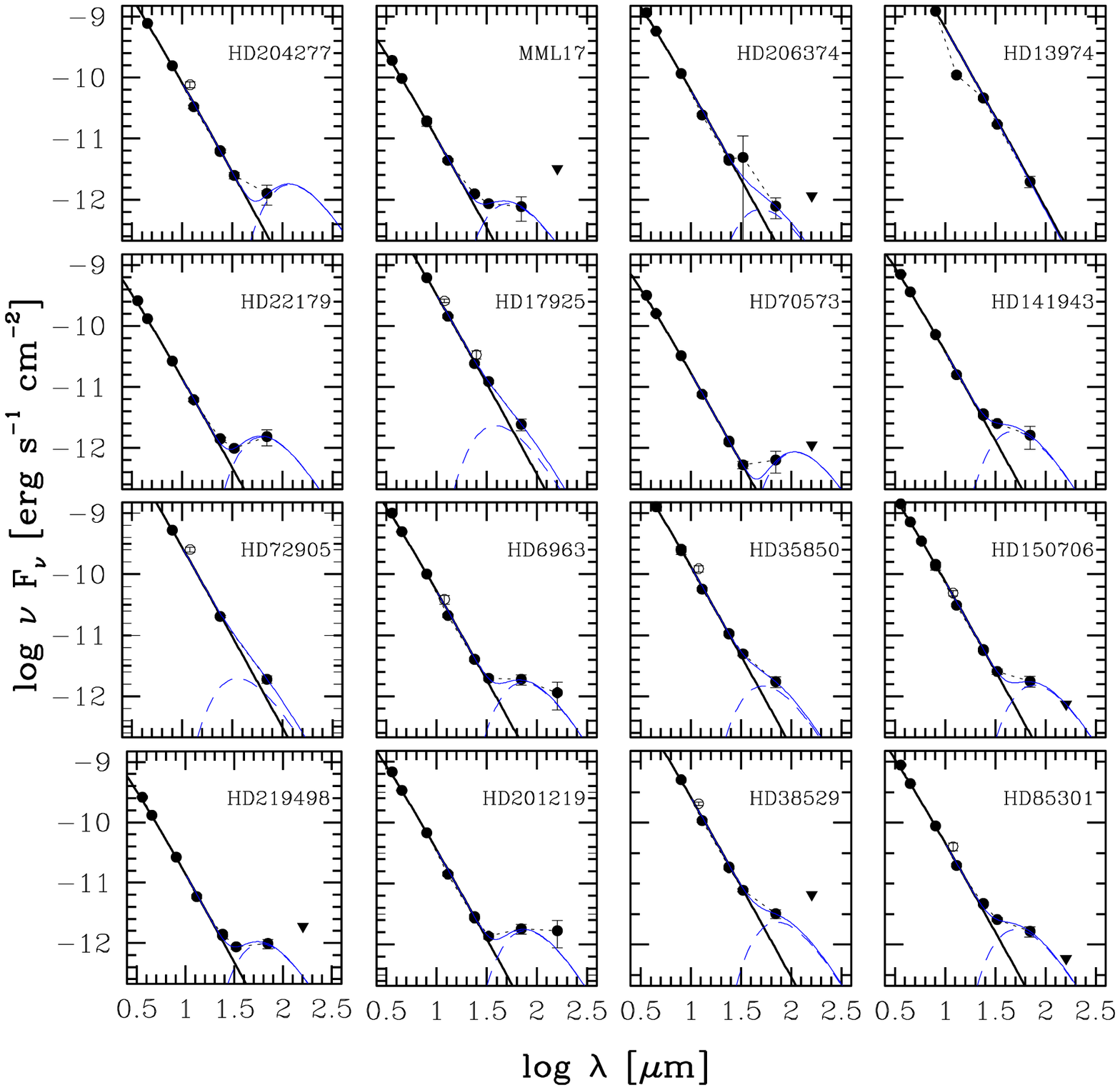}{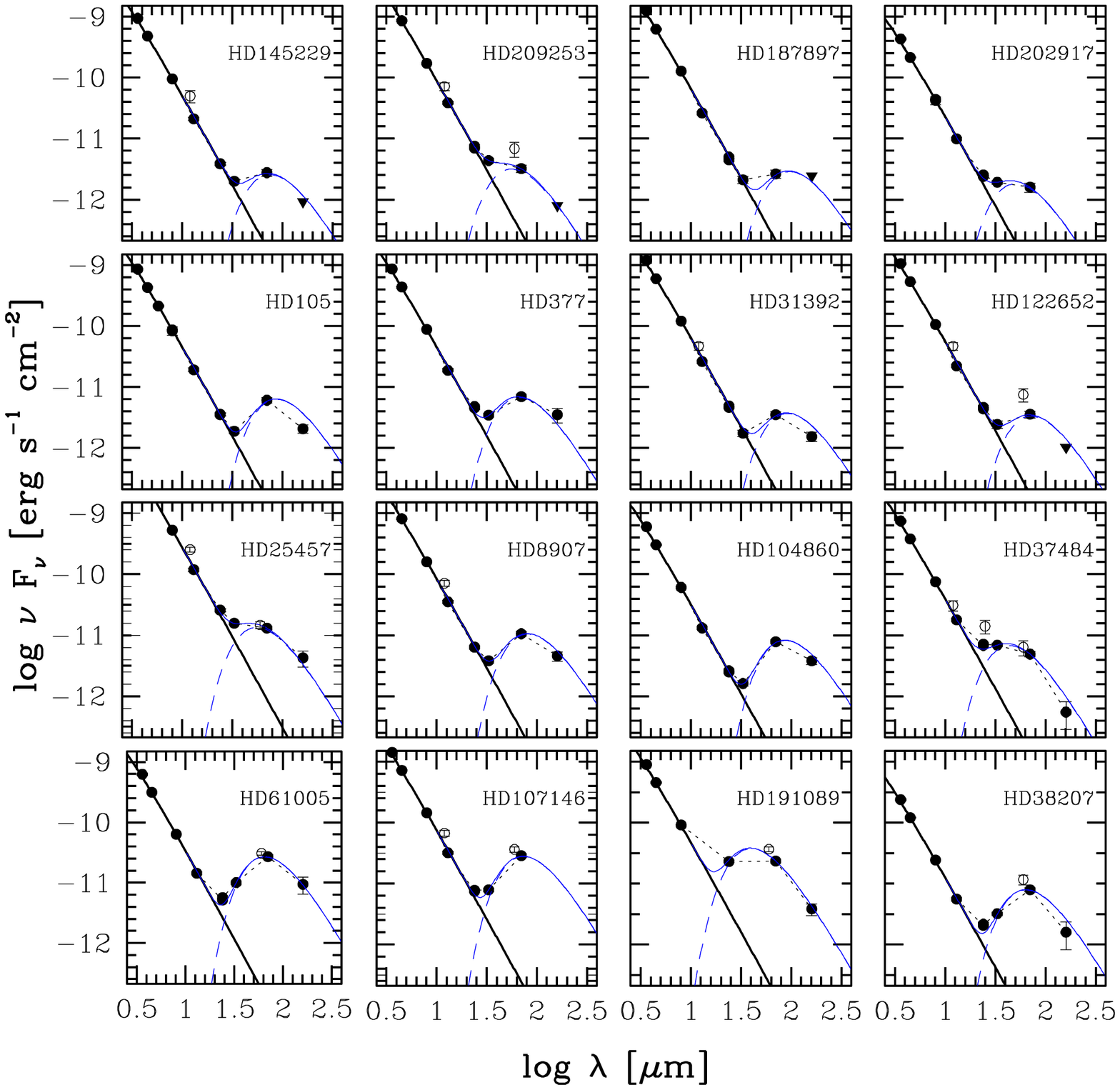}
    \epsscale{1.0}
  \end{center}
  \caption{
  Spectral energy distributions for the 70 $\mu$m excess sources 
  in the $FEPS$ debris disk sample, having $>3\sigma$ significance
  in 25 cases and $>$2 but $<3\sigma$ signficance in 6 cases. 
  Objects are ordered most significant excess at bottom left  
  to least significant excess in top right (see Table 3).
  Symbols and lines are as in Figure~\ref{fig:sedsprim}.  
  Single temperature blackbody fits are generally to the 33-70 $\mu$m 
  color excess and slightly under-predict the 24 $\mu$m excess while
  over-predicting the 160 $\mu$m excess (when detected).  
  The fits are to the 24-70 $\mu$m color excess
  for HD 191089 and HD 72905 due to the absence of 33 $\mu$m
  photometry, and for HD 206374 due to the poor quality
  of IRS spectrum just around 33 $\mu$m.
  \label{fig:seds}
  }
\end{figure}

In Figures~\ref{fig:sedsprim} and \ref{fig:seds} we present spectral energy
distributions for the $FEPS$ 70 $\mu$m excess sources.
Included are ground-based data from Tycho and
2MASS along with newly presented Spitzer IRAC, IRS, and MIPS photometry
and IRS spectrophotometry.  Simple blackbody dust models as discussed
in the next section (\S4) are also overplotted.

There are six $FEPS$ 70 $\mu$m excess sources shown in Figure~\ref{fig:sedsprim}
with large excesses that are broad in wavelength and
associated with some of the youngest stars in our sample. These
are likely primordial disks
(see Silverstone et al. 2006 and Bouwman et al. 2008).
While the four strongest (in terms of monochromatic excess) of these six
are in fact the largest 70 $\mu$m excess sources amongst the entire
$FEPS$ sample, two of the six
($[$PZ99$]$ J161411.0-230536 and RX J1842.9-3532) are weaker at 70 $\mu$m
and have inferred $L_{disk}/L_*$ lower than several (much) older debris disks.

\begin{figure}[t]
  \begin{center}
    \epsscale{1.2}
    \plottwo{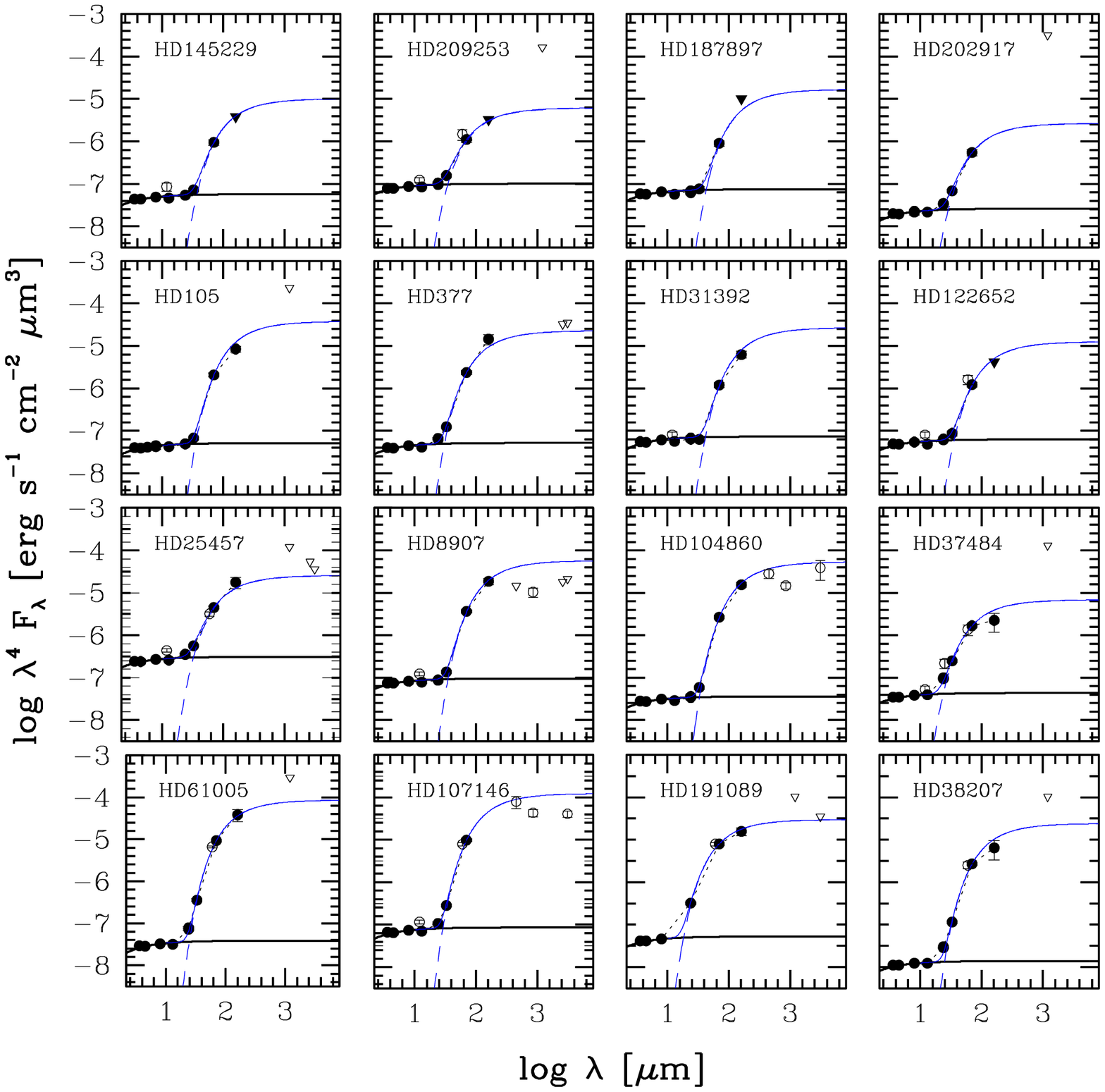}{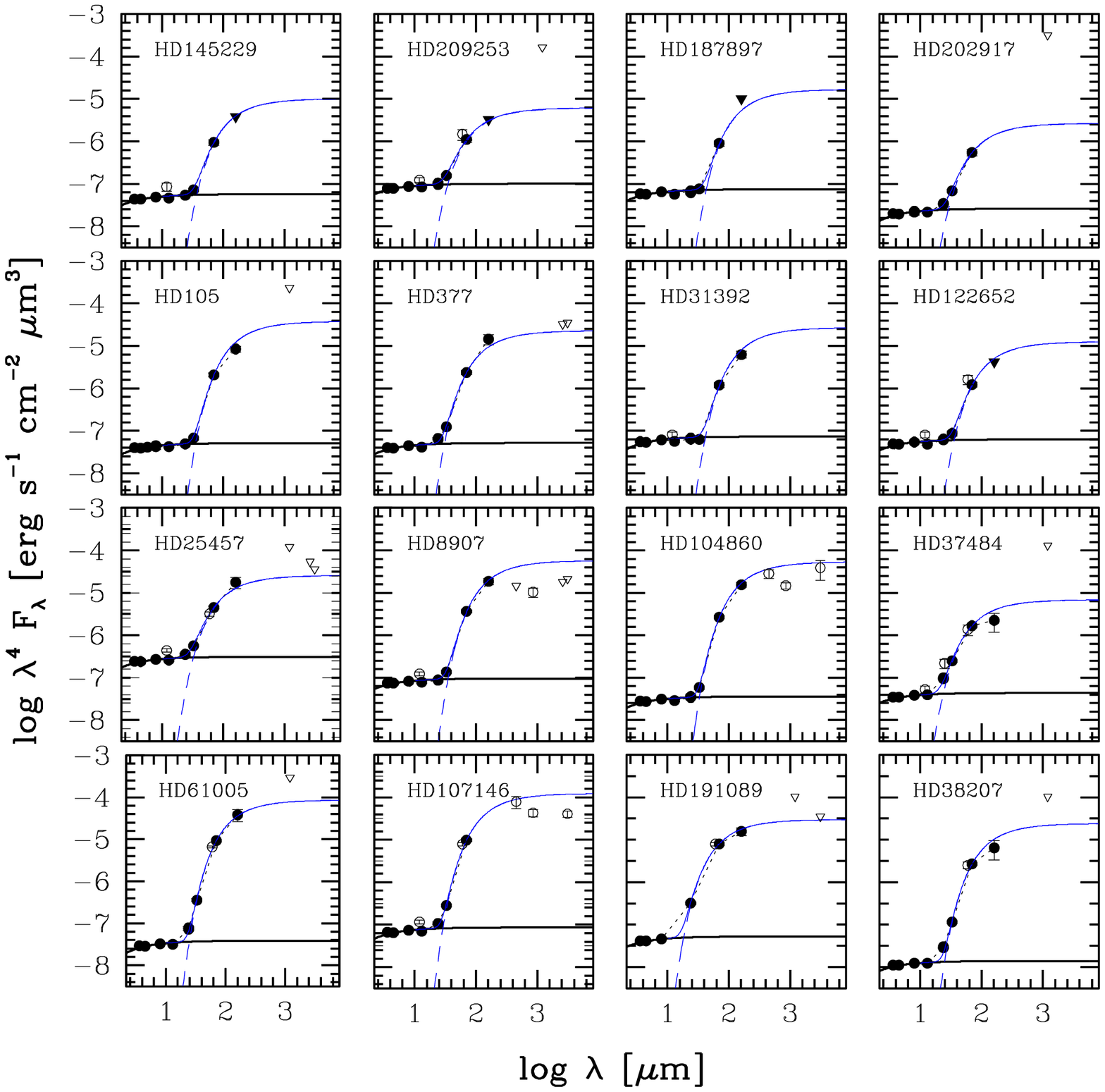}
    \epsscale{1.0}
  \end{center}
  \caption{
  Spectral energy distributions for the same Spitzer 70 $\mu$m excess sources 
  of Figure~\ref{fig:seds}  now plotted in units such that the long wavelength
  Rayleigh-Jeans regime of
  a blackbody function is flat.  Additionally, the abscissa has been extended 
  to mm wavelengths so as to demonstrate the need for models more complex
  than single temperature blackbodies 
  to match available long wavelength photometry. 
  In cases where the flux is over-predicted, 
  this is generally achieved using modified blackbodies for 
  which the temperatures would be similar to those derived 
  from our mid-infrared fitting, but the
  far-infrared to mm slope could be used to constrain the spectral index $\beta$.
  }
  \label{fig:flatseds}
  \end{figure}

There are 25 $FEPS$ 70 $\mu$m excess sources shown in Figure~\ref{fig:seds}
which are debris disk candidates having $>3\sigma$ significance in
the 70 $\mu$m excess.  A further 6
have $>2\sigma$ but $<3\sigma$ significance (see Table 1).
The spectral energy distributions are photospheric over several octaves
in wavelength with evidence of infrared excess only longward of 13 $\mu$m.
Of these 31 total sources, 7 were presented in various earlier $FEPS$ papers
and 10 in literature previous to that; thus 14 debris disks
are newly announced here from $FEPS$.
For this ensemble, detections and upper limits at sub-mm and mm
wavelengths, where available from other investigations of $FEPS$ targets
(e.g. Williams et al. 2003; Carpenter et al. 2005, Najita et al.  2005), 
are included in Figure~\ref{fig:flatseds} which plots the energy distributions
in units where the long wavelength Rayleigh--Jeans tail of the Planck function 
is flat.


\section{Debris Disk Modeling}

Having selected a sample of objects likely to be surrounded by cool dusty 
material, we proceed in this section to model the plausible radial 
distribution of the dust around these stars using basic assumptions.
We take the simplest possible approach to modeling the data and
add complexity only as warranted.  
We consider the scenario in which a dust grain of given size and composition
is in thermal equilibrium with the stellar radiation field. We assume emission
from optically thin ensembles of grains, 
which we justify {\it post facto} by the resulting low
fractional excess luminosities (L$_{dust}/L_* < 10^{-3}$).

First we consider single temperature 
blackbody fits, which have the minimum number of free parameters,
to the observed excess emission (\S{4.1}), then we explore 
multi-temperature models for selected sources (\S{4.2}).  
In \S{4.3} we summarize results from more detailed modeling 
pursued elsewhere within the $FEPS$ program using sophisticated 
radiative transfer dust models with many free parameters, including
grain size distributions. 
In \S{4.4} we discuss upper limits on the amount of dust potentially
located interior to our inferred inner disk annuli.

\subsection{Single Temperature Models}

The observed excesses are most prominent 
at wavelengths around 70 $\mu$m, as illustrated in Figure~\ref{fig:seds}. 
High precision Spitzer photometry at shorter wavelengths 
generally samples the Wien side of the blackbody function. 
$FEPS$ data include IRS spectrophotometry which represent a higher 
resolution sampling of the spectral energy distribution
from 5-35 $\mu$m; these data allow accurate determination of the wavelength 
at which the departure from a photospheric model occurs as
presented in Carpenter et al (2008a). 
Here, we use simple blackbody fitting to color temperatures
(including synthetic IRS-13 and IRS-33 $\mu$m bands). 
Many of the debris disk sources are detected at 160 $\mu$m as well,
providing information past the excess peak (Rayleigh-Jeans regime). 

\subsubsection{Dust Temperature}

We calculate color temperatures $T_{color}$, 
defined as the blackbody temperature required to fit the flux ratios
in the excess above the photosphere, at 24-33, 33-70, and 70-160 $\mu$m. 
We also tabulate 13-33 $\mu$m color temperatures, which we choose over
13-24 $\mu$m for two reasons: first, data from a single instrument are used, 
avoiding systematics due to calibration, and second, in practice
the 13-33 $\mu$m flux ratio provides a tighter upper limit on the maximum
color temperature than the 13-24 $\mu$m flux ratio.
The color temperatures are considered measured values 
when the excess is $\ge2\sigma$ at both the shorter 
and longer wavelengths, and limits when 
the excess is $<2\sigma$ at one of the two wavelengths but 
$\ge2\sigma$ at the other.  For example,
some stars have 33 $\mu$m photometry consistent with purely 
photospheric emission; in calculating a 33-70 $\mu$m color temperature 
we are assuming, therefore, that the infrared excess begins 
just longward of 33 $\mu$m.  In such cases we determine 
the maximum color temperature from the minimum
wavelength of infrared excess onset.  The significance values include 
total uncertainty on the photometry (internal measurement plus calibration  
error) and the formal uncertainty on the photosphere (2-3\% 
is typical).

The various $T_{color}$ fits are given 
in Table 4, along with the $\chi^2_\nu$ values resulting 
from a fit of a blackbody having this temperature to the broader 
excess spectral energy distribution from 13-160 $\mu$m; the number 
of data points used to calculate reduced $\chi^2$ is 4 or 5 in almost all cases.  
Mean (median) color temperatures for 24-33, 33-70, and 70-160 $\mu$m
are 101.6 (92.5)K, 73.1 (59)K, and 61.6 (56.5) K, respectively, 
including the limits, which implies that the mean (median) values
above are also upper limits.  As cited in the MIPS data
Handbook (Table 3.11 in version 3.2), color corrections 
for source temperatures of 50-100K are in the range 2-11\% 
depending on photometric band (24, 70, or 160 $\mu$m). The color 
temperatures calculated for color-corrected photometry are different by only
0.5-3 K from those calculated without the inclusion of color terms; in most
cases these are within or comparable to the formal errors on the 
color temperature as calculated
from the photometric/photospheric uncertainties (see Table 4). As our
blackbody analysis is meant to be illustrative of the dust properties 
characterizing our debris disk sample rather than definitive, we have
not applied color corrections to individual sources. 
Given the systematic differences in color temperature across the
spectral energy distribution of some of the excesses, this seems prudent.

As is evident from the Table, short wavelength excesses are rare among
the $FEPS$ sources with 70 $\mu$m excess. None (among the debris disk sample)
exhibit $<$8 $\mu$m excess within our errors.  Only HD 202917 exhibits 
possible 13 $\mu$m excess at a level $>2\sigma$ (just 2.2$\sigma$). 
All other 13-33 color temperatures in Table 4 are upper limits, 
and produce very poor $\chi^2_\nu$ values when used to fit the overall 
energy distribution as might thus be expected.
At 24 $\mu$m, approximately 1/2 of the 70 $\mu$m excess sources are also 
in excess, while at 33 $\mu$m approximately 2/3 of the 70 $\mu$m excess sources 
are also in excess; thus about 1/2 of the 24-33 $\mu$m 
and 1/3 of the 33-70 $\mu$m color temperatures are upper limits.  

In some cases the color temperatures derived from the flux ratios
at different wavelengths agree quite well, for example HD 104860, HD 8907
and HD 209253.  In other cases, such as HD 377, HD 38207 and HD 85301, 
the three color temperatures 
are very discrepant and none produces an adequate fit to the overall spectral 
energy distribution.  The general trend among our sources is of cooler 
color temperatures inferred from the longer wavelength data and hotter
temperatures derived from the shorter wavelength data. 
\footnote{Of note is that the (modified)
blackbody dust temperatures fitted by Carpenter et al. (2008a) 
to 5-35 $\mu$m IRS spectrophotometry (as opposed to just the 
synthetic 13 and 33 $\mu$m ``photometry" points used here) 
are in every case intermediate between those of our 13-33 $\mu$m 
and 24-33 $\mu$m values.} 
Furthermore, Carpenter et al. (2008a) find that the fits to
5-35 $\mu$m IRS data underpredict the 70 $\mu$m excess by $>3\sigma$
for 3/4 of the sources for which data are available.
The systematic discrepancies are suggestive 
of a physical effect rather than resulting from data errors; 
further, the phenomenon of inconsistent color temperatures 
is not due to the presence of prominent spectral features, 
as none are detected in our SNR $>30$ spectra from IRS.
It should also be noted that we would expect the same blackbody that fits
the shorter wavelength points to also fit the longest wavelength 160 $\mu$m
point only if the grains are as large as 20 $\mu$m;  
smaller grains that produce blackbody times emissivity ($Q_\lambda<$1)
would underpredict this flux density.  

We offer an explanation for the temperature discrepancies in terms of
multi-temperature dust located over a range of radii, in the next section.  
In the remainder of this section we interpret the color temperatures
derived from the 33-70 $\mu$m flux ratio as the fiducial, characteristic, 
dust temperature (T$_{dust}$) that represents the bulk of 
the excess spectral energy distribution. 
In several cases noted in Table 4 
(e.g. HD 22179, HD 35850, HD 37484, HD 85301, MML 17
in addition to HD 141943 and HD 209253)  
the $\chi^2_\nu$ values resulting from the
(hotter) 24/33 color temperature models are $<$2 and are comparable to, 
or better in some cases, than those for the 33/70 color temperature models.
In one case (HD 31392) we adopt the 70/160 color temperature model.

\subsubsection{Dust Location and Luminosity}

For the assumed blackbody case, simple radiative balance suggests
$$(R_{dust}/50~AU) = 0.62(L_*/L_\odot)^{1/2} (T_{dust}/50~K)^{-2}$$ 
where $R_{dust}$ is the radial distance of the dust 
from the star, $T_{dust}$ is the dust temperature
and $L_*$ is the stellar luminosity 
\footnote{
Allowing for smaller, non-blackbody grains with emissivity 
$Q_\lambda \propto \lambda^{-\beta}$ ($\beta$ is in the range 0.5-2 whereas
$\beta=0$ for blackbody grains) would increase the grain temperature 
at a given distance from the star.  This would mean that derived dust radii 
would increase relative to the blackbody case having 
$$L_*/L_\odot = 2.62(R_{dust}/50~AU)^2 (T_{dust}/50~K)^4$$ 
for the same fitted 
dust temperature.  Specifically, for graybody grains which are efficient
absorbers and inefficient emitters
$$L_*/L_\odot = 3.47\times 10^{-2}(R_{dust}/50~AU)^2 (T_{dust}/50 K)^5 (<a>/\mu m)$$ 
and for those which are inefficient absorbers as well as inefficient emitters,
as is the case for very small ISM-like grains,
$$L_*/L_\odot = 2.10\times 10^{-3} (T_*/ T_{\odot})^{-1.5} (R_{dust}/50~AU)^2 (T_{dust}/50~K)^{5.5}$$ 
derived from formulae in Backman \& Paresce (1993).  
}.
With $L_*$ from the Kurucz model and $T_{dust}$ assumed to be
the color temperature of the infrared excess as derived above, 
$R_{dust}$ the dependent variable can be calculated.  
The dust luminosity, L$_{dust}$, is then estimated
using T$_{dust}$, R$_{dust}$, and the Stefan-Boltzmann relation.  
For single temperature blackbody emission this is a more precise method than
trapezoidal integration of the measured excess flux densities which results in 
only a minimum value for L$_{dust}$.
Finally, $f=L_{dust}/L_*$, the fractional infrared excess, is derived.

Table 5 lists 
the inner radii corresponding to the assumption of large (relative to
wavelength) grains along with corresponding values of fractional dust excess.
Formal error propagation from the dust temperature and 
stellar luminosity uncertainties into those
for the dust radii, R$_{dust}$, reveals uncertainties 
of $\sim$10-35\% but we emphasize that these radii are only notional minimum
values derived under the strong assumption of blackbody grains.  They are
{\it lower limits} as smaller grains would achieve the estimated temperatures 
at larger radii  
\footnote{Specifically, for a typical source such as HD 105, 
the formal uncertainty in the fitted
dust temperature ($\sim$10\%) corresponds to an uncertainty of $\sim$20\%
in the dust inner radius estimated under the assumption of blackbody emission
from the equation above (with a best-fit value of 42 AU listed in
Table 5).  Under different assumptions regarding the nature of the
emitting grains, this same temperature would correspond to much larger radii
of $\sim$ 400 AU (efficient absorbers and inefficient
emitters with a mean grain-size of 0.95 $\mu$m)
or $>$1000 AU for ISM-like grains smaller than the blowout size (0.59 $\mu$m).
}.
The uncertainties on dust luminosity, $f$, are more complex to quantify.

\begin{figure}[t]
  \begin{center}
    \plotone{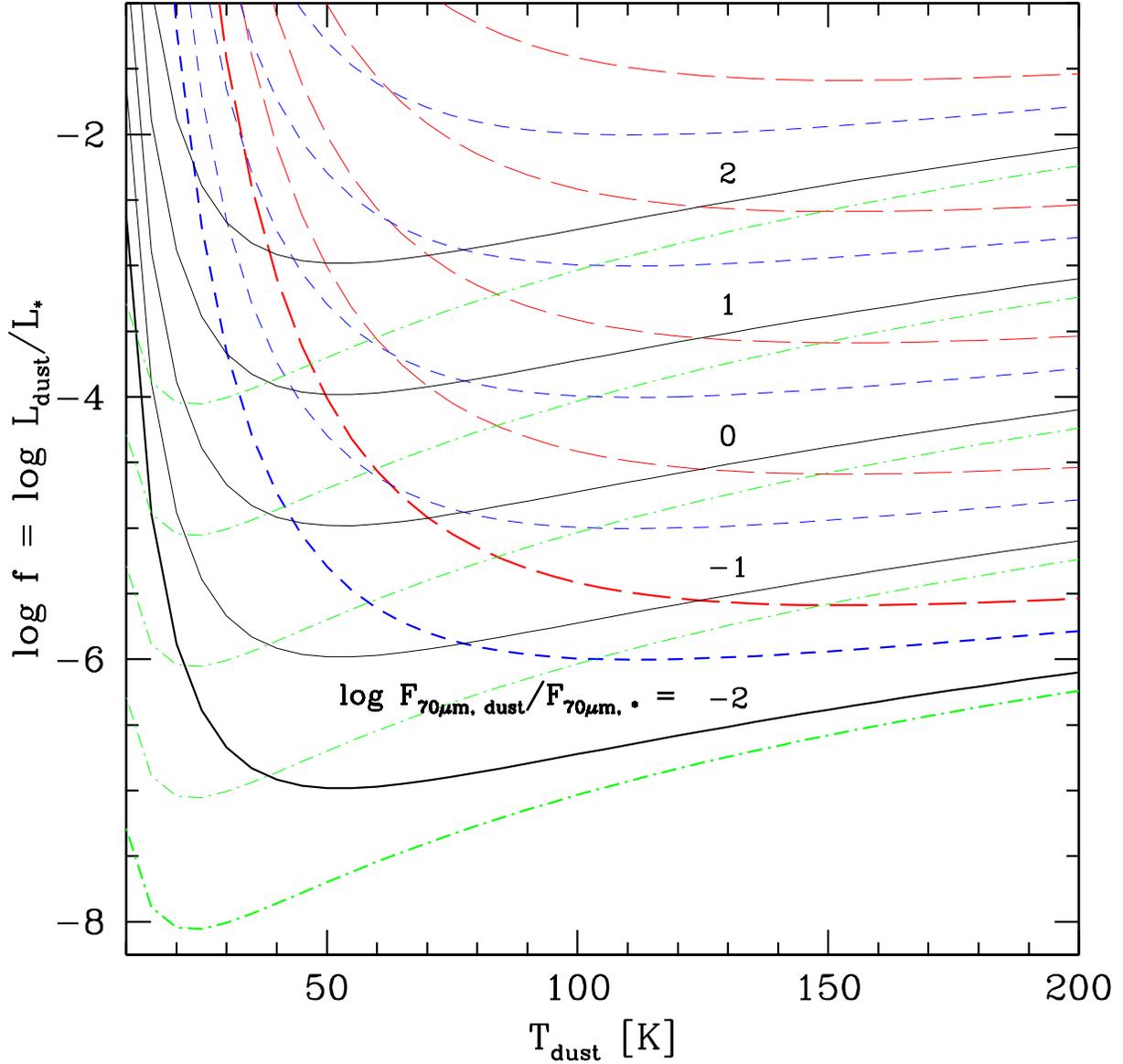}
  \end{center}
  \caption{
  Fractional infrared excess as a function of T$_{dust}$ for different values
  of the observed quantity $F_\lambda(dust) / F_\lambda(\ast)$ 
  at $\lambda=70 \mu$m (black). Comparable curves are shown for the same
  fractional excess values at $\lambda=160 \mu$m (green),  
  $\lambda=33 \mu$m (blue), and $\lambda=24 \mu$m (red).  Note that at the
  same monochromatic contrast level the fractional luminosity of the excess
  to which we are sensitive goes down towards longer wavelengths
  (along with the instrumental capability to achieve those monochromatic 
  contrast levels).  The curves also become broader towards
  shorter wavelengths which can probe a wider dust temperature range to the
  same fractional excess luminosity level.  Observations at 70 $\mu$m 
  and longer are uniquely suited to detection of Kuiper Belt like dust
  distributions.  The $FEPS$ observations discussed here detect primarily 
  50-100 K dust with log f = -5 to -3.
  \label{fig:f70}
  }
\end{figure}

In the pure blackbody assumption, the maximum contribution at 70um 
(if in $\nu~F_\nu$) comes from dust at $T=3675/71.4 = 51.5$ K. 
For a dust emission peak near 70 $\mu$m, the factor $h\nu/kT$ in the blackbody
flux density equation is constant and, 
from further consideration of the contrast with the
Rayleigh-Jeans tail of the underlying stellar spectral energy distributions, 
one finds a minimum value for $f$ 
$$(L_{dust}/L_*)_{minimum} = 10^{-5} (5600/T_*)^3 (F_{70, excess}/F_{70, *}).$$
For a dust excess peaking shortward or longward of 70 $\mu$m, the resulting
dust luminosity is higher for the same monochromatic excess; 
this is illustrated in Figure~\ref{fig:f70} 
which shows the run of $L_{dust}/L_*$ with $T_{dust}$ for constant values 
of the measured quantity $F_{70, excess}/F_{70, *}$.  
In many cases we have, rather than a measurement of $T_{dust}$,
only a limit on $T_{dust}$. This leads to a limit on $f$ which is an
upper or lower limit depending on the value and sign of the $T_{dust}$ 
limit.  For solar-luminosity stars, the dust temperature maxima that are 
much larger than 51.5 K result in inferred $f$ values that are 
likely upper limits while for dust temperature maxima smaller than 51.5 K 
the $f$ values are definitely {\it lower} limits, all in the blackbody
situation.  

In practice, the values of $f$ derived in the blackbody scenario
from the inferred $T_{dust}$
are in fact quite close to the $f$ minima for an assumed blackbody
radiation peak at 70 $\mu$m, within 0.1-0.2 dex in most cases.

Returning now to the case in which the grains are non-blackbody (i.e. smaller)
and the dust inner radii ($R_{dust}$) inferred via the blackbody assumption
are thus lower limits,
the dust cross-sectional areas ($A$) are then also lower limits.
In other words, if the dust is actually smaller than the assumed blackbody
size, in order to achive the same
$T_{dust}=T_{color}$, the observed 70 micron flux density
would require more total dust surface area by a factor
$\lambda/2\pi a = 11.4\mu$m$/a$.
Dust luminosity scales with $T_{dust}^4\times Q$
but also with $A$ where $A/R_{dust}^2 << 1$ for optically thin emission.
In contrast to the pure blackbody case above, here
$f = f_{blackbody}\times T_{dust}/51.5$ K. Thus, dust temperature maxima
larger than 51.5 K result in $f$ values that are larger than
blackbody and the assumed blackbody case produces a
lower limit on $f$, while for dust temperature maxima smaller than 51.5 K,
the $f$ values are smaller than blackbody and the assumed blackbody
case is an upper limit on $f$.


\subsubsection{Dust Mass}

The dust mass, M$_{dust}$, can be determined by assuming a grain material 
density and estimating an average grain size to compute the mass per particle,
which is then multiplied  by the number of particles.  
We consider 2.5 g/cm$^3$ an appropriate average density for silicate dust,
though acknowledge a 50\% range in the values inferred among asteroids
and asteroidal IDP's.
For the grain size there are several options depending on the dominant
physical process that is controlling the removal of grains from the dust disk.
We assume that the production of dust grains is through the collisional 
cascade of larger parent bodies, though this detail is not important just yet.

One option is to use a 10 $\mu$m grain size.
For efficient (i.e. blackbody) emission at a wavelength of 70 $\mu$m, the 
radiative absorption and emission efficiency factors $Q_\lambda(abs,~emis)$ 
are close to unity, 
implying for $1 < 2\pi a/\lambda$, grains larger than a$\approx10 \mu$m.  
Such large sizes are also consistent with, though not necessarily 
demanded by, a lack of observed spectral features in the shorter wavelength 
IRS data, which for our sources generally trace well the stellar photospheric 
or photosphere-plus-dust continuum levels 
(Bouwman et al. 2008; Carpenter et al. 2008a).  Although the 
details of this argument depend on the temperature structure of the disk, 
emission from hot dust of any size, including small amounts of 
moderate sized ($\sim 1 \mu$m) equilibrium silicates or very small 
($<0.05 \mu$m) non-equlibrium grains/PAH's, is not evident.

A second option
is to estimate the average grain size from the minimum grain size
expected to survive in a dust disk in which stellar radiation pressure removes
grains not balanced by gravitational (and P-R drag)
forces working to keep them in orbit about the star (i.e. $\beta = 0.5$). 
Based on Burns et al. (1979) and Artymowicz et al. (1988),
the minimum grain size in a (gas-poor) disk is 
$${a_{min}\over \mu{\rm m}}=0.52\times{2.5 {\rm g/cm^3}\over \rho}\times
{L_*/L_\odot\over {(M_*/M_\odot)(T_*/5780)}}$$
(which can be scaled as ${{1+albedo}\over1.1}$ if an alternate is desired to the
assumed 0.1 albedo of Solar System silicate dust). 
Smaller grains are blown out while bigger grains are retained and
subject to collisions with other grains in a sufficiently dense disk.
Among our sample stars, the range in the predicted $a_{min}$ is 0.3-2.5 $\mu$m.
The average grain size $<a>$ is close to the minimum grain size, 5/3 $a_{min}$,
for a distribution going as a power law with exponent -3.5 as is appropriate 
for either the interstellar distribution (Mathis et al. 1977, though 
extrapolated to larger sizes than typically populate the ISM) or to
to a self-similar infinite collisional cascade 
\footnote{
For collisionally dominated disks, such as we think dominate our sample,
it can be argued that the number of small grains at the inner edge 
of the debris disk is actually {\it higher} than predicted by such a power law 
since those just below the blowout size are preferentially removed via 
radiation pressure and therefore not available to collide with those just above
the blowout size, e.g. Krivov et al. (2000) leading to a ``wavy" size
distribution, e.g. Th\'ebault et al. (2003, 2007), with more grains at
about 1.5$\times a_{min}$ and fewer grains at 10-50 $\times a_{min}$, 
relative to the Dohnanyi distribution. We do not consider such complexity here.
For P-R dominated disks, on the other hand,
there may be {\it fewer} small grains at the inner
edge and overall, since a shallower power law may be more appropriate as in the
Solar System, where $n(a)\propto a^{-2.4}$ e.g. Fixsen \& Dwek (2002).
}
(Dohnanyi, 1969; Durda \& Dermott 1997).  
In this case the typical grain size is thus a few $\mu$m. 

A third option would be to assume that corpuscular drag (due to the effects of
stellar winds on orbiting dust particles, rather than to those of 
stellar radiation as in P-R drag) is responsible for grain removal.
Neither P-R drag nor corpuscular drag effects appear to dominate in our disks
however (see \S5.3) and so we do not consider this case in detail.

Dust masses, M$_{dust}$, can be calculated from the assumed grain density,
the grain size, and the total number of particles at that size,
which we estimate by considering the fractional infrared luminosity 
($L_{dust}/L_*$) divided by the fractional solid angle intercepted 
by a single dust grain ($\pi a^2/(4 \pi R_{dust}^2)$).  
For simplicity, we consider only the average grain size $<a>$.  Because any
smaller grains that are present provide more surface area, and hence opacity, 
we thus calculate minimum dust masses.  The dust mass is thus
$$M_{dust} >  {16\over 3} \pi (L_{dust}/L_*) \rho <a> R_{dust}^2   $$
or
$$M_{dust}/M_{\Earth} >  1.59\times10^{-4}  (L_{dust}/L_*) (\rho/2.5 g~cm^3) (<a> / \mu m)  (R_{dust}/ AU)^2 $$
(see Backman et al. 1993 ; Jura et al. 1995).

The results of our simple modeling can be found in Table 5 where we have 
adopted from the above discussion a value of 10 $\mu$m for $<a>$.  
The uncertainty in the dust masses is significant, not only
because of the linear scaling with assumed $<a>$, but also because 
our values of $R_{dust}$ are always lower limits in the blackbody assumption. 

A separate point is that much mass can be hidden in larger grains, pebbles,
and rocks that, given their ratio of surface area to mass,
do not radiate strongly at even the longer Spitzer 
wavelengths. For an assumed grain size distribution going as 
$n(a)\propto a^{-3.5}$,
the mass in larger grains can be accounted for, yielding a total mass 
$$M_{total} =  M_{dust, min}\times \sqrt{a_{max}/a_{min}}$$
(see Wyatt et al. 2006 for a more general formula for an arbitrary
particle size distribution).
Because $a_{max}$ generally is not known, we quote {\it dust} masses 
calculated for the average grain size $<a>$ only, which in our case is
relatively close to $a_{min}$. 


\subsection{Multi-Temperature Disk Models}

For many of our 70 $\mu$m excess detections, single temperature blackbody 
models fit to the excess emission do a poor job according to the $\chi^2_\nu$ 
values in Table 4, of reproducing the observed spectral energy distributions. 
We identify for further investigation
those sources with $\chi^2_\nu > 1.2$ in the 
33-70 $\mu$m color temperature fit.  The probability that such 
high $\chi^2_\nu$ values are a good fit to the data is $<$25\%.
There are 12 systems in our excess sample, more than 1/3 of our excess sample, 
which we propose in Table 6 as having evidence for material with 
(at least) two different temperatures. 

\begin{figure}[t]
\vspace*{-10mm}
  \begin{center}
  \plotone{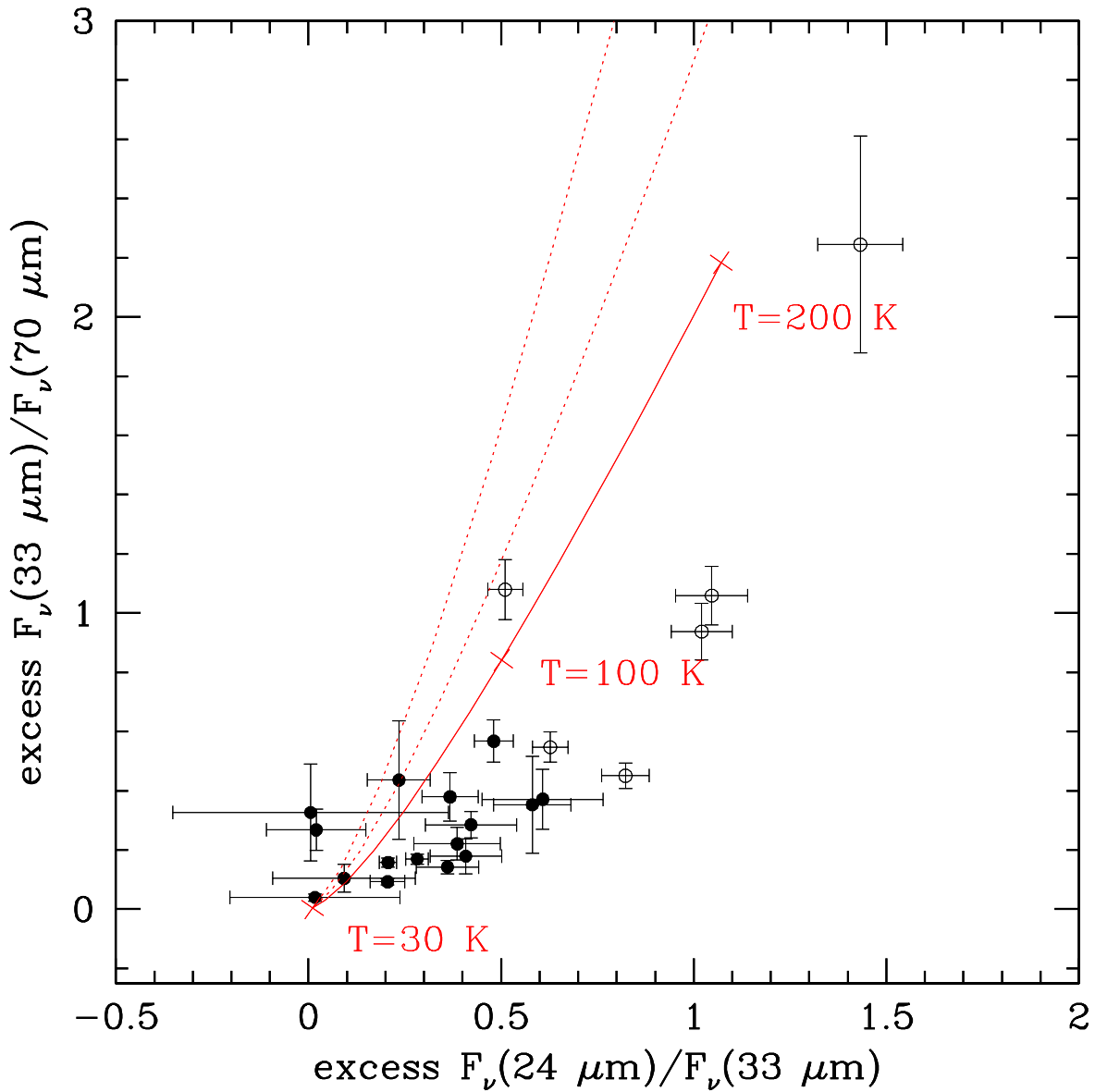}
  \end{center}
  \caption{
  Demonstration of the failure of single temperature blackbodies to explain 
  simultaneously the observed 24, 33, and 70 $\mu$m photometric excesses.  
  Points with errors represent excess flux density ratios, 
  that is, observed flux density ratios corrected for 
  their underlying stellar photospheric contributions.  
   Error bars include
   the observational errors but no error in the photosphere.
  Filled symbols are the 
  debris disks of Figure~\ref{fig:seds} while open symbols 
  are the primordial disks of Figure~\ref{fig:sedsprim}.
  Solid line (colored red in the electronic edition)
  is a blackbody temperature sequence from 30-200 K
  while dotted lines (colored red in the electronic edition)
  are the same for modified blackbodies 
  (optically thin dust having additional multiplicative factors of
  $\lambda^{-1}$ and $\lambda^{-2}$ which may be important depending
  on grain size relative to wavelength).  Although most objects are within 
  1-2$\sigma$ of the expected blackbody relationship, the systematic
  offset suggests that such single temperature blackbody models may not be 
  the most appropriate models.
  \label{fig:excesscolor}
  }
\end{figure}

For these sources, we quantify 
the disparity in the color temperatures derived from the 24-33 $\mu$m 
vs the 33-70 $\mu$m excess flux density ratios in the second column of Table 6.
We illustrate in Figure~\ref{fig:excesscolor} the color temperatures
for all sources in our 70 $\mu$m-selected excess sample having 33 $\mu$m
and 24 $\mu$m excesses as well.  Regardless of the temperature, 
no single temperature model can fit simultaneously the measured 
24, 33, and 70 $\mu$m excesses for many (those listed in Table 6)
of these sources.  
So-called modified blackbodies (or graybodies, having optical depth
$\tau=\tau_0(\lambda/\lambda_0)^{-\beta}$;
$\beta = 0$ for a $\tau>1$ blackbody) that represent analytically
the case of inefficient small, compared to the wavelength
of observation, grain emission are also illustrated in 
Figure~\ref{fig:excesscolor}.  
Modified blackbody models are in even less agreement with
the data, which suggests that other effects (perhaps dust geometry)
trump any inaccuracies in our treatment of grain properties.
Observatory calibration errors of a systematic nature 
could potentially improve the agreement in terms of fitting the mean of the
distribution of points. 
However, such errors would have to be large, about 50\% too high 
for either of MIPS-24 or MIPS-70 and about 30\% too low for IRS, 
much larger than the current calibration precision.
We emphasize based on Figure~\ref{fig:excesscolor} and Table 6
that for any individual source 
the disagreement of the data and the single temperature blackbody
is generally only a 1-2$\sigma$ effect and any conclusion would be 
marginal at best.  However, we interpret 
the systematic trend as indicative of a real effect that characterizes 
the ensemble of stars.  

We are thus motivated to consider multi-temperature dust models.  
While primordial gas and dust rich disks offer
ample evidence for multi-temperature disks (e.g. Dullemond
et al. 2007), it is unusual for debris disks to exhibit spectral energy
distributions with emission at a wide range of temperatures.


Detailed discussion of the source HD~107146 can clarify
the logic.  Using the 1-sigma extremes on the photometry,    
the 24/33 micron color temperature is nominally 72 K but could be in the
range 64 K to 81 K (72$^{+9}_{-8}$ K) 
while the 33/70 micron color temperature, nominally 52 K, 
could be in the range 50 to 54 K  ($52^{+2}_{-2}$ K).
Those two temperature ranges are inconsistent at the $>$2-sigma level.   
Fitting the 24/33 color excess with the nominal color temperature 
that goes exactly through the data points
requires a solid angle of dust $\Omega= 2.62\times 10^{-14}$ sr; this model
then predicts a 70 $\mu$m flux density 10$\sigma$ below the observed data.
If, instead, we fit the 33/70 color excess with the nominal color temperature
we require $\Omega = 2.87\times 10^{-13}$ sr or 11 times larger solid angle
than for the hot dust source;  
this model predicts a 24 $\mu$m excess which is low by 3.6$\sigma$.
As a compromise one could consider an intermediate temperature set by fitting
the 24/70 micron excess with a 58.5 K blackbody.  To also fit the intermediate
33 $\mu$m point then requires a source size $\Omega = 1.09\times 10^{-13}$ and leads to
under-predictions at both 24 $\mu$m (by 2.6$\sigma$) and 70 $\mu$m (by 5.7$\sigma$).
For several other stars the 24 micron excess amplitude is higher 
than it is for this star, and the required color temperatures 
are even farther apart.

One could postulate under the blackbody assumption that the range 
of temperatures inferred for our debris disks is caused either 
by a range in grain locations, or a range in grain sizes. 
These location and/or size distributions may be distinct or continuous.
In either scenario, if the different temperatures 
emanate from different components, with only a few flux density points 
measured in the excess we can not determine grain location or size 
as well as the temperature.  A continuum of temperatures, indicating in 
the simple blackbody assumption material over a continuous set of distances
from the star (a.k.a. a disk) or having a continuous distribution of sizes,
encompass the case of two or more distinct temperature components and 
so we adopt this more general model in what follows.

The grain size scenario, in which a range of small
grain sizes from the blowout size  
upward maintain different temperatures at the same physical distance from
the star (due to the different absorption/radiation efficiencies), would 
result in the observed photometry reflecting a weighted mean of the emission.  
For dust sizes 1-10 $\mu$m, a temperature dependency 
$T\propto a^{-1/5}$ (\S4.1.2 footnote) or even allowing for something 
as strong as $T\propto a^{-1/2}$ from the grain absorption/emission efficiencies
implies a factor of $\sim$1.5-3 range in temperature.  
In an idealized size distribution such as $n(a)\propto a^{-3.5}$, the
smaller 1 $\mu$m grains absorb $\sim$3 times more starlight than the
10 $\mu$m grains.  These smaller grains will then dominate the 
(non-blackbody) emission but will have only $\sim$1/3 of their luminosity
coming out in the longer wavelength tail we observe, which is emitted primarily
by the larger cooler grains. 
Thus we consider more worthy of exploration
the radial range scenario, in which there are multiple grain locations 
leading to the temperature ranges.  In support of this interpretation,
the evidence from debris disks detected in scattered light seems 
to be that multiple rings or extended structures indeed are present
(e.g. Stapelfeldt et al. 2004), which is an 
existence theorem only that may or may not apply to our particular debris disks
(though it does apply to at least two of them; see \S5.1).

To model radially extended disks for the 12 candidate multi-temperature
systems, we consider excesses in a photometric band
as significant if they are $>3\sigma$,
or if they are only $>2\sigma$ when the excess in an adjacent band 
is $>3\sigma$ and the inferred color temperatures are decreasing 
with increasing wavelength.  We  assume blackbody grains and a flat 
surface density distribution with radius, 
$\Sigma(r)= \Sigma_0 r^{\alpha}$ with $\alpha = 0$. 
Although $\alpha=0$ is thought most appropriate to radiation dominated
disks and $\alpha= -1$ or $-1.5$ 
perhaps more descriptive of collisionally dominated disks,
the radial optical depth per logarithmic
interval of $r$ goes as $\alpha$, so it is the dust at the inner
radius which is responsible for most of the absorption and re-emission.
Results for lower $\alpha$ thus should be close to those for
the uniform temperature ring.
Further, Th\'ebault \& Augereau (2007, Figure 10) show specifically 
for a model $\alpha= -1.5$ initial distribution undergoing collisional
evolution, that the 
micron to sub-mm grains quickly establish a flat density distribution,
and it is only the larger bodies which retain the steeper distribution. 

With the surface density exponent fixed, we step through a grid of
$R_{inner}$ and $R_{outer}$, calculating under the blackbody assumption
the fractional surface density $\sigma_o$ (a dimensionless quantity) 
at the disk inner edge that exactly matches the 70 $\mu$m 
excess flux density. We then find the combination of the above 3 parameters 
that produces the lowest residuals when fit to the overall spectral energy
distribution of the excess, letting the 70 $\mu$m point be fit freely in this 
second stage.  Our disk models certainly are 
non-unique, but they do allow estimates of disk parameters;  
$R_{inner}$ can be constrained (minimum value) from the warm color excess while
$R_{outer}$ can be constrained (minimum value) by finding the smallest radius 
that satisfies the (F$_{70 \mu{\rm m}}$ - error) 
or (F$_{160 \mu{\rm m}}$ - error) constraints. 
We present our extended disk modeling results in Table 6.  We list lower limits 
to the range of radii inferred from these (blackbody) temperature estimates, 
and refer the reader to section 4.1.2 for the caveats
in interpreting these radii as physical constraints on the
true location of the dust. 
Taken at face value, the values imply fractional disk widths 
$\delta~R/R$ $\approx (R_{outer}-R_{inner})/(R_{outer}+R_{inner})/2$ 
of at least factors several.  
We emphasize that we are unable to constrain the dust outer radii
very well as the sensitivity and wavelength coverage 
of our observations does not fully probe the coolest dust.
However, we are quite confident in our general result of extended disks.

Returning to HD~107146, our results derived to match the overall
spectral energy distribution indicate dust from 14 to at least 200 AU 
for $\alpha=0$ (or 12 to 92 AU for $\alpha=-0.5$). 
One could truncate the $\alpha=0$ disk at 130 AU to undershoot 
the 70 $\mu$m flux density by 1$\sigma$. 
For comparision, the single-temperature model for this system predicts dust 
at 15.5 AU for the 72 K grain temperature and 30 AU for the 52 K 
grain temperature.  

As an illustration of the limited application  of our approach,
we call attention to the case of HD 141943. 
Fitting the 24/33 color excess results
in a nominal color temperature of 90 K while fitting the 33/70 color excess
gives 81 K.  Both temperatures produce an equally good fit to the overall
spectral energy distribution (Table 4) 
and the hotter temperature leads to a derived inner disk radius of 15.7 AU
while the cooler temperature gives 19.3 AU.  An extended disk model, which is
fitted for illustration rather than because of poor $\chi^2_\nu$ 
from the single temperature fit, spans $\sim$9--40 AU (Table 6) 
with better $\chi^2_\nu$ but also more free parameters in the model
\footnote{
The reason an extended disk model leads to a smaller inner radius than the
single temperature blackbody matched to the 24/33 color excess is because 
the former is a fit to the broader spectral energy distribution including
errors, rather than a calculation specific to the exact 24/33 flux ratio.
}.
An independently fit disk model using the more sophisticated methods
referred to in \S4.3 below also produces an extended structure, from 9.5-42 AU
and having $\Sigma_o(R_{inner}) \approx 2\times10^{-5}$ g/cm$^2$, 
impressively close to the simple model though with the 1$\sigma$ confidence 
contour exceeding 50\% of the nominal best fit values.
Conversely, the only other source with similarly consistent single
temperature blackbody fits to the 24/33 and 33/70 color temperatures
is HD 209253, with derived temperatures of 77 and 70 K; 
a disk model fit to this source spans only 18.6 to 18.8 AU.  

As evidenced from the $\chi^2_\nu$ results,  
small differences in color temperature with wavelength
are probably consistent within the 2-3$\sigma$ errors. 
For the majority of sources not presented in Table 6, extended disk fits lead,
as for HD 209253 mentioned above, to $<$1AU wide rings; exceptions 
are HD 145229 and HD 201219 to which $>$10 AU wide disks can be fitted
(albeit with lower significance than for the sources in Table 6). 
We note that choosing a surface density exponent $\alpha$ other than zero,
either positive (e.g. a disk having low density warm and high density cold 
components) or negative (e.g. a disk having highest density at its
inner edge),
would lead to an increase in the number of disks with inferred broad 
radial ranges.  For example, a large negative value of $\alpha$
would place most of the particles (i.e. mass) near the inner edge of the disk 
and the spectral energy distribution would resemble a single temperature / 
narrow ring model.
In summary, we interpret the larger, most significant color temperature 
differences with wavelength as the most compelling spectral energy
distribution evidence for extended disks.

\subsection{Potential for More Detailed Modelling}

In most cases, our 70 $\mu$m excess sources exhibit 
only a limited number of photometry points (sometimes just one) in excess,
and simple blackbody models either with or without geometric complexity,
are sufficient. More detailed modeling may be warranted in several cases, 
however.  Candidates include HD~8907 (e.g. Kim et al. 2005), HD~104860, 
and HD~107146 which have multi-wavelength sub-mm photometry 
(see Figure~\ref{fig:flatseds}), 
and HD~61005 and HD~107146 (again) which are both
spatially resolved in scattered light
at optical/infrared wavelengths (HD~107146 is resolved as well 
at sub-mm/mm wavelengths, measuring thermal emission).
We can use any such more detailed modeling results to inform our
strong assumptions made above in deriving values of dust temperature,
location, luminosity, and mass.

As an example of what is possible,
HD 38529 has been analyzed in some detail by Moro-Mart\'{\i}n et al. (2007b). 
This source is of particular interest due to the presence of multiple planets 
detected via the radial velocity method.  The characteristic dust temperature 
derived here 
from the ratio of 33 $\mu$m to 70 $\mu$m excess emission is $<$48 K, 
implying dust at $>$98 AU in the blackbody assumption (see Table 5). 
The temperature derived from fitting a photosphere plus a single temperature 
blackbody to the shorter wavelength IRS spectrum is 79 K, implying dust 
at 31 AU assuming blackbody emission (Carpenter et al., 2008a). 
As with the sources in section 4.2, this difference in HD 38529 dust
temperatures derived for different wavelength ranges indicates that the dust 
probably is not confined to a narrow ring.  
Moro-Mart\'{\i}n et al. (2007b) explored the complexity and degeneracy 
of debris disk spectral energy distribution modeling 
in the non-blackbody grain case using
the radiative transfer code developed by Wolf \& Hillenbrand (2003).
They found for 10 $\mu$m astronomical silicate grains in a dust annulus having
free parameters R$_{inner}$, R$_{outer}$, M$_{dust}$, and 
a $\Sigma_0$ (initially assumed constant with radius) 
that the derived R$_{inner}$ increases as: 
(1) R$_{outer}$ decreases, because for a given dust mass, smaller 
R$_{outer}$ means a larger $\Sigma_0$, and hence more warm dust 
needs to be eliminated in order to be consistent with 
lack of 24 $\mu$m-emitting dust in this particular source; 
(2) as $\Sigma$ becomes steeper (e.g. $\Sigma \propto r^{-1}$ instead 
of constant); and (3) as smaller grains are considered.
Because the outer radius of the disk, R$_{outer}$, can not be constrained with 
data currently available, it was found that a wide range of over-all
disk properties (dust location, total mass, and luminosity) 
are consistent with the sparsely
sampled spectral energy distribution. 

Similar modeling of the other sources presented in this paper would have
comparably uncertain results and we do not attempt it here.
However, in the case of HD 38529, one can move beyond 
spectral energy distribution degeneracies by using dynamical simulations 
that take into account the role of mean motion and secular resonances 
of the two known planetary companions, to study the location 
of stable niches of potential dust-producing planetesimals. 
Moro-Martin et al.  concluded from such dynamical modelling
that the planetesimals responsible for most of the 
dust emission are likely located within the 20--50 AU region, one of the
possible results from the spectral energy distribution modeling and
consistent with the $\sim$100 AU inner dust edge in the simple
blackbody scenario adopted here. 
Similar procedures may become possible for other $FEPS$ targets
if the planetary systems are discovered directly.

\subsection{Inner Cleared Regions}

Neither the spatially resolved imaging of inner disk holes nor the
detection of planets that would enable inference of
inner clearings based on dynamical analyis, 
are available yet for most of our sources.  
However, our simple modeling procedure has led to the result 
that the dust excess is dominated by a cold component which contributes 
prominently to the spectral energy distribution at 70 $\mu$m and is 
typically located exterior to $\sim$10 AU.  
P-R radiation drag causes dust at large radii to spiral in towards the 
central star on time scales of only millions of years. 
Even in collisionally dominated disks such as we think dominate 
our sample (\S 5.3), some grains will avoid collisions and migrate 
to the inner disk.  

We can ask the question of whether the inferred values of $R_{inner}$ 
imply a lack of substantial amounts of warmer dust closer to the star,
by testing how much mass could be hidden interior to $R_{inner}$ without
producing detectable radiation at the shorter Spitzer wavelengths.
To do so, we adopt the same 10 $\mu$m 
average grain size as above, such that the opacity scales only with
surface density. We also assume the flat surface density profile
($\Sigma(r)= \Sigma_0 r^{0}$) appropriate for the radiation-dominated, 
relatively cleared inner region that we postulate could 
extend from $R_{inner}$ to an $R_{0}$
that corresponds to the dust sublimation temperature at 1500 K.  
We then find the corresponding dust mass such that the most stringently
confining flux density not observed in excess among the
13, 24, and 33 $\mu$m measurements, is not violated by more than 1$\sigma$; 
we note that it is usually the 24 $\mu$m point that provides the best limit. 

The resulting dust masses are 
small, roughly $10^{-6}$ to $10^{-4} M_\Earth$ and would decrease if we
decreased the assumed grain size (\S4.1). This corresponds to roughly a
single asteroid mass pulverized into micron-sized grains.  
The surface density contrast between any such 
low mass inner dust disk and the outer dust disk that we in fact observe 
can be constrained by fitting a two-component model with surface density  
$\Sigma_{o,inner}(r/r_{o,inner})^0$ in the hypothetical inner P-R dominated disk 
and $\Sigma_{o,outer}(r/r_{o,outer})^{-1 {~~\rm or~~} 0}$ in the outer collision-dominated disk
(recall the Th\'ebault \& Augereau 2007 result noted earlier regarding the 
quick establishment of a flat surface density profile for the dust even in
a collisionally dominated disk). 
Resulting values of $\Sigma_{0,outer}$/$\Sigma_{0,inner}$ range from 
$>$30 at the minimum to $>>10^5$ depending on model choices.  For example, 
if we assume a very large (200 AU) outer disk it requires
relatively little surface density to match the 70 $\mu$m measurement, 
compared to a narrower disk or belt which would require substantially
more (factor of $\sim 10^3$) surface density and hence produce larger 
outer/inner disk contrast than the minimum quoted above. Similarly, 
a declining surface density profile for the outer disk also requires more 
(factor of $\sim 10^2$) surface density relative to the flat surface density 
profile, and hence would also enhance the above minimum contrast numbers.

\section{Physical Implications}

\subsection{Extended Dust Disks}

Tables 5 and 6 show the characteristic location of the dust for 
single temperature blackbody models and for extended disk models, 
respectively. Of note is that for a number of systems with 
high $L_{dust}/L_*$, we have inferred the existence 
of dust disks of wide radial extent. 

The best evidence for extended disks around other dusty stars
has come from spatially resolved imaging in both scattered light 
at short wavelengths and thermal emission at longer wavelengths.
Prominent nearby examples of non-narrow ring sources include the 
very young (10-20 Myr) debris disks AU~Mic (M-type star) 
and $\beta$~Pic (A-type star) as well as the somewhat older systems 
$\epsilon$~Eri and HD 53143 (K-type stars), HD 32297 (G-type star), 
and Vega and 49 Ceti (A-type stars).  
The spatially resolved images indicate dust over a
wide range of radii.  In most cases the data are contrast-limited 
at the inner edges, implying widths $>$50 AU (Ardila et al. 2004, 
Kalas et al. 2006), roughly consistent with our understanding 
of the Solar System's dust distribution having
width 25-30 AU beginning outside 30 AU. 
From the spectral energy distributions, alone, one would not have 
inferred the presence of multi-temperature material for these 
particular sources.  Indeed, it is rare to infer extended debris
dust from spectral energy distributions. 

There are two $FEPS$ sources, both in our
70 $\mu$m excess sample, which have been spatially resolved. 
HD~107146 (first discussed as an infrared excess object 
by Metchev et al. 2004 and Williams et al. 2004) is the first 
spatially resolved disk associated with a G-type star 
(Ardila et al. 2005, Carpenter et al. 2005, Metchev et al. 2008), 
and extends from 80-185 AU optically and 30-150 AU at sub-mm wavelengths. 
HD~61005 (first discussed as an excess object here) 
is spatially resolved in scattered light (Hines et al. 2007).
Both objects also appear in our Table 6 of candidate extended disks.  
That we infer extended dust geometries based on spectral energy distributions 
for several additional $FEPS$ sources indicates that they are 
prime targets for high spatial resolution, high contrast observations
that might succeed in imaging the disks.

\subsection{Steady State vs. Stochastic Collisions}

The debris disk systems discussed here have dust at temperatures and
locations roughly comparable to the inner regions of the Solar System's 
own Kuiper Belt. However, the $f=L_{dust}/L_*$ values that result 
from our simple blackbody modeling indicate much higher levels of dust: 
$L_{dust}/L_*$ $\sim 10^{-4.5}$ to $\sim 10^{-3}$ 
compared to the $\sim 10^{-7} - 10^{-6}$ inferred 
for the Kuiper Belt (Fixsen \& Dwek 2002; Backman et al. 1995; Stern 1996a). 
Higher values of $L_{dust}/L_*$ at the same location suggest 
that our disks contain more dust than our present day Solar System.

By experimental design, the {\it FEPS} 
sources typically are younger than our Solar System (only 6/328
are comparably aged or older according to our most recent age estimates). 
A more appropriate comparison of the dust luminosities and masses might 
be made, therefore, to models of the earlier dust content in the Solar System. 
Because the relevant processes is dissipative, we can not extrapolate 
backwards in time.  However, we can use forward modelling that assumes
$\tau^{-1}$ (for a collision-dominated dust disk) or $\tau^{-2}$ 
(for a radiation-dominated dust disk) scaling 
as may be appropriate during different stages of Solar System dust evolution  
(e.g. Dominik \& Decin 2003).  See \S3.1 and Meyer et al. (2007)
for brief discussion of such a model. Our data would thus be explained  
in the context of our solar system
by a more massive planetesimal belt/s (e.g. Wyatt 2006). 

Alternate to the more massive and perhaps younger debris disk scenario, 
we could be witnessing the effects of transient phases of high dust production 
due to recent massive collisional events in these particular 70 $\mu$m-bright
systems (e.g. Jura 2004). If all the observed disks were transient, our 
observations could be used to assess the duty cycle 
of such short-lived events given the rapid blowout times for small
grains, once produced.  If, for example, we are detecting 10\% of systems
in states that should disperse in 1\% of the system lifetimes, 
and assuming that all stars go through this process,
then we would be seeing a phenomenon that occurs 10 times in the lifetime
of the system, rather than single, unique catastrophes.

To assess whether the observed dust could be produced by the steady
grinding down of planetsimals or, on the contrary, if a transient event is
required, we can compare the observed excess ratio, $f$, to that
corresponding to the maximum dust production rate that could be
sustained for the age of the system, an $f$(maximum). 
Following Wyatt et al. (2007) 
and using the same parameters for debris belt width (50\%), 
planetesimal strength, maximum planetesimal size, and orbital parameters, 
we find that:
$$f{\rm(maximum)} = 0.00016(R/AU)^{7/3}(\tau/Myr)^{-1}(M_*/M_\odot)^{-5/6}(L_*/L_\odot)^{-0.5}.$$
This equation represents equilibrium evolution of a standard 
$a^{-3.5}$ grain size distribution with no grain growth
or planetesimal accretion. Most of the {\it FEPS} 70 $\mu$m-selected 
debris disks 
appear below the predicted line, by up to two orders of magnitude,
though some are very close to it.
Exceptions for which $f>f$(maximum) are HD 206374 and HD 85301. 
The former is a marginal excess detection.
The later is a factor of a few above the collisional
prediction (which is notably the lowest $f$(maximum) among our sample stars)
and could therefore be a rare stochastic system. 
With $f<f$(maximum) in general, the $FEPS$ debris disks appear consistent 
with a steady grinding down of planetesimals.  
This is also the conclusion reached by L{\"o}hne et al. (2008) who find,
unlike Wyatt (2007), a dependence of $f$ on initial disk mass and
an evolutionary behavior of $f$ with shallower slope
($\tau^{-0.3}$ rather than $\tau^{-1}$).

Another comparison that can be made is to the planetesimal formation and
early debris ``self-stirring" models of Kenyon \& Bromley (2005) which predict 
a peak in $f$ between 10-100 Myr, rather than the monotonic steady 
decay of Dominik \& Decin (2003),
Wyatt et al. (2007) or L{\"o}hne et al. (2008).  
The $FEPS$ data may be more consistent with this
genre of collisional evolution at the young ages; 
see Carpenter et al. (2008a) for in-depth comparison 
to these particular models and discussion
of the evolution of debris having a range of temperatures and locations
(regardless of detectability at 70 $\mu$m).  That our highest-$f$ values 
occur roughly around 100 Myr (\S6.1) may also be indicative of consistency
with these models.  

In summary, we find no strong evidence for transiently bright dust
among our sample of 70 $\mu$m selected disks.  Rather, our brightest
70 $\mu$m-selected debris systems seem consistent with massive, 
youthful debris disks undergoing collisional evolution.


\subsection{Radiation-dominated vs. Collision-dominated Disks}

To further evaluate the possibilities regarding the steady state evolution 
vs. stochastic event interpretation of debris around solar type stars,  
we address in this section
whether our detected debris disks are collision-dominated or 
radiation-dominated. We consider the time scales for various processes 
(following Backman \& Paresce 1993) and then evaluate their relation
and appropriateness to our debris disks.
\begin{itemize}

\item {\it Collisional Lifetime:}
The time between collisions involving a single grain can be estimated simply 
as $1/n\sigma v$.  Under the presumptions of circular orbits and completely 
destructive collisions between grains of the same size, this becomes
$${\tau_{collisions}\over{\rm yr}} > ({R\over AU})^{1.5} {1\over 9\sigma(R) \sqrt{M_*/M_\odot}},$$
where $\sigma(R)$ is the face-on fractional surface density, in units of 
cm$^2$ of grain cross section per cm$^2$ of disk area, 
also termed radial optical depth; see below.
For a constant surface density, $\Sigma(r) \propto r^{0}$, 
$\sigma(R) = 2f/ln(R_{outer}/R_{inner})$, with $R_{inner}$ the 
inner disk boundary and $R_{outer}$ the outer disk radius (e.g. Backman, 2004). 
Then in the case of a broad belt with $R_{outer}/R_{inner} \sim 7$, 
$ln(R_{outer}/R_{inner}) = 2$  and $\sigma(R)$ is simply 
$\approx f = L_{dust}/L_*$.
Th\'ebault \& Augereau (2007) propose a significant modification
to the above formula which accounts for the grain size distribution as
a scaling factor of $[(a/1.2a_{min})^{-2} + (a/100a_{min})^{2/7}]$; this 
lengthens the collision lifetime by up to a factor of 10 for grains 
very near the blowout size, and shortens it by up to a factor of 100
for grains less than 100 times the blowout size (maximally so for grains
$\sim$10 times blowout size), and again lengthens the collision lifetime 
for even larger grains. Because we generally consider grains of several
times the blowout size, we note that from the simple formula above the
derived collision times are likely overestimates, i.e. the true collision
time scales for average grains in our disks are even shorter than the
values we calculate, roughly $10^{4}$ to $10^{5.5}$ years.

\item{\it Radiative Blowout Lifetime:}
For very small grains, the relevant time scale is the travel time for
removal from the disk under the influence of radiation pressure. This occurs
when the radiative force exceeds the gravitational force. The time it takes 
for a grain to go from R$_{inner}$ to say 4$\times$R$_{inner}$ where,  
under the blackbody assumption, the grain temperature is reduced by half 
so that the grain no longer contributes significantly to the flux density 
measured at R$_{inner}$,  is given by
$${\tau_{blowout} \over{\rm yr}} = 0.5 \sqrt{(R/AU)^3 \over{(M_*/M_\odot)}}.$$
The so-called blowout time is generally many orders of magnitude 
smaller than the other time scales, roughly $10^{1}$ to $10^{2.5}$ years.

\item{\it P-R Lifetime:}
For grains larger than the blowout size (formula given in \S 4.1.3), 
the time it takes for a dust grain 
to spiral inward under the effect of P-R drag from a distance $R$ all the 
way to the star  $(R\approx 0)$ is given by 
$${\tau_{P-R} \over{\rm yr}} = 720 {(\rho/g~cm^3) (a/\mu m) (R/AU)^2 \over{(L_*/L_\odot)(1 + albedo)}}.$$
To estimate the time it would take for a grain to drift from a distance $R_2$ to 
a distance $R_1$ we substitute $R^2$ by ($R_2^2 - R_1^2$). 
Replacing $R^2$ by ($R_{outer}^2 - R_{inner}^2$) gives the
time scale for a particle to move through the entire debris belt, which can be
compared to the time for particles in the belt to collide, as given above.
For our disks we calculate radiative drift times ranging from
$10^{5.5}$ to $10^{7.5}$ years.

\item{\it Corpuscular Drag Lifetime:}
The effect of stellar wind (or corpuscular) drag scales with the P-R lifetime as
$$\tau_{wind} = { {L_*} \over {\dot M_{wind} c^2}} \times \tau_{P-R},$$
assuming comparable coupling efficiencies or Q's for wind and P-R drag
(Jura 2004). While less important for relatively old stars like the Sun, 
for which $\dot M_{wind}=3\times10^{-14}M_\odot/yr$ and hence
$\tau_{wind} \approx 3\tau_{P-R}$, wind drag may be relevant for young stars 
such as {\it FEPS} targets which may have much higher mass loss rates 
at the same or only slightly higher stellar luminosity L$_*$. 
There is still significant uncertainty in the appropriate values, however
(c.f. Wood et al. 2005, Matt et al. 2007) 
and so we do not consider further the details 
of this potentially relevant dust removal mechanism.


\end{itemize}

At issue is whether the dust grains we observe will disappear by 
moving outward from the planetesimal belt (which would be the case in a 
collision-dominated systems where dust particles erode and fragments blow 
out), or whether they spiral inward towards the star
producing a zodiacal-like dust cloud located between the inner edge of the 
planetesimal belt ($R_{inner}$) and the star (like would happen in a P-R 
dominated system such as the inner solar system).  As the processes
are statistical in nature, the question can be partially addressed by comparing
the collisional lifetime to the time it takes a grain to migrate.
%
In the above equations, when the surface density profile exponent $\alpha$= 0, 
$\tau_{collisional}$ increases as $r^{1.5}$ and 
$\tau_{P-R}$ increases as $r^2$; thus
if $\tau_{collisional} << \tau_{P-R}$ at $R = R_{inner}$
then this condition will hold thoughout the disk, at all points $R>R_{inner}$.
In an $\alpha$= -1 disk, 
$\tau_{collisional}$ increases as $r^{2.5}$ and, if as above, 
$\tau_{P-R} >> \tau_{collisional}$, but now at R$_{outer}$ then this condition 
will hold towards the inner disk, at all points $R<R_{outer}$.
In what follows we demonstrate that the disks are collisionally dominated 
at $R_{inner}$ (and therefore throughout, if $\alpha = 0$) as well as 
at an assumed $R_{outer}$ (and therefore throughout, if $\alpha = -1$). 


We can consider migration between a 
characteristic point in the middle of the belt ($R_{mid}$) and either
$R_{inner}$ or $R_{outer}$.  Those reaching $R_{outer}$ from the interior
are lost from the system and it is assumed
that all grains small enough to undergo blowout are thus quickly removed.
Those reaching $R_{inner}$ from 
original locations between $R_{mid}$ and $R_{inner}$ 
would create a zodiacal cloud and be
able to drift past $R_{inner}$ before colliding. 
A reasonable value for  $R_{mid}$ is
$\sqrt{2}$$R_{inner}$, and in this case the P-R drift time scale is 
numerically equal to the P-R lifetime derived above for R = $R_{inner}$. 
Alternately, Moro-Mart\'{\i}n et al. (2007b) calculate the time it takes 
grains to fill the inner gap as
$\tau_{fill} = (1-(1-x/100)^2)\times \tau_{P-R}$, where $x$ 
is a percentage scaling (assumed to be 10\%) of $R_{inner}$ 
over which the dust density decreases.


The disk optical depth is roughly $f$, with vertical optical depth 
$\approx 1/3$ radial optical depth, that in the disk plane, 
which above was called the dimensionless face-on 
surface density; see also Backman 2004. For small values of 
the fractional infrared luminosity (i.e. optical depth)
$f=L_{dust}/L_* < 10^{-4}$, the primary effects on the dust population
are radiative (e.g. P-R drag and blowout) and/or mechanical (e.g. stellar wind). 
When dust removal is dominated by these mechanisms, the radial distribution 
of the existing dust is expected to extend over a larger radial range 
than the location of the parent bodies generating the dust (such as may be the
case for Vega; Su et al. 2005 and $\epsilon$~Eri; Backman et al. 2008).  
Conversely, for large values 
of $L_{dust}/L_* > 10^{-4}$, such as we report here, the disks 
are expected to be collisionally dominated (e.g. Krivov et al. 2000)
with larger grains cascading into smaller grains on time scales shorter
than they are affected by the above-mentioned grain removal processes.  
In this case, the radial distribution of the dust is expected to mimic that 
of the parent bodies colliding to produce the dust, and to
undergo further collisions {\it in situ}.  

\begin{figure}[t]
  \begin{center}
    \plotone{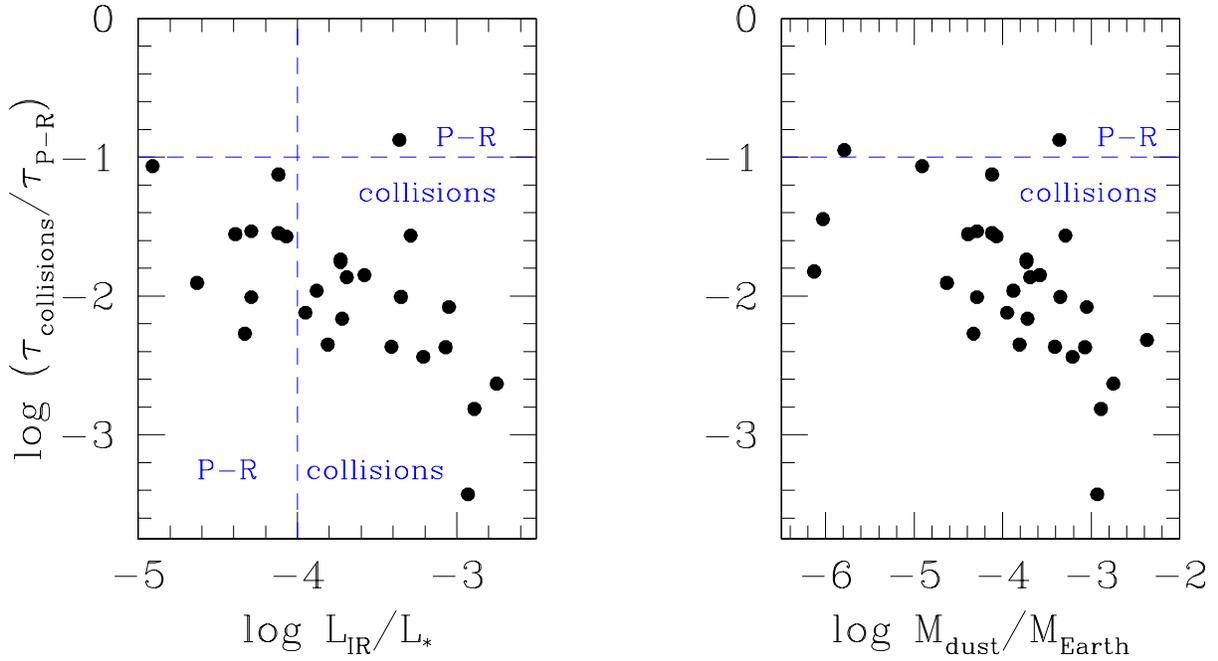}
  \end{center}
  \caption{
  Correlation of the time scales for collisional vs radiative processes in
  the disks, with measured fractional dust luminosity
  $f = L_{dust}/L_*$ and inferred dust mass $M_{dust, min}/M_\Earth$.
  Divisions between radiation-dominated and collision-dominated regimes 
  (Wyatt et al. 2005) are shown as dashed lines 
  (colored blue in the electronic edition), for rough guidance only. 
  The time scales are calculated relative to the inner radius for an
  $\alpha = 0$ surface density profile; using the outer radius and
  $\alpha = -1$ increases the time scale ratio by about an order of magnitude.
  \label{fig:time}
  }
\end{figure}

In Figure~\ref{fig:time} 
we compare the collisional and radiative removal time scales 
with both the $L_{dust}/L_*$ and M$_{dust}$ values as computed in {\S 5.3}.  
In Table 5 we consider both situations described above, the $\alpha = 0$ disk 
evaluated at the $R = R_{inner}$ inferred from the assumed blackbody scenario
as given in Table 5, 
and the $\alpha = -1$ disk evaluated at an $R = R_{outer}$ which is 
unconstrained by the observations but assumed to be 200 AU for purposes 
of illustration.  In the former case,
$\tau_{collisions} / \tau_{P-R} \propto 1/\sqrt{R}$
and so the 10-30\% uncertainties in $R_{inner}$ are not a large effect.
Our calculations consider only a single grain size, 10 $\mu$m, with
$\tau_{collisions} / \tau_{P-R} \propto 1/\sqrt{a}$ (Wyatt 2005). 
We conclude that for the debris disks discussed in this paper, 
the observed fractional infrared luminosities,
$L_{dust}/L_* \gtrsim 10^{-4}$ imply $\tau_{collisions} / \tau_{P-R}$ 
in the range $10^{-3} - 10^{-1}$ near the inner edge of the disk, 
i.e., the dynamics of the dust particles 
in these disks are indeed dominated by collisions. 
Similar analysis by Dominik \& Decin (2003) and Wyatt (2005) 
of previously known bright debris disks has led to similar conclusions: 
that they are all collision-dominated.  
Even for our evaluation at a fabricated outer disk radius of 200 AU and
a falling surface density profile, the values of
$\tau_{collisions} / \tau_{P-R}$ are an order of magnitude higher than at
$R_{inner}$ and a flat surface density profile, but generally less than unity;
a few cases have ratios higher than unity by factors of several.
Among our sample, HD 38529 stands out with the highest ratio
of $\tau_{collisions} / \tau_{P-R},$ (in either scenario) suggesting 
a long collision time scale likely because of the large inner radius 
of the disk; hence P-R effects may play a more prominent role 
for this particular system relative to the others.  

Although the small grains produced in collisionally dominated disks are 
blown out by radiation pressure, grains larger than the blowout size 
cascade into smaller grains before they have time to migrate 
far from the dust-producing planetesimals under the effect of P-R drag. 
Even in collisionally dominated disks in which each collision preserves
say roughly 1/e of the grains, some small percentage 
(5\% of grains survive 3 collisions while
$\sim$1\% of grains survive 5 collisions) 
of the dust particles may survive long enough to have a chance to
undergo P-R drag and thus permeate the inner disk. The debris disks around 
49~Ceti (Wahhaj et al. 2007) and $\epsilon$~Eri (Backman et al. 2008), 
with small grains in the inner disk contributing to spatially resolved 
mid-infrared emission and little to the spectral energy distribution, 
but large grains in the outer disk contributing most of the excess 
in the spectral energy distribution, may be examples of exactly 
this phenomenon.  In \S4.4 we reported $<10^{-6}-10^{-4} M_\Earth$
in dust within the inner cleared regions of our debris systems.

\subsection{Location of the Dust-Producing Planetesimals and Potential Planets}

We concluded above from
examination of the time scales involved in different dust production
and dust removal mechanisms, that the disks we observe are
collisional.  Hence, the dust and the parent bodies are expected
to be co-located.

We consider now as 
a general question whether the wide radial extent inferred for the dust 
implies a similarly wide planetesimal belt.
When dust particles are released from their parent 
planetesimals, their semimajor axes increase due to the effect of radiation
pressure, instantaneously from     
$r$ to $r' =  r {1-\beta \over 1-2\beta r/r_{rel}}$
(Burns, 1979),  
where $\beta$ is the dimensionless ratio between radiation pressure force 
and gravitational force and $r_{rel}$ is the radius of the dust release point. 
For $\sim$10 $\mu$m 
silicate grains with optical constants from Weingartner \& Draine (2001) 
and assuming $\beta\sim 0.025$ as in the Solar System, ${r'=1.026r}$. 
Other combinations of larger $\beta$ values (0.05, 0.2, and 0.4)
which imply smaller grains for the same radiation field 
(4.5, 1.3, and 0.7 $\mu$m, respectively) lead to ${r'=1.05, 1.3, 
{\rm ~~and~~} 3\times r}$. 
The increase in semimajor axis is therefore small for the big grains that 
we expect dominate the 70 $\mu$m emission, but can be large as the
blowout size is approached, consistent with intuition.
Assuming the large grain case, if the planetesimals were located 
in just a narrow ring, they therefore
can not account for the large difference between found between 
R$_{inner}$ and R$_{outer}$ in the dust modeling.  This suggests 
that our disks likely harbor wide planetesimal belts roughly comparable
in size to their wide dust belts.  

Separately, the strong depletion of warm dust, inferred from lack
of short wavelength excess emission, also has implications if it can be
argued that the inner edge of the dust distribution betrays
an inner edge to the planetesimal distribution. 
Possible processes that would create such inner edges are 
the presence at early times of large (1000 km) planetesimals 
that could have stirred up and ground away the inner region 
of the planetesimal disk, or the existence today of 
gravitational perturbations due to one or more planetary companions 
on the planetesimals, as argued for HD 38529 by Moro-Mart\'{\i}n et al. (2007b).

For radiation-dominated disks, an inner cleared disk geometry is often used to 
suggest the presence of a planet that is sufficiently massive to
not only stir the exterior planetesimals, increasing their velocity 
dispersion such that dust-producing collisions occur, but also then 
either eject efficiently any dust particles that cross its orbit as they spiral
in due to P-R drag ($>$80\% of the particles are ejected by a 
1--10 M$_{Jup}$ mass planet in a circular orbit at 1--30 AU; 
Moro-Mart\'{\i}n \& Malhotra 2002, 2003, 2005), or/and trap them into 
temporary resonances as they migrate inward (e.g. Ozernoy et al. 2000).  
Such scenarios prevent or at least impede material from reaching radii 
much smaller than those where the dust is in fact detected. 

For collision-dominated disks, such as we infer here, the location of
the planetesimal-stirring planets is less obvious, and requires detailed
modeling of individual systems once they can be spatially resolved.

\section{Cool Dust Trends with Stellar Parameters}

Overall, $FEPS$ finds 37/328 stars with 70$\mu$m excess. Six of these are 
young primordial disks with excess emission extending from $<8 \mu$m 
to $>70 \mu$m.  Removing these 6 results in an overall incidence of 70 $\mu$m
excess indicative of cool debris, of 10\% within $FEPS$.  Further accounting 
for the fact that 14 of the $FEPS$ sources were chosen for the program based 
on previous suspicion of having infrared excess (only 12 of which
are confirmed as such) yields a 6\% excess fraction at 70 $\mu$m.
We consider these percentages lower limits due to the sensitivity-limited
observations.

The majority of the 70 $\mu$m excess
sources also have 33 $\mu$m excess and/or 24 $\mu$m excess 
indicating that the dust is within 5-30 AU.
Meyer et al. (2008) and Carpenter et al. (2008a) present a more complete
picture of excess vs wavelength and vs age/mass.
Here, in Figure~\ref{fig:excessamp}, we present 
detected and 3-sigma upper limits to fractional excess luminosities
based on our 70 $\mu$m data analysis, 
compared to both stellar age and stellar temperature/luminosity. 
The $f$ values are determined as above for the 70 $\mu$m detections
and as 3-sigma maxima to (the minimum) $f$ for the 70 $\mu$m upper limits
 assuming a hypothetical dust spectral energy distribution
that peaks at 70 $\mu$m (using the formula in \S4.1.2).
Note that these values could be factors of several 
higher for hotter dust or orders of magnitude higher for cooler dust 
not peaking at 70 $\mu$m; see Figure 9.  Note also that these values 
are still upper limits because the flux densities are upper limits.

\subsection{Stellar Age}

The primary goal of the {\it FEPS} Legacy survey is to trace the time evolution 
of dusty debris around solar type stars.  Proper assessment requires
consideration of the full spectral energy distribution.  This would 
allow physical parameters such as T$_{dust}$ and R$_{dust}$ 
to drive the discussion rather than empirical or
wavelength driven constraints, such as are imposed here. 
However, mindful of the observational biases 
which render the great majority of {\it FEPS} 70 $\mu$m observations
upper limits and the 70 $\mu$m detections almost all excess objects, 
we summarize a first analysis of cold debris disk evolution with age.  
Stellar age determinations for the {\it FEPS} sample are discussed in detail
by Hillenbrand et al. (2008).  In brief, a variety of age indicators
such as coronal activity, chromospheric activity, stellar rotation, 
lithium abundance, and the Hertzsprung-Russell diagram are calibrated
to open clusters and used to assess stellar ages between 3 Myr and 3 Gyr.
Rough ages for the sample of 70-$\mu$m excess sources are provided in Table 1.

In Figure~\ref{fig:excessamp}, there is no apparent trend in 70 $\mu$m dust
detection frequency with stellar age, other than
a dearth of strong debris type excesses in our youngest age bins 3-10 Myr
\footnote{There are 6 {\it primordial} disk excesses in this age range but
the remainder of this young sample is undetected at 70 $\mu$m.
}. 
Further, comparison of the age distribution for the detected (using either
2$\sigma$ or 3$\sigma$ threshold) and the non-detected objects 
is insignificant using the K-S test, 
suggesting that we can not distinguish them.
The excess amplitude at 70 $\mu$m does, however, appear to decline with age, 
most obviously in the upper bound; this could be significant 
when coupled with an invariant excess frequency with age. 
Analysis of a more physical quantity like L$_{dust}$/L$_*$, 
as illustrated, reveals that
the average value and the upper bound (including measured as well as 
upper limits on L$_{dust}$/L$_*$) 
decrease with age. Given the large scatter in L$_{dust}$/L$_*$ values 
(both measured and upper limit) at all ages, 
and the intermingling of detections and non-detections in $f$, we
do not draw any strong conclusions regarding the physical implications
of this trend.  

Of interest is that the four highest values of $f$ among our debris disk
sample (HD 61005, HD 38207, HD 191089, and HD 101746) 
all occur within the narrow age range 80-200 Myr. Note that each of these 
high $f$ debris systems is suggested in Table 6 as a radially extended disk.
There are no trends in either the evidence for an extended disk, or in
the derived $\delta R/R$ values among the extended disks, with stellar age.
For the entire debris sample detected at 70 $\mu$m, there are no 
apparent trends in derived T$_{dust}$, R$_{dust}$, or M$_{dust}$ with age.
An outlier object in such scatter plots is HD 38529 with a very large 
inferred inner dust radius, perhaps driven by its larger than average
stellar luminosity or the presence of a sculpting planet.




\subsection{Stellar Mass}

\begin{figure}[t]
  \begin{center}
    \plotone{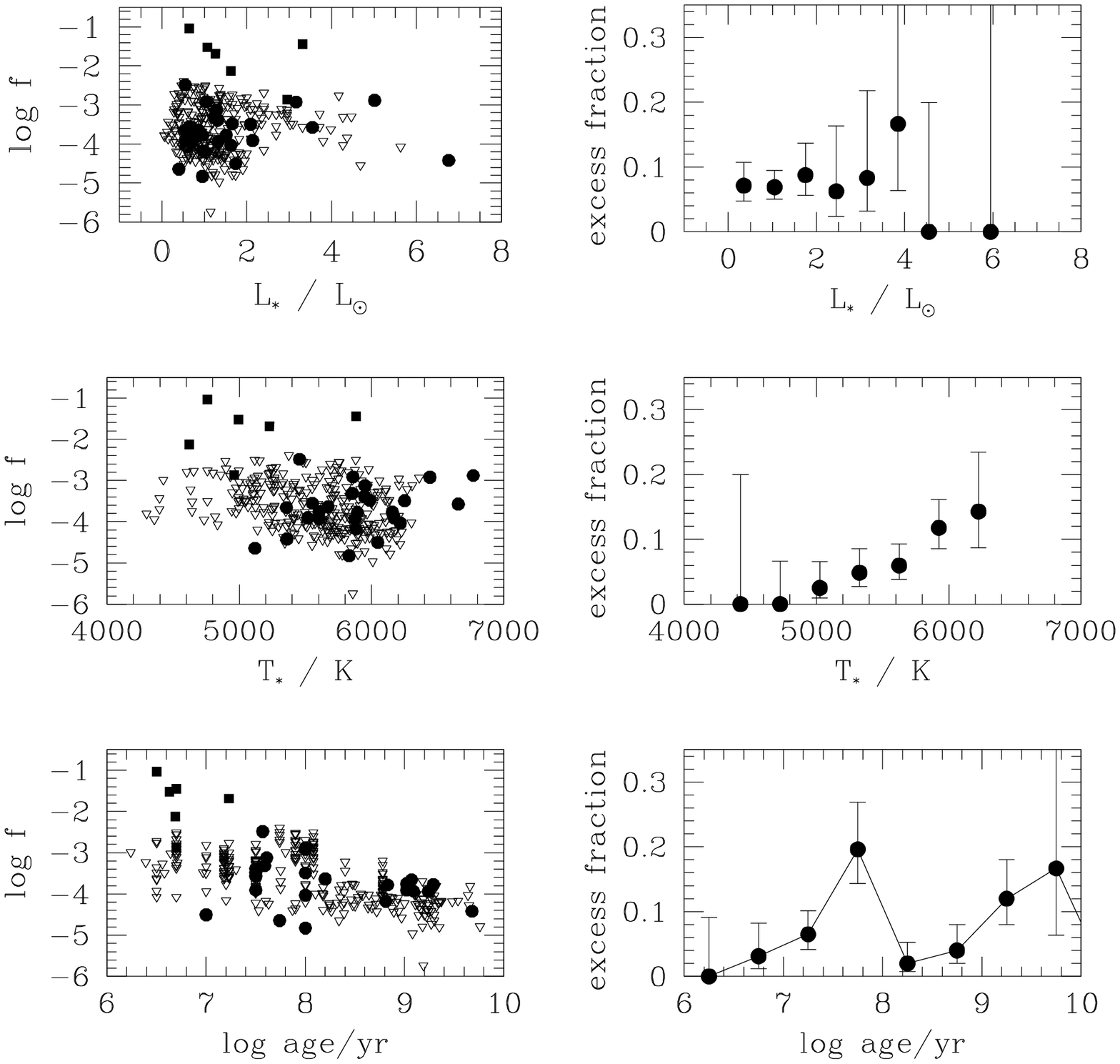}
  \end{center}
  \caption{70 $\mu$m excess statistics as a function of stellar age 
  (bottom panels), and stellar temperature/luminosity (middle/top panels).
  On the left side is the fractional excess luminosity due to dust 
  ($f=L_{dust}/L_*$);
  triangles are sources with no detected 70um excess, plotted at
  $3\sigma$ upper limits, filled circles are detected debris disks,
  and filled squares are detected primordial disks. 
  On the right side is the frequency of excess detection. 
We suggest a correlation in the upper bound
of $f$ with stellar age, but no trends in $f$ with stellar mass or luminosity.
We do not find trends in the 70 $\mu$m dust detection frequency 
with stellar age or luminosity, but do suggest that there may be a trend 
with stellar mass below 6400 K where the sample is unbiased.
Note that there are two bins at high temperature and high luminosity
which exceed the range of the plot along the ordinate.
  \label{fig:excessamp}
  }
\end{figure}

In addition to the lack of T$_{dust}$, R$_{dust}$ or M$_{dust}$
dependence with age among our detections, there are, similarly, 
no trends of these parameters (nor of $f$) with T$_*$ or L$_*$. 
Note that because our sample spans a range of ages from $\sim$3 Myr to 3 Gyr
there is not a 1:1 correlation between T$_*$ and L$_*$ as would be true for
a purely main sequence sample lacking young objects; 
thus we show both in Figure~\ref{fig:excessamp}.  In contrast to
the situation for stellar ages where we found a trend in $f$ with age,
but no evidence for a trend in detection frequency with age,
here for masses we do not find any correlation with $f$ (or other parameters)
but we do find a trend in the dust detection {\it frequency} at 70 $\mu$m. 

Among our sample of 70 $\mu$m debris disks, 10 are F stars,
20 are G stars, and just 2 are K stars. 
The relative search samples in the total $FEPS$ program are 42, 181, and 77 
stars with spectral types F, G, and early K, corresponding to
detection percentages of 24\%, 10\% and 3\%, respectively.
Only 3 of the excess G stars are G5 and later,
rendering the above percentages 24\%, 12\%, 7\%, and 3\% 
for F, G0-G4, G5-G8, and early K types.  
Finally, excluding those sources with 70 $\mu$m excesses that were known
from previous work with IRAS and ISO (see Table 1) leaves among 
F stars: 3/35 ( 9\%), G stars: 16/178, ( 9\%) and early K stars: 1/76 ( 1\%)
detected.  The trend of less frequently detected debris dust around 
the later type stars is very clear when detection frequency is correlated
directly with the inferred T$_*$ (see Figure~\ref{fig:excessamp}, noting only
the points below 6400 K where the sample is not biased by excess objects 
chosen for the probe of disk gas evolution.  We emphasize that the trend 
is not driven by any trend in debris frequency with L$_*$, which is flat for
the bulk of our sample even at fixed T$_*$. 

These findings are consistent with the relative detection frequencies 
of debris among A star samples, FGK samples and M star samples having 
a wide range of ages
(e.g. Rieke et al. 2005, Meyer et al. 2008, Gautier et al. 2007).  
For example, Beichman et al. (2006) claim that they did not detect in their 
more limited survey any debris disks around 23 stars later than K1,
``a result that is bolstered by a lack of excess around
any of the 38 K1-M6 stars in two companion surveys."  The $FEPS$ sample
shows similar behavior for stars typically younger the Beichman sample.
However, it is not yet clear whether all relevant variables --
observational (including selection effects)
as well as astrophysical -- have been normalized properly 
among the various samples, as would be required before such claims can
be validated.

To assess the significance of the apparent trend reported here amongst our
FGK sample, we consider again our lack of sensitivity at 70 $\mu$m 
to stellar photospheres.  First, the distance range of the $FEPS$ sample 
is peaked at d$<$50 pc for all spectral types, with a long tail 
in the distance distribution out to 180 pc.  The F star sample contains
no objects at distances between 70 and 120 pc but the 10 excess F stars
include objects with both d$<$70 and d$>$120 pc.  The G and K star samples
cover the full distance range $\sim$10-180 pc with the two excess K stars
both nearby (26 and 10 pc).  We conclude that distance effects do not
bias the apparent trend with spectral type.  
A second consideration is the relative age distributions.  
The K star sample peaks at younger age than the F and G samples, 
but there are similar numbers of F, G, and K stars at all ages older 
than 100 Myr.  Thus age effects (which we have not claimed for the sample
as a whole) also do not seem to bias the results on debris vs 
stellar temperature/mass since, if anything, the younger K stars
might be expected to have a higher incidence of detectable disks
rather than a lower incidence.


A final consideration is whether we can reach the same value of
$f=L_{dust}/L_*$ for later type (generally less luminous and cooler)
as for earlier type stars.  A monochromatic debris detection trend with mass 
such as we report might be expected if it is driven by 
temperature or luminosity effects given that hot/luminous stars are capable 
of illuminating the same amount of dust to produce higher dust luminosity
relative to cooler and less luminous stars in a flux-limited survey.  
There are no trends either in the limits or in the observed (detected) values 
of $f$, with either $L_*$ or $T_*$.
Again, we stress the several orders of
magnitude spread in $f$ and the intermingling of detections and upper limits
over this range.  It is only the {\it frequency} of detection which
varies with the stellar temperature but not the luminosity.  
We conclude, though admittedly have not proven,
that decreasing detection frequency towards cooler, less luminous,
and typically less massive stars is not a result of varying sensitivity limits
or luminosity effects.  
We believe that this again confirms our earlier assertion
(recall discussion in \S3.1 and \S3.2) 
that the dominant sensitivity limitation for $FEPS$
70 $\mu$m observations is primarily infrared background.

%


\section{Comparison to the Solar System and Context Relative to other Work} 

We have identified 25 secure (Tier 1) 
and 6 candidate (Tier 2) debris disks in the $FEPS$ sample. 
Considering their 33-70 $\mu$m color temperatures, approximately 25\%
of the systems have cold ``Kuiper Belt-like" temperatures, 
$T < 50$ K, $R > 30$ AU.  Another 45\% of the systems have ``Jovian-like" 
temperatures, 60 K  $< T <$ 120 K, R in the range $5-20$ AU.  
About 30\% of our sample have characteristic temperatures 50-60 K, 
which would be analagous to ``Uranus-Neptune-zone" temperatures at 20-30 AU.
Interestingly, there are no planetesimals in this region of the Solar System, 
due to the earlier migration of the outer planets
(in particular Neptune, e.g. Morbidelli et al. 2007).
Only a small percentage (5-10\%) of our 70 $\mu$m-selected sources 
have evidence at shorter wavelengths for ``Asteroidal" belts with temperatures 
above 125 K, roughly corresponding to dust at $R <$ 5 AU.  

In $>$1/3 of the debris objects, disk models having broad 
temperature/radius ranges rather than single temperature/radius models
can be fitted.  For these 12 objects (see Table 6)
the data typically  indicate material lying across more than a factor 
of 2-5 in radius.  

The wide belts can be compared to the Solar System's 2-4 AU asteroid belt 
having $\delta~R/R$ = (2~AU)/3~AU = 66\% or the 40-65 AU (estimated) Kuiper belt 
with $\delta~R/R$ = (25~AU)/50~AU = 50\%.  These relatively
narrow belts or rings in our Solar System and elsewhere
would be well-described by a single temperature blackbody model, in
contrast to the subset of $FEPS$ debris disks we propose as extended disk
candidates. Assuming small grains instead of blackbody grains 
would move both the inner and the outer radii of the wide disks 
to larger values but would not change the fundamental result 
of disk breadth.  

Although the dust temperatures and radii that we have derived 
from our debris disk modeling compare well to those
characterizing the cold outer dust in our own Solar System, there is
an important difference.
At 70 $\mu$m the signal from our Kuiper Belt would be
a few percent of the stellar photosphere while that from our Asteroid belt
would be less than 0.1\% of the stellar photosphere. 
The dust luminosities inferred for our {\it FEPS} sources 
are several orders of magnitude above these levels.
We have argued based on the age distribution of our sample that 
some of these systems may be younger analogs
of our own cold outer dust distribution, but the majority appear to have
higher dust levels relative to our evolving solar system.

We can compare our work to analyses of solar-type stars previously detected 
in the IRAS survey and/or in ISO pointed observations 
(e.g. the systematic studies by Mannings \& Barlow 1998, 
Silverstone 2000, Decin et al. 2003, or the modern re-analyses 
of data from both space observatories by e.g. Moor et al. 2006, Rhee et al. 2007
\footnote{These two papers
reach conflicting results on the debris disk status of at least 10 stars.}). 
Comparative plots of $L_{dust}$ vs $T_{dust}$ and $R_{dust}$ vs $T_{dust}$ 
show that $FEPS$ is finding somewhat warmer, 
closer in, and slightly lower luminosity disks relative to previous work.
Numbering only a few tens before Spitzer, the sample of debris disks 
surrounding {\it solar type field stars} has increased 
by more than 4-fold, collating results from various Spitzer programs.  
Approximately $\sim$45 of these new debris disk detections come from 
the {\it FEPS} survey of 328 young FGK stars using IRAC, IRS, and MIPS data, 
as summarized in Carpenter et al. (2008a). 

Another large debris disk survey is the MIPS GTO {\it FGK Survey}, 
designed to search for excesses around 150 stars using IRS and MIPS. 
These targets are generally older and located at smaller distances
relative to the {\it FEPS} targets.  Results presented in Bryden et al. (2006) 
and Beichman et al. (2006, 2007) indicate a 13$\pm$3\% detection rate 
of 3-$\sigma$ confidence level emission
at 70 $\mu$m (derived from 12/88 stars detected plus accounting for 
the stars with large excesses intentionally left out of the survey).  
Of the 12 stars with 70 $\mu$m excess, most have longer wavelength 
IRS spectra rising above the photosphere at the red end and 4 also show 
weak 24 $\mu$m excess, indicating that the dust is located beyond 5--10 AU. 
We can compare the above with the 6-10\% detection rate of 70 $\mu$m
excess from the younger but typically more distant (and hence more limited
by sensitivity) $FEPS$ sources.  
There are additional long wavelength Spitzer results 
within the younger age range of $FEPS$ stars 
(e.g. Low et al. 2005, Smith et al. 2006,
Padgett et al. 2006, Cieza et al. 2007).  A complete analysis
of dust temperature/location and luminosity vs stellar age
may soon be possible.

\section{Concluding Remarks}

In the present work we have identified 25 likely and 6 possible debris disk 
systems plus 6 primordial disks based on 70 $\mu$m excesses observed
with Spitzer.  In addition to confirming previously known/suspected debris
systems, we have newly discovered 14 systems (see Table 1).

Rather than the selection of 70 $\mu$m excess sources resulting in physically
similar dust belts, we find from simple blackbody modelling of the debris
systems, factors of more than several in the dynamic range
of the physical parameters.
T$_{dust}$ ranges from 45 to $>$100 K, R$_{dust, inner}$ from 7-90 AU,
$f = L_{dust}/L_*$ from $10^{-4.75}-10^{-2.75}$, 
and M$_{dust}$(minimum) $= 10^{-6}-10^{-2.5} M_\Earth$.
We also place limits on the amount of dust in the relatively cleared regions 
interior to our derived R$_{dust, inner}$, finding $<10^{-6}-10^{-4} M_\Earth$. 
We argue for approximately 1/3 of our systems that extended disk
models are more appropriate than single ring models.  Such models imply disks
with inferred (but poorly constrained) R$_{dust, outer}$ values
ranging from 35 to beyond 200.  The debris disks are thus 
factors of several to tens wide in $\delta R/R$, compared 
to the Solar System's two debris belts which are each only about 50-75\% wide.

The above characteristics suggest that the massive disks we see are
collisionally dominated and thus potentially earlier analogs of the 
present-day Solar System's dusty debris system
(in which the dynamics are controlled by radiation and mechanical wind effects
rather than by collisions, and in which there are massive planets located 
in the debris-free zones).  The large radial extent of the dust implies
either wide planetesimal disks as well, or multiple narrow planetesimal belts,
given the collisional nature of the debris.
From our survey of 328 solar type stars ranging in age from 0.003-3 Gyr, 
at least 6-10\% of solar-type stars appear to have cold debris.   
We direct the reader to Carpenter et al. (2008a) for discussion 
of the complete set of cold, warm, and hot debris disks from the $FEPS$ program.

Future investigations of the debris systems identified here
will include high spatial resolution imaging in scattered light
and thermal emission, in order 
to determine the dust geometry, particle size distribution,
and temperature structure, as well as the application of 
various planet detection techniques, in order
to detect directly the large bodies responsible for inducing the
collisional cascade and leading to the dust that we infer from Spitzer data.
One of our debris systems, HD 38529, is known already to harbor 
such a planetary system.

Spitzer in general and $FEPS$ in particular is
dramatically increasing the sample of nearby cold debris disks.
Indeed, this is a ``sweet spot" for the Spitzer Space Telescope.  
The new objects are typically fainter in terms of fractional infrared
luminosity than those found from studies with previous generation satellites
such as IRAS and ISO.  
True analogs to our {\it inner} Solar System debris, with its asteroid belt 
that is controlled dynamically by massive Jupiter, remain elusive; we 
are still several orders of magnitude above required observational sensitivity 
and precision for detection of the current or even earlier (higher)
inferred dust levels 
(including with the state-of-the-art Spitzer telescope).  However, we have
approached the observational sensitivity needed to detect
current {\it outer} Solar System dust values, found in the Kuiper Belt 
region which is sculpted by Neptune. 

Probing effectively the formation and
evolution of solar systems within this Spitzer Legacy Program,
we are even more sensitive to extrapolations backwards in time to a younger 
version of our Solar System's cold dust.
Based on the architecture of our own Solar System, we would not expect
to see as much dust at cool temperatures (corresponding to location between
the Kuiper Belt and Asteroid Belt) at any point, even allowing
for various planetary orbital migration scenarios (e.g. Bottke et al. 2005,
Levison \& Morbidelli 2003).  That we see a range of dust temperatures 
and dust luminosities/masses in other debris disk systems may not be 
too surprising given the diversity of proto-planetary disk properties
as well as planetary architectures 
found amongst the known exo-solar planet population.

\acknowledgements
Acknowledgements:
We thank all members of the $FEPS$ team for their contributions to this effort.
We also acknowledge with appreciation the long-term contributions of
the Spitzer instrument teams and Spitzer science center staff.
Our work is based on observations made with the Spitzer Space Telescope
which is operated by JPL/Caltech under NASA contract 1407.  $FEPS$ is supported 
through NASA contracts 1224768, 1224634, and 1224566 administered through JPL.

{}

\clearpage

\begin{deluxetable}{lrrrrrll}
\tabletypesize{\scriptsize}
\rotate
\tablewidth{0pt}
\tablecolumns{8}
\setlength{\tabcolsep}{0.06in}
\tablecaption{Stellar Properties\label{properties}}
\tablehead{
\colhead{Source} & \colhead{d/pc} & \colhead{log Age/yr} &
\colhead{SpT} & \colhead{T$_{eff}$/K} & \colhead{log L/L$_\odot$} &
\colhead{Previous Discussion of Mid-Infrared Excess} &
\colhead{Comment} \\
}
\startdata
\multicolumn{8}{c}{\it Tier 1 debris disks (excess SNR$_{70 \mu m} \ge 3$)}\\
HD 105    &40 &7.5       &  G0V    &    5948   &     0.12 &   Silverstone 2000; Meyer et al. 2004          &  \nodata \\
HD 377     &40 &7.5       &  G2V    &    5852   &     0.09   & this paper          &  $FEPS$ excess discovery  \\
HD 6963    &27 &9.0      &   G7V    &    5517   &     -0.26  &Kim et al. 2005         &   $FEPS$ excess discovery \\
HD 8907\tablenotemark{a}   &34 &8.5      &  F8     &    6250   &     0.32 &   Silverstone 2000; Kim et al. 2005        & \nodata  \\
HD 22179   &100&8.0     &  G5IV    &    5986   &     0.36   & this paper          &   $FEPS$ excess discovery \\
HD 25457\tablenotemark{a}  &19 &8.0      &  F7V    &    6172   &     0.32 &   Silverstone 2000    & \nodata  \\
HD 31392   &26 &9.0       &  K0V    &    5357   &     -0.26  & this paper         &   $FEPS$ excess discovery \\
HD 35850\tablenotemark{a}  &27 &7.5     &  F7/8V   &    6047   &     0.25  &  Silverstone 2000        &  \nodata \\
HD 37484\tablenotemark{a}  &60 &8.0      &  F3V    &    6656   &     0.55 &   Patten \& Willson 1991; Spangler et al. 2001           &  \nodata \\
HD 38207\tablenotemark{a}   &127&8.0      &  F2V    &    6769   &     0.72    &Silverstone 2000  & \nodata  \\
HD 38529   &42 &9.5      &G8III/IV  &    5361   &     0.82   & this paper; Moro-Martin et al. 2007b &   $FEPS$ excess discovery \\
HD 61005   &35 &8.0       & G3/5V   &    5456   &     -0.25  & this paper   & $FEPS$ excess discovery  \\      
HD 72905\tablenotemark{a,b}& 14  &  8.0    & G1.5 & 5831 & -0.04 & Spangler et al. 2001 & \nodata\\
HD 85301   &32 &9.0      &   G5     &    5605   &     -0.15  & this paper &   $FEPS$ excess discovery \\
HD 104860  &48 &7.5       &  F8     &    5950      &  0.12 &   this paper          &   $FEPS$ excess discovery \\
HD 107146  &29 &8.0        & G2V     &    5859   &     0.04   & Metchev et al. 2004; Williams et al. 2004  &   $FEPS$ precursor work\\
HD 122652  &37 &9.5       &  F8     &    6157   &     0.18   & Kim et al. 2005          &   $FEPS$ excess discovery \\
HD 145229  &33 &9.0       &  G0     &    5893   &     -0.02  & Kim et al. 2005         &   $FEPS$ excess discovery \\
HD 150706  &27 &9.0      &  G3(V)   &    5883   &     -0.02  & Meyer et al. 2004         &   $FEPS$ excess discovery \\
HD 187897  &33 &9.0       &  G5     &    5875   &     0.10   & this paper &   $FEPS$ excess discovery \\
HD 191089\tablenotemark{a} &54 &8.5      &  F5V    &    6441   &     0.50  &  Sylvester \& Mannings 2000 &  \nodata \\
HD 201219\tablenotemark{c}  &36 &9.0       &  G5     &    5604   &     -0.16  & this paper &   $FEPS$ excess discovery; small positional offset \\
HD 202917\tablenotemark{a,b}& 46  &   7.5     &  G5  & 5553    & -0.18  & Silverstone 2000 &  \\
HD 209253\tablenotemark{a} &30 &8.0      & F6/7V   &    6217   &     0.21 &   Silverstone 2000  &  \nodata \\
HD 219498  &150&8.5    &   G5    &    5671   &     0.69   & this paper &   $FEPS$ excess discovery \\
\hline
\multicolumn{8}{c}{\it Tier 2 debris disks (excess SNR$_{70 \mu m} \ge 2 {\rm ~and~} < 3$)}   \\
HD 17925\tablenotemark{a,b}&  10   &  8.0 & K1V    & 5118 & -0.43     & Habing et al 2001  & \nodata  \\
HD 70573   &46 &8.0   &   G1/2V    &  5841   &     -0.23 & this paper &   moderate positional offset\\   
HD 141943  &67 &7.5       & G0/2V      &  5805   &     0.43  & this paper &   moderate positional offset; excess detected with IRS\\
HD 204277  & 34  &  8.5    &   F8       &  6190   &  0.29     & this paper & small positional offset\\
HD 206374& 27  &  9.0    &   G6.5     &  5580   &  -0.17    &   Kim et al. 2005, & $FEPS$ excess discovery; moderate positional offset \\
MML 17     &124&7.0      &  G0IV      &  6000   &     0.43  & this paper & some concern upon visual inspection\\   

\hline
\multicolumn{8}{c}{\it Primordial Disks}  \\
HD 143006\tablenotemark{a}           &   145&6.5   &     G6/8   &     5884   &     0.39 &Sylvester et al. 1996& \nodata  \\
PDS 66               &   86 &7.0   &     K1IVe  &     5228   &     0.10 &   Gregorio-Hetem et al. 1992&  \nodata  \\  
$[$PZ99$]$ J161411.0-230536& 145&6.5     &    K0      &   4963     &   0.50   &   Mamajek et al. 2004&  \nodata  \\
RX J1111.7-7620\tablenotemark{c}    &   153&6.5   &      K1    &     4621   &     0.21 &   Spangler et al. 2001 &   \nodata \\   
RX J1842.9-3532      &   130&6.5   &      K2    &     4995   &    -0.01 &    Neuh{\"a}user et al. 2000 & \nodata  \\
RX J1852.3-3700      &   130&6.5  &      K3    &     4759   &    -0.23 &    Neuh{\"a}user et al. 2000 & \nodata  \\  
\hline
\multicolumn{8}{c}{\it Unconfirmed debris disks suggested in previous literature}\\
HD 41700\tablenotemark{a}  &27 &8.0 &     F8/G0V  &     6140  &      0.24 &  Decin et al. 2000 &   undetected by $FEPS$ but 2.9$\sigma$ 70um source in GO-2 program\\
HD 104467\tablenotemark{c}  &118&6.5  &    G5III/IV &     5690  &      0.75 &  candidate in this work&   70 $\mu$m source is offset by 12.8" and likely unassociated\\
HD 134319\tablenotemark{a}&44  &8.0   &  G5           &   5656      & -0.14   &  Silverstone 2000 &   \\
HD 216803\tablenotemark{a,b}& 7.6  & 8.5 & K4 & 4625  &  -0.71 & Fajardo-Acosta 1999 &   \\
ScoPMS 214&145&6.5  &      K0IV   &     5318  &      0.26 &  Spangler et al 2001 & possible excess at IRS but not detected at MIPS-70\\
\enddata
\tablenotetext{a} {
Previously known/suspected excess source placed on $FEPS$ program for purpose of gas detection experiment. 
}
\tablenotetext{b} {
70um and some other data for this $FEPS$ target derives from a GTO program; see \S2 in text. 
}
\tablenotetext{c} {
See Carpenter et al. 2007a for caveats regarding Spitzer 70 $\mu$m photometry of this source.
}
\end{deluxetable}


\begin{deluxetable}{lrrrrrrrr}
\tabletypesize{\scriptsize}
\rotate
\tablewidth{0pt}
\tablecolumns{9}
\setlength{\tabcolsep}{0.06in}
\tablecaption{Spitzer Photometry in mJy \tablenotemark{a}\label{fluxes}}
\tablehead{
\colhead{Source} & \colhead{3.6 $\mu$m} & \colhead{4.5 $\mu$m} &
\colhead{8.0 $\mu$m} & \colhead{13 $\mu$m\tablenotemark{b}} & \colhead{24 $\mu$m} & \colhead{33 $\mu$m\tablenotemark{b}} &
\colhead{70 $\mu$m}  & \colhead{160 $\mu$m} \\
}
\startdata
\multicolumn{8}{c}{\it Tier 1 debris disks (excess SNR$_{70 \mu m} \ge 3$)}\\
 HD  105    &    1022.7 $\pm$      22.0 &       645.4 $\pm$       14.8 &       230.7 $\pm$        4.9 &    83.3 $\pm$ 5.1  &   28.3 $\pm$        1.2 &        20.5 $\pm$        1.6 &       141.2 $\pm$       14.3 &      110.1 $\pm$       16.7 \\ 
 HD  377\tablenotemark{c}  &  1029.1 $\pm$       22.1 &       648.6 $\pm$       14.9 &       234.7 $\pm$       5.0 &     81.6 $\pm$ 5.0  &  36.6 $\pm$        1.5 &        37.8 $\pm$        2.7 &       162.0 $\pm$       16.9 &      187.5 $\pm$       50.4 \\ 
 HD  6963   &   1211.3 $\pm$       26.0 &       752.9 $\pm$       17.3 &       271.5 $\pm$        5.8 &    91.2 $\pm$ 5.6  &     32.5 $\pm$        1.3 &        21.8 $\pm$        2.0 &        44.0 $\pm$        8.6 &       61.1 $\pm$       29.1 \\ 
 HD  8907\tablenotemark{c} &  1918.2 $\pm$       41.2 &      1223.7 $\pm$       28.1 &       427.3 $\pm$      9.1 &  154.1 $\pm$ 9.4  &   51.3 $\pm$        2.1 &        41.8 $\pm$        3.5 &       247.4 $\pm$       19.7 &      243.8 $\pm$       42.3 \\ 
 HD  22179  &     311.7 $\pm$        6.7 &       196.2 $\pm$        4.5 &        71.0 $\pm$        1.5 &   26.7 $\pm$ 1.7  &   11.1 $\pm$        0.5 &        10.8 $\pm$        0.8 &        35.9 $\pm$       10.6 &      \nodata              \\ 
 HD  25457  &    6259.7 $\pm$      134.4 &      3956.3 $\pm$       90.8 &      1412.3 $\pm$       30.1 &  513.4 $\pm$ 31.2 &  205.8 $\pm$        8.4 &       173.5 $\pm$       10.8 &       307.2 $\pm$       23.4 &      229.4 $\pm$       67.5 \\ 
 HD  31392  &    1431.9 $\pm$       30.7 &       891.7 $\pm$       20.5 &       321.8 $\pm$        6.9 &  112.4 $\pm$ 6.8  &   36.9 $\pm$        1.5 &        19.1 $\pm$        2.2 &        81.6 $\pm$       10.1 &       82.0 $\pm$       14.4 \\ 
 HD  35850  &    3030.3 $\pm$       65.0 &      1917.9 $\pm$       44.0 &       690.7 $\pm$       14.7 &  246.6 $\pm$ 15.0 &   83.5 $\pm$        3.4 &        54.6 $\pm$        4.2 &        40.3 $\pm$        8.0 &      \nodata              \\ 
 HD  37484  &     893.7 $\pm$       19.2 &       568.1 $\pm$       13.4 &       202.2 $\pm$        4.4 &   77.8 $\pm$ 4.7  &   54.6 $\pm$        2.2 &        76.2 $\pm$        4.7 &       114.4 $\pm$       11.2 &       29.3 $\pm$       14.0 \\ 
 HD  38207  &     287.0 $\pm$        6.2 &       181.3 $\pm$        4.2 &        64.6 $\pm$        1.4 &  24.16 $\pm$ 1.5  &   16.5 $\pm$        0.7 &        35.2 $\pm$        2.1 &       184.6 $\pm$       13.8 &       84.8 $\pm$       40.7 \\ 
 HD  38529  &    5893.1 $\pm$      126.5 &      3634.0 $\pm$       83.4 &      1340.0 $\pm$       28.6 &  467.3 $\pm$ 28.4 &  149.6 $\pm$        6.1 &        85.7 $\pm$        5.4 &        75.3 $\pm$       12.4 &       84.8 $\pm$      176.3 \\ 
 HD  61005  &     753.5 $\pm$       16.2 &       472.3 $\pm$       10.8 &       169.2 $\pm$        3.6 &   62.3 $\pm$ 3.8  &   41.5 $\pm$        1.7 &       110.0 $\pm$        6.7 &       628.7 $\pm$       45.4 &      502.6 $\pm$      160.1 \\ 
 HD  72905\tablenotemark{d} &   6226.5 $\pm$      133.6 &      3915.2 $\pm$       89.9 &      1411.5 $\pm$       30.1 &  \nodata   &   163.5 $\pm$        6.7 &       \nodata              &        44.5 $\pm$        6.3 &      \nodata              \\ 
 HD  85301   &    1050.8 $\pm$       22.6 &       652.1 $\pm$       15.0 &       234.2 $\pm$        5.3 &  86.6 $\pm$ 5.3  &   36.8 $\pm$        1.5 &        28.5 $\pm$        2.2 &        38.5 $\pm$        7.5 &       -2.6 $\pm$       15.8 \\ 
 HD  104860  &     724.8 $\pm$       15.6 &       455.3 $\pm$       10.5 &       162.5 $\pm$        3.5 &  57.3 $\pm$ 3.5  &   19.9 $\pm$        0.8 &        17.8 $\pm$        1.8 &       183.1 $\pm$       14.8 &      202.7 $\pm$       27.0 \\ 
 HD  107146  &    1711.3 $\pm$       36.7 &      1074.8 $\pm$       24.7 &       384.4 $\pm$        8.2 & 138.9 $\pm$ 8.5  &   59.8 $\pm$        2.5 &        86.7 $\pm$        5.7 &       669.1 $\pm$       47.8 &      \nodata              \\ 
 HD  122652  &    1260.6 $\pm$       27.1 &       795.3 $\pm$       18.3 &       283.1 $\pm$        6.0 &  96.1 $\pm$ 5.8  &   35.2 $\pm$        1.4 &        26.7 $\pm$        3.5 &        83.1 $\pm$       10.8 &       35.3 $\pm$       27.3 \\ 
 HD  145229  &    1128.8 $\pm$       24.2 &       717.4 $\pm$       16.5 &       254.0 $\pm$        5.4 &  91.7 $\pm$ 5.6  &   31.0 $\pm$        1.3 &        22.1 $\pm$        1.9 &        64.4 $\pm$        8.6 &       33.9 $\pm$       24.7 \\ 
 HD  150706  &    1715.1 $\pm$       36.8 &      1077.0 $\pm$       24.7 &       388.1 $\pm$        8.3 & 135.7 $\pm$ 8.3  &   44.9 $\pm$        1.8 &        28.0 $\pm$        2.7 &        41.3 $\pm$        8.5 &      -28.7 $\pm$       19.9 \\ 
 HD  187897  &    1495.5 $\pm$       32.1 &       934.1 $\pm$       21.4 &       338.7 $\pm$        7.2 & 113.5 $\pm$ 6.9  &   39.8 $\pm$        1.6 &        23.2 $\pm$        3.2 &        61.6 $\pm$        9.3 &      -40.2 $\pm$       65.3 \\ 
 HD  191089\tablenotemark{c,d}  &   1071.7 $\pm$       23.0 &       678.4 $\pm$       15.6 &       242.2 $\pm$        5.2 & \nodata   &   185.6 $\pm$        7.6 &       \nodata              &       544.3 $\pm$       40.1 &      204.6 $\pm$       44.6 \\ 
 HD  201219  &     816.4 $\pm$       17.5 &       508.5 $\pm$       11.7 &       181.3 $\pm$        3.9 &  62.3 $\pm$ 3.8  &   22.0 $\pm$        0.9 &        15.0 $\pm$        1.6 &        42.4 $\pm$        7.8 &       89.9 $\pm$       42.2 \\ 
 HD  202917  &     519.2 $\pm$       11.1 &       320.8 $\pm$        7.4 &       117.3 $\pm$        2.8 &  43.0 $\pm$ 2.6  &     19.2 $\pm$        0.8 &        21.1 $\pm$        1.5 &        37.1 $\pm$        6.5 &      \nodata              \\ 
 HD  209253  &    2008.4 $\pm$       43.1 &      1285.3 $\pm$       29.5 &       454.5 $\pm$        9.7 & 167.6 $\pm$ 10.2 &     55.9 $\pm$        2.3 &        48.1 $\pm$        3.8 &        75.0 $\pm$       10.6 &       17.6 $\pm$       21.4 \\ 
 HD  219498  &     313.2 $\pm$        6.7 &       196.1 $\pm$        4.5 &        70.5 $\pm$        1.5 &  25.5 $\pm$ 1.8 &     10.5 $\pm$        0.4 &         9.5 $\pm$        0.8 &        22.8 $\pm$        4.0 &      -16.7 $\pm$       50.7 \\ \hline
\multicolumn{8}{c}{\it Tier 2 debris disks (excess SNR$_{70 \mu m} \ge2 {\rm ~and~} < 3$)}   \\
 HD  17925\tablenotemark{d}  &    7280.6 $\pm$      156.3 &      4520.6 $\pm$      103.8 &      1644.6 $\pm$       35.1 & 624.0 $\pm$ 38.0 &   193.6 $\pm$        7.9 &       134.7 $\pm$  8.2     &        57.0 $\pm$       12.3 &      \nodata              \\ 
 HD  70573   &    381.6 $\pm$        8.2 &       239.7 $\pm$        5.5 &        86.4 $\pm$        1.8 &  32.7 $\pm$ 2.1  &    10.4 $\pm$        0.4 &         5.8 $\pm$        0.8 &        14.8 $\pm$        5.8 &       29.7 $\pm$       30.0 \\ 
 HD  141943  &    849.1 $\pm$       18.2 &       541.0 $\pm$       12.4 &       193.1 $\pm$        4.2 &  68.4 $\pm$ 4.2  &    27.3 $\pm$        1.1 &        27.9 $\pm$        2.0 &        37.6 $\pm$       15.3 &      \nodata              \\ 
 HD  204277  &   1847.5 $\pm$       39.7 &      1172.2 $\pm$       26.9 &       415.8 $\pm$        8.9 & 143.8 $\pm$ 8.8  &    48.9 $\pm$        2.0 &        27.3 $\pm$        3.1 &        29.6 $\pm$       10.8 &      \nodata              \\ 
 HD  206374  &   1389.1 $\pm$       29.8 &       871.1 $\pm$       20.0 &       312.1 $\pm$        6.7 & 106.5 $\pm$ 6.5  &    35.2 $\pm$        1.4 &        54.1 $\pm$       67.4 &        18.1 $\pm$        6.8 &       -1.3 $\pm$       30.8 \\ 
 MML  17     &    228.1 $\pm$        4.9 &       145.0 $\pm$        3.3 &        52.2 $\pm$        1.1 &  19.2 $\pm$ 1.2  &     9.8 $\pm$        0.4 &         9.5 $\pm$        0.8 &        18.0 $\pm$        7.7 &      -11.7 $\pm$       85.7 \\ \hline
\multicolumn{8}{c}{\it Primordial disks}\\
 HD  143006  &    1069.4 $\pm$       23.0 &       929.9 $\pm$       21.4 &       792.1 $\pm$       16.9 &  737.7 $\pm$ 44.9 &  2130.0 $\pm$      136.0 &      4111.6 $\pm$      250.1 &      3795.1 $\pm$      267.7 &     3228.1 $\pm$      400.2 \\ 
 PDS  66\tablenotemark{c}  &   656.8 $\pm$       14.1 &       521.4 $\pm$       12.0 &       470.6 $\pm$   10.0  &  720.9 $\pm$ 43.9  &  1874.0 $\pm$      120.0 &      1779.2 $\pm$      108.2 &      1672.0 $\pm$      118.0 &     2138.4 $\pm$      290.4 \\ 
$[$PZ99$]$ J161411.0-230536\tablenotemark{c}  &   498.3 $\pm$  10.7 &  401.9 $\pm$  9.2 &    363.5 $\pm$    7.8  &  349.1 $\pm$ 21.2   &   304.0 $\pm$       12.5 &       209.3 $\pm$       12.7 &        91.1 $\pm$       13.3 &       43.1 $\pm$      100.4 \\ 
 RX  J1111.7-7620\tablenotemark{c}  &   447.9 $\pm$   9.6 &   363.5 $\pm$   8.3  &    198.7 $\pm$    4.4         &  160.9 $\pm$ 9.8  &   229.6 $\pm$        9.4 &       217.5 $\pm$       13.2 &       224.3 $\pm$       17.8 &      445.4 $\pm$      183.6 \\ 
 RX  J1842.9-3532\tablenotemark{c}  &   269.5 $\pm$   5.8 &   216.2 $\pm$   5.0  &    157.3 $\pm$    3.4         &  100.7 $\pm$ 6.1  &   358.9 $\pm$       14.7 &       429.4 $\pm$       26.1 &       942.6 $\pm$       67.4 &      844.0 $\pm$      114.8 \\ 
 RX  J1852.3-3700\tablenotemark{c}  &    88.6 $\pm$   1.9 &    58.5 $\pm$   2.3  &     33.6 $\pm$    1.2         &   35.0 $\pm$ 2.1  &   472.2 $\pm$       19.4 &       749.4 $\pm$       45.6 &      1367.0 $\pm$       96.8 &     1490.1 $\pm$      195.7 \\ 
\enddata
\tablenotetext{a} {Uncertainties include both internal and calibration 
		   terms.  See \S2 in text for details.}
\tablenotetext{b} {13 and 33 $\mu$m synthetic photometry is derived from IRS low resolution spectra. For HD~206374 the 33 $\mu$m spectrum has significant noise features appearing within the synthetic passband, resulting in low signal-to-noise.}
\tablenotetext{c} {160 $\mu$m photometry is from a GO-2 or GO-3 program following up the 70 $\mu$m detections reported here.  Integration times are longer than standard $FEPS$ observing procedures; see \S2 in text.}
\tablenotetext{d} {13 and 33 $\mu$m synthetic photometry is derived from IRS high resolution rather than low resolution spectrum.}
\end{deluxetable}

\begin{deluxetable}{lrrrrrrrrrr}
\tabletypesize{\scriptsize}
\rotate
\tablewidth{0pt}
\tablecolumns{5}
\setlength{\tabcolsep}{0.06in}
\tablecaption{Fractional Excess Flux and Excess Signal-to-Noise Ratio \label{excessSNR}}
\tablehead{
\colhead{Source}  & \multicolumn{2}{c}{13$\mu$m} & \multicolumn{2}{c}{24 $\mu$m} & \multicolumn{2}{c}{33 $\mu$m} & \multicolumn{2}{c}{70 $\mu$m}  & \multicolumn{2}{c}{160$\mu$m} \\
\colhead{} &
\colhead{F$_{excess}$/F$_*$}  & \colhead{F$_{excess}/\sigma$ \tablenotemark{a} } & 
\colhead{F$_{excess}$/F$_*$}  & \colhead{F$_{excess}/\sigma$ \tablenotemark{a,b} } & 
\colhead{F$_{excess}$/F$_*$}  & \colhead{F$_{excess}/\sigma$ \tablenotemark{a,b} } & 
\colhead{F$_{excess}$/F$_*$}  & \colhead{F$_{excess}/\sigma$ \tablenotemark{a} } & 
\colhead{F$_{excess}$/F$_*$}  & \colhead{F$_{excess}/\sigma$ \tablenotemark{a} } 
}
\startdata
\multicolumn{11}{c}{\it Tier 1 debris disks (excess SNR$_{70 \mu m} \ge 3$)} \\
      HD 105  &  0.06&     0.8 &    0.13&    2.4  &   0.55  &    4.5&       49.34  &    9.7  &     194.6~~ &     6.6   \\
      HD 377  &0.00&  0.0 &    0.41  &    6.5  &    1.75  &    8.8&       54.77  &    9.4  &  320.6~~   &     3.7    \\
     HD 6963  & -0.06 &   -0.9 &   0.05&   0.9  &   0.33  &    2.6&       11.73  &    4.7  &     87.00  &     2.1    \\
     HD 8907  & 0.01&    0.1 &   0.05&    0.9  &   0.63  &    4.5&       44.57  &    12.3  &   222.4~~   &     5.7    \\
    HD 22179  &  0.09&     1.2 &     0.42  &    6.1  &    1.62  &    7.9&       40.12  &    3.3  &     --    &     --     \\
    HD 25457  &  0.04&    0.6 &      0.31  &    4.6  &    1.10   &    8.0&       16.54  &    12.4  &     64.18&     3.3   \\
    HD 31392  & -0.003&  -0.1 &   0.02&   0.4  & 0.01  &  0.1&       19.27  &    7.7  &     100.3~~&     5.6   \\
    HD 35850  &  0.07&     1.1 &    0.14&     2.6  &   0.41  &     3.7&       3.94  &    4.0  &     --    &     --     \\
    HD 37484  &   0.10&     1.3 &     1.43  &    13.6  &    5.43  &    13.5&       44.67  &    10.0  &     57.2 &    2.1   \\
    HD 38207  &  0.08&     1.0 &     1.31  &    13.0  &    8.36  &    14.7&       231.5~~ &    13.4  &   532.1~~   &     2.1    \\
    HD 38529  &  -0.04&   -0.5 &  -0.04&   -0.6  &  0.04  &    0.5&       3.33  &    4.7  &   23.29   &    0.5     \\
    HD 61005  &  0.07&     1.0 &     1.24  &    12.6  &    10.24  &    14.9&       302.6~~ &    13.8  &  1207     &     3.1    \\
    HD 72905  &    --    &        -- &   -0.16  &   -2.3  &    --     &    --   &       1.68  &    4.4  &     --    &     --     \\
    HD 85301  &  0.02&    0.3 &    0.36  &    5.8  &   1.00    &    6.4&       11.73  &    4.7  &   -5.21   &    -0.2    \\
   HD 104860  &  0.01&    0.2 &     0.10  &    1.9  &   0.87  &    4.6&       89.83  &    12.2  &     499.6~~ &     7.5  \\
   HD 107146  &  0.03&    0.5 &    0.40  &    6.3  &    2.84  &    11.2&         139.0~~  &    13.9  &     --    &     --     \\
   HD 122652  & -0.05&   -0.8 &   0.08  &    1.5  &   0.57  &    2.7&       22.56  &    7.4  &    48.82  &     1.3  \\
   HD 145229  &  0.02&    0.3 &   0.09  &    1.6  &   0.46  &     3.7&       19.19  &     7.2  &    51.86  &     1.3   \\
   HD 150706  &  0.02&    0.2 &    0.05  &    1.0  &   0.24  &    2.0 &       7.69  &    4.3   &    -31.01 &  -1.5  \\
   HD 187897  & -0.06&   -0.9 &   0.03&    0.6  &   0.14  &   0.9&       13.29  &    6.2  &   -47.42  &   -0.6     \\
   HD 191089  &    --    &        -- &      5.94  &    20.8  &    --     &    --   &       181.7~~ &     13.5  &    340.6~~ &    4.6     \\
   HD 201219  & -0.03 &   -0.4 &   0.07&    1.4  &   0.39  &    2.7&       17.56  &    5.2  &   194.9~~   &     2.1    \\
   HD 202917  &   0.16&     2.2 &    0.63&    8.8  &    2.39  &    9.9&       27.19  &    5.5  &     --    &     --     \\
   HD 209253  &  0.09&     1.2 &    0.14  &    2.4  &    0.86  &     5.7&       12.69  &    6.5  &    14.96  &    0.8     \\
   HD 219498  &-0.005&  -0.1 &    0.29  &    4.4  &    1.19  &    6.2&       23.95  &    5.4  &   -91.78  &    -0.3    \\
\hline
\multicolumn{11}{c}{\it Tier 2 debris disks (excess SNR$_{70 \mu m} \ge 2 {\rm ~and~} <3$)} \\  
    HD 17925  &  0.07    &        1.0&   -0.21  &   -4.2  &    0.37   &   4.2   &       1.74  &    2.9  &     --    &     --     \\
    HD 70573  &  0.10   &     1.2 &   0.09  &     1.8  &   0.15  &   0.9&       12.95  &    2.4  &    138.3  &     1.0\\   
   HD 141943  &  0.08 &      1.1 &    0.35  &    5.6  &    1.61  &    8.6&       15.66  &    2.3  &     --    &     --    \\
   HD 204277  &  -0.05 &   -0.7 &   0.02  &   0.3  &  0.08  &   0.6&       4.53  &    2.3  &     --    &     --    \\
   HD 206374  & -0.05&   -0.8 &  -0.03  &  -0.7  &    1.86  &   0.5&       3.61  &    2.1  &    -2.71 &     -0.1  \\
      MML 17  &  0.07&    0.9 &    0.72  &    8.6  &    2.13  &    8.1&       27.17  &    2.3  &  -92.5    &   -0.1    \\
\enddata
\tablenotetext{a} {Excess signal to noise ratio is calculated using
                   total uncertainty on the photometry.  
		   }
\tablenotetext{b} {
Some sources indicated here as having low SNR in the excess
are reported as significant excesses by Carpenter et al. 2007a
based on empirical colors rather than Kurucz model analysis.
Specifically, that paper reports for those stars in this table that
were not selected for $FEPS$ based on previous suspecion of mid-infrared 
excess (19 of the 31) {\it only} HD 187897 and HD 206374 lack 33 um excess
and {\it only} HD 206374, HD 187897, HD 150706, HD 38529, and HD 31392 
lack 24 um excess, with excess defined at the $>3\sigma$ level.
Of the remaining 11 previously suspected mid-infrared excesses,
the majority are confirmed at 33 $\mu$m from our own
analysis (just HD 70573 is indeterminate).
}
\end{deluxetable}

\begin{deluxetable}{lccrcrcr}
\tabletypesize{\scriptsize}
\tablewidth{0pt}
\tablecolumns{8}
\setlength{\tabcolsep}{0.06in}
\tablecaption{Color Temperatures\tablenotemark{a}
and Reduced $\chi^2$ Values\tablenotemark{b} \label{temps}}
\tablehead{
\colhead{Source} &
\colhead{T$_{color}$(13/33$\mu$m)} &
\colhead{T$_{color}$(24/33$\mu$m)} &
\colhead{$\chi^2_\nu$(13-160$\mu$m)} &
\colhead{T$_{color}$(33/70$\mu$m)} &
\colhead{$\chi^2_\nu$(13-160$\mu$m)} &
\colhead{T$_{color}$(70/160$\mu$m)} &
\colhead{$\chi^2_\nu$(13-160$\mu$m)} \\ 
}
\startdata
\multicolumn{8}{c}{\it Tier 1 debris disks (excess SNR$_{70 \mu m} \ge 3$)}\\
HD 105	 &      $<$212 	&93$^{+50 }_{-28 }$	& 23.9&46$^{+3}_{-3}$	&2.97  & 57$^{+12}_{-8}$  & 13.0       \\	
HD 377	&        $<$69  &93$^{+17}_{-13}$	& 13.2&58$^{+3}_{-3}$	& 3.92 &   46$^{+12}_{-7}$	    &   14.0 \\	
HD 6963	&         --    &$<$76& 2.76&56$^{+9}_{-8}$	& 0.36 &   42$^{+25}_{-9}$  &   1.13   \\
HD 8907	&       $<$123 	&$<$59& 14.0&48$^{+3}_{-3}$	& 0.08 &   50$^{+8}_{-6}$	  &  0.24    \\	
HD 22179&	$<$175  &100$^{+22}_{-16}$   & 1.92&61$^{+9}_{-6}$	& 4.14 &    --	    &    --        \\	
HD 25457&	$<$157  &106$^{+30}_{-21}$  & 15.0&70$^{+5}_{-5}$	& 1.41 & 58 $^{+23}_{-10}$ &  6.67      \\	
HD 31392&	  --	& --    &  -- &$<$47   & 0.80 & 49$^{+10}_{-6}$   &  1.02  \\	
HD 35850&	$<$265  &120$^{+132}_{-42}$	& 1.59&82$^{+19}_{-13}$	& 0.65 &   --	    &  --	    \\
HD 37484&	$<$149  &101$^{+11}_{-9}$  & 2.19&86$^{+6}_{-6}$	& 2.87 &291$^{+10}_{-194}$   & $>$100    \\	
HD 38207&	$<$133  &76$^{+6}_{-5}$	& 12.5&59$^{+3}_{-2}$	& 7.9  &   88$^{+212}_{-29}$	    &       $>$100   \\	
HD 38529&	  --    &  --	& --  &$<$48	& 0.10 & $>$27    &	  3.41  \\
HD 61005&	$<$111	&68$^{+5}_{-4}$   &7.7&58$^{+3}_{-2}$	& 3.93 &   57$^{+23}_{-10}$     &   7.32         \\	
HD 72905&	  --    &  --	& --  &$<$103\tablenotemark{c}      &  --      & --	  &       --    \\
HD 85301&	$<$139	&127$^{+49}_{-27}$	& 2.17&76$^{+11}_{-8}$ & 2.57 & $>$53&  10.4 \\	
HD 104860&	$<$124 	&$<$67& 29.6&45$^{+3}_{-3}$	&0.55 & 47$^{+6}_{-4}$  &  0.58      \\	
HD 107146&	$<$122 	&72$^{+9}_{-8}$	& 25.1&52$^{+2}_{-2 }$	&3.34  &    --	    &    --        \\	
HD 122652&	 --     &$<$71	& 4.50&55$^{+7}_{-7}$	&0.53  &$>$63 &    1.61      \\	
HD 145229&	$<$169  &$<$82	& 6.41&54$^{+5}_{-5}$	&0.37  &$>$56      &  1.52        \\	
HD 150706&	$<$181  &$<$91	& 2.78&58$^{+11}_{-11}$ &0.12  &  $>$48 &	0.42\\
HD 187897&	 --     &$<$92& 8.40&$<$45	& 0.28 &  $>$36 &  0.33 \\
HD 191089&	 --     &  --	& --  &$<$92\tablenotemark{c}      & 0.41  &113$^{+113}_{-32}$    & $>$100\\	
HD 201219&	 --     &$<$83& 4.24&53$^{+7}_{-7}$	& 0.53 &    36$^{+15}_{-6}$   &1.56   \\
HD 202917&	   185$^{+33}_{-33}$	&101$^{+16}_{-12}$	& 2.99&77$^{+9}_{-7}$	& 4.18 &   --	  &         --     \\	
HD 209253&	$<$211  &77$^{+27}_{-19}$	& 0.86&70$^{+8}_{-6}$	& 0.59 &$>$68   &   7.05     \\	
HD 219498&	 --     &95$^{+29}_{-19}$	& 3.44&65$^{+7}_{-6}$	& 1.44 & $>$29   & 11.4         \\	
\multicolumn{8}{c}{\it Tier 2 debris disks (excess SNR$_{70 \mu m} \ge2 {\rm ~and~} < 3$)}   \\
HD 17925&	  --    &   --	& --  &110$^{+57}_{-25}$      &0.32&   --	  &    --          \\	
HD 70573&	  --    & --   	&  -- &$<$47	& 0.88 &$>$29 &    1.38    \\
HD 141943&	$<$167  &90$^{+18}_{-13}$	& 0.35&81$^{+27}_{-12}$&0.58  &    --	    &   --         \\	
HD 204277&	  --    &  --    &  --   &$<$50	& 0.14 &  --	  &	    --  \\
HD 206374&	 --     &  --   & --  & $<$74\tablenotemark{c}&0.46 &$>$29 &  0.20   \\
MML 17	&       $<$153  &120$^{+27}_{-18}$	& 0.70&74$^{+22}_{-10}$	& 6.23 &   $>$24 &    28.3  \\	
\enddata
\tablenotetext{a}{ 
Temperatures are in Kelvin.  Uncertainties are based on formal uncertainties
in flux density ratios and are rounded to nearest degree.
Upper limits are noted when the shorter wavelength from which the color 
temperature is calculated is in excess of the photosphere by $<2\sigma$.
Lower limits are noted in some instances at 70/160 $\mu$m and are 
based on $2\sigma$ upper limits to 160 $\mu$m flux densities.
}
\tablenotetext{b}
{Chisq values are calculated for usually 4-5 bandpasses within 13-160 $\mu$m,
which requires efficient grains over a factor of 10 in wavelength for good fits.
}
\tablenotetext{c}{
For HD 191089, HD 72905, and HD 206374 the quoted color temperature 
between 33-70 $\mu$m is actually from 24-70 $\mu$m due to lack of or
poor quality IRS data at 33 $\mu$m.  The quoted values are considered
upper limits to the 33/70 color temperature, but are upper limits
to the measured 24-70 $\mu$m color temperature only in the case of HD 206374.
		   }

\end{deluxetable}

\begin{deluxetable}{lrrrrrrr}
\tabletypesize{\scriptsize}
\tablewidth{0pt}
\tablecolumns{7}
\setlength{\tabcolsep}{0.06in}
\tablecaption{Dust Properties \label{dust}}
\tablehead{
\colhead{Source} &
\colhead{$T_{dust}/K$\tablenotemark{a}} &
\colhead{$R_{inner}/AU$} &
\colhead{log $L_{dust}/L_*$} &
\colhead{log $M_{dust, min}/M_\Earth$}  &
\multicolumn{2}{c}{$\tau_{collisions}/\tau_{P-R}$} \\
\\
& & & & &
assuming  &
assuming  \\
& & & & &
\colhead{$\alpha=0$ at $R_{inner}$} &
\colhead{$\alpha=-1$ at 200 AU} \\
}
\startdata
\multicolumn{7}{c}{\it Tier 1 debris disks (excess SNR$_{70 \mu m} \ge 3$)}                            \\
HD 105      & 46   &	42   &-3.5 &	-3.1 &	       0.004         & 0.4\\
HD 377      & 58   &	23  &	-3.4 &	-3.4 &	       0.004         & 0.3\\
HD 6963     & 56   &	18   &-4.0 &	-4.3 &	       0.010          & 0.6\\
HD 8907     & 48   &	49   &-3.6 &	-3.1 &	       0.008         & 0.9\\
HD 22179    & 61\tablenotemark{e}   &	31   &-3.6 &	-3.4 &	       0.010         & 0.8\\
HD 25457    & 70   &	23   &-4.0  &	-4.1 &	       0.027         & 1.8\\
HD 31392    & 49\tablenotemark{d}   &	24   &-3.8 &	-3.8 &	       0.005         & 0.3\\
HD 35850    & 82   &	15  &	-4.5 &	-4.9 &	       0.086         & 4.8\\
HD 37484    & 86\tablenotemark{e}   &	 20    &-3.5 &	-3.7 &	       0.018         & 1.2\\
HD 38207    & 59   &	51   &-3.0  &	-2.4 &	       0.005         & 0.5\\
HD 38529    & $<$48\tablenotemark{b}   &	86   &-4.6 &	-3.5 &	       0.150         & 30.7\\
HD 61005    & 58   &	17   &-2.6 &	-2.9 &	       $<$0.001         & $<$0.1\\
HD 72905    & 103  &	7   &	-4.7 &	-5.8 &	       0.110         & 4.2\\
HD 85301    & 76\tablenotemark{e}   &	11  &	-3.9 &	-4.6 &	       0.012         & 0.6\\
HD 104860   & 46\tablenotemark{c}   &	42   &-3.2 &	-2.8 &	       0.002         & 0.2\\
HD 107146   & 52   &	30     &-3.1 &	-2.9 &	       0.001         & 0.1\\
HD 122652   & 55   &	31   &-3.9 &	-3.7 &	       0.014         & 1.1\\
HD 145229   & 54   &	26   &-3.9 &	-3.9 &	       0.010         & 0.8\\
HD 150706   & 58   &	23   &-4.3 &	-4.4 &	       0.027         & 1.9\\
HD 187897   & $<$45\tablenotemark{b}   &	43   &-4.0 &	-3.6 &	       0.014         & 1.4\\
HD 191089   & 92   &	16   &-2.8 &	-3.2 &	       0.004         & 0.2\\
HD 201219   & 53   &	23   &-3.9 &	-4.0  &	       0.007         & 0.5\\
HD 202917   & 77\tablenotemark{e}   &	11   &-3.6 &	-4.3 &	       0.005         & 0.3\\
HD 209253   & 70   &	20   &-4.1 &	-4.3 &	       0.030         & 1.9\\
HD 219498   & 65   &	41   &-3.7 &	-3.3 &	       0.027         & 2.5\\

\multicolumn{7}{c}{\it Tier 2 debris disks (excess SNR$_{70 \mu m} \ge2 {\rm ~and~} < 3$)}  \\ 
HD 17925    & 110  &	4   &	-4.4 &	-6.0 &	       0.036         & 1.0\\
HD 70573    & 41\tablenotemark{c}   &	35   &-4.0 &	-3.7 &	       0.007         & 0.6\\
HD 141943   & 85\tablenotemark{c}   &	18   &-3.8 &	-4.1 &	       0.014         & 1.7\\
HD 204277   & $<$50\tablenotemark{b}   &	43  &	-4.6  &	-4.1 &	       0.074         & 7.2\\
HD 206374   & $<$74\tablenotemark{b}   &	12  &	-4.5 &	-5.1 &	       0.042         & 2.0\\
MML 17      & 74\tablenotemark{e}   &	23  &	-3.7 &	-3.7 &	       0.018         & 1.2\\
\enddata
\tablenotetext{a}{Adopted T$_{dust}$ from among values in Table 4, typically the 33/70 color temperature, but other cases as footnoted.}
\tablenotetext{b}{In the cases of temperature upper limits, R$_{inner}$ 
$L_{dust}/L_*$, and  $M_{dust, min}/M_\Earth$ are all minimum values
while $\tau_{collisions}/\tau_{P-R}$ are maxima.
}
\tablenotetext{c}{Adopted T$_{dust}$ is an average of consistent values from Table 4}
\tablenotetext{d}{Adopted T$_{dust}$ is from the 70/160 flux ratio rather than 33/70.}
\tablenotetext{e}{While the adopted T$_{dust}$ is from the 33/70 flux ratio, the hotter 24/33 value produces a lower $\chi^2$ to the overall SED; see Table 4.
In these cases, contrary to situation of Tablenote $b$, 
the R$_{inner}$, $L_{dust}/L_*$, and  
$M_{dust, min}/M_\Earth$ may be lower than we quote
while $\tau_{collisions}/\tau_{P-R}$ may be higher, all in the blackbody assumption.}
\end{deluxetable}

\begin{deluxetable}{lrrrrr}
\tabletypesize{\scriptsize}
\tablewidth{0pt}
\tablecolumns{6}
\setlength{\tabcolsep}{0.06in}
\tablecaption{Extended Disk Models\tablenotemark{a}\label{disk}}
\tablehead{
\colhead{Source} &
\colhead{$\Delta T_{color}$/K \tablenotemark{b}}&
\colhead{$R_{inner}/AU$ \tablenotemark{c}}&
\colhead{$R_{outer}/AU$ \tablenotemark{d}}&
\colhead{$\sigma_{0}$ \tablenotemark{e}}&
\colhead{RMS Deviation \tablenotemark{f}} \\ 
}
\startdata
HD 85301    &   51$\pm 15$ &  0.7  &$>$200 &  6.63$\times 10^{-5}$  &  0.12 \\
HD 105      &   47$\pm 39$ &  36.8 &   38  &  2.80$\times 10^{-2}$   &   1.27   \\
MML 17      &   47$\pm 28$ &  1.9  &$>$200 &  1.24$\times 10^{-4}$   &  0.15 \\
HD 22179    &   39$\pm 20$ & 6.4   &$>$200 &  1.66$\times 10^{-4}$   &  0.66  \\
HD 377      &   36$\pm 15$ &  8.5  & 137   &  3.15$\times 10^{-4}$   &  1.05  \\
HD 25457    &   36$\pm 26$ &  5.0  &  92   &  8.50$\times 10^{-5}$   &  0.44  \\
HD 219498   &   31$\pm 25$ & 11.6  &$>$200 &  1.57$\times 10^{-4}$   &  0.40 \\
HD 202917   &   24$\pm 16$ & 2.5   & 40    &  2.08$\times 10^{-4}$   &  0.07 \\
HD 107146   &   20$\pm  9$ & 13.6  & $>$200 &  9.52$\times 10^{-4}$   &  0.93  \\
HD 38207    &   17$\pm  6$ & 21.2  & 130   &  1.15$\times 10^{-3}$   &  1.22  \\
HD 37484    &   15$\pm 12$ & 8.2   & 34    &  4.35$\times 10^{-2}$   &  0.87  \\
HD 61005    &   10$\pm 5 $ &  8.6  & 41    &  3.35$\times 10^{-3}$   &  0.35  \\
HD 141943   &   9$\pm  25 $  & 8.6 &  40   &  1.95$\times 10^{-4}$   &  0.01 \\
%
\enddata
\tablenotetext{a}{
Included in this table are all sources
from Table 4 with $\chi_\nu ^2 >$ 1.2 from the 33/70 $\mu$m flux ratio.  
In addition we present for comparision, a disk model of HD 141943 
which is well fit by a single temperature blackbody matched to 
the 33/70 $\mu$m flux ratio.
List is sorted inversely by $\Delta T_{color}$. 
}
\tablenotetext{b}{
$\Delta T_{color}$ is the difference between the 24/33 and 33/70 $\mu$m color
temperatures in Table 4, i.e. the temperature range that must be explained
in the extended disk model.  Quoted error is the root-sum-squared of the 
individual color temperature errors, ignoring the covariance term suggested by 
the appearance of the 33 $\mu$m flux density in both color temperatures;
asymmetric errors have been simply averaged.  
}
\tablenotetext{c}{
$R_{inner}$ is the inferred inner disk radius.
}
\tablenotetext{d}{
$R_{outer}$ is the inferred outer disk radius.
A value of 200 means that the outer boundary is indeterminate, 
even for $\alpha$= 0 models; for the fits it 
is held constant and only $R_{inner}$ and $\sigma_0$ are varied.
}
\tablenotetext{e}{
$\sigma_{0}$ is the inferred surface density at $R=R_{inner}$, 
constant with radius in these $\alpha=0$ models.  
The units of $\sigma$ are dimensionless, in cm$^2$ of grain cross
section per cm$^2$ of disk area, distinguished from the usual definition of
$\Sigma_{0}$ which is in per cm$^2$ of disk area.  Note that high values are
required in the narrower disk cases.
}
\tablenotetext{f}{
RMS deviation of the spectral energy distribution from the 
constant surface density blackbody disk 
having $R_{inner}$, $R_{outer}$, and $\sigma_0$.
This is the square root of the summed squared deviations divided
by number of points (usually 4 from 24-160 $\mu$m).
}

\end{deluxetable}

\clearpage

\begin{thebibliography}{}
\bibitem[]{2693} Ardila, D. R., Golimowski, D. A., Krist, J. E.  et al. 2004, ApJL, 617
\bibitem[]{2694} Artymowicz, P. 1988, ApJ, 335, L79
\bibitem[]{2697} Backman, D. \& Paresce, F. 1993, in Protostars and Planets III ed. E.H. Levy \& J.I. Lunine (Tucson: Univ. Arizona Press), p 1253

\bibitem[]{2699} Backman, D., Dasgupta, A., \& Stencel, R. E. 1995, ApJL, 450, 35
\bibitem[]{2700} Backman, D. 2004, in ASP Conf. Ser 324, Debris Disks and the Formation of Planets: A Symposium in Memory of Fred Gillett (ed. L. Caroff, L.J. Moon, D. Backman, and E. Praton), p 9
\bibitem[]{} Backman, D. et al. 2008, in preparation
\bibitem[]{2708} Beichman, C. A., Bryden, G., Stapelfeldt, K. R., Gautier, T. N., Grogan, K., et al. 2006, ApJ, 652,1674
\bibitem[]{2709} Beichman et al. 2007, Protostars and Planets V, B. Reipurth, D. Jewitt, and K. Keil (eds.), University of Arizona Press, Tucson, 951
\bibitem[]{2710} Bottke, W.F., Durda, D.D., Nesvorny, D., Jedicke, R., Morbidelli, A., et al. 2005, Icarus, 179, 63
\bibitem[]{2712} Bouwman et al. 2008, ApJ, accepted  
\bibitem[]{2713} Bryden, G., Beichman, C., Trilling, D., Rieke, G., Holmes, E., et al. 2006, ApJ, 636, 1098
\bibitem[]{2714} Burns, J.A., Lamy, P.L., \& Soter, S. 1979, Icarus, 40, 1 
\bibitem[]{2715} Carpenter, J. M.,  Wolf, S., Schreyer, K., Launhardt, R., \& Henning, T. 2005, AJ, 129, 1049
\bibitem[]{2716} Carpenter et al. 2008a, (omnibus disks paper) in preparation 
\bibitem[]{2717} Carpenter et al. 2008b, (data paper) in preparation 
\bibitem[]{2721} Cieza, L., Padgett, D. L., Stapelfeldt, K. R., Augereau, J.-C., Harvey, P., et al. 2007, ApJ, 667, 308
\bibitem[]{2722} Cohen, M., Megeath, S. T., Hammersley, P. L., Marti{\'i}n-Luis, F., \& Stauffer, J. 2003a, AJ, 125, 2645
\bibitem[]{2723} Cohen, M., Wheaton, W., \& Megeath, S. T. 2003b, AJ, 126, 1090
\bibitem[]{2724} Cortes, S., et al., 2008, in preparation. 
\bibitem[]{2725} Cutri, R. M., Skrutskie, M. F., van Dyk, S., Beichman, C. A., Carpenter, J. M., et al. 2003
\bibitem[]{2727} Decin, G., Dominik, C., Malfait, K., Mayor, M., \& Waelkens, C. 2000, A\&A, 357, 533  
\bibitem[]{2728} Decin, G., Dominik, C., Waters, L. B. F. M., \& Waelkens, C. 2003, ApJ, 598, 636 
\bibitem[]{2729} Dohnanyi, J.W., 1969, J.G.R., 74, 2531  
\bibitem[Dole et al.(2004a)]{2004ApJS..154...87D} Dole, H., et al.\ 2004a, \apjs, 154, 87 
\bibitem[Dole et al.(2004b)]{2004ApJS..154...93D} Dole, H., et al.\ 2004b, \apjs, 154, 93 
\bibitem[]{2732} Dominik, C. \& Decin, G., 2003, ApJ, 598, 626 
\bibitem[Downes et al.(1986)]{1986MNRAS.218...31D} Downes, A.~J.~B., Peacock, J.~A., Savage, A., \& Carrie, D.~R.\ 1986, \mnras, 218, 31
\bibitem[]{2735} Durda, D. D., \& Dermott, S. F. 1997, Icarus, 130, 140
\bibitem[]{2737} Fajardo-Acosta, S. B., Stencel, R. E., Backman, D. E., \& Thakur, N. 1999, ApJ, 520, 215
\bibitem[]{2738} Fixsen, D. J., \& Dwek, E. 2002, ApJ, 578, 1009  
\bibitem[]{2739} Frayer, D.T. et al., 2006 AJ, 131, 250
\bibitem[]{2740} Gautier, T. N., Rieke, G. H., Stansberry, J., Bryden, G. C.,
 Stapelfeldt, K. R., et al. 2007, ApJ, 667, 527
\bibitem[]{2742} Gladman, B., Kavelaars, J.J., Petit, J.M., Morbidelli, A., Holman, M.J., Loredo, Y. 2001, AJ, 122, 1051    
\bibitem[]{2743} Gordon, K.  et al. 2005, PASP, 117, 503
\bibitem[]{2749} Gregorio-Hetem, J., Lepine, J. R. D., Quast, G. R., Torres, C. A. O., de La Reza, R., 1992, AJ, 103, 549
\bibitem[]{2750} Habing, H. J, Dominik, C.,  Jourdain d. M., Laureijs, R. J., et al. 2001, A\&A, 365, 545
\bibitem[]{2751} Hillenbrand et al. 2008 in preparation
\bibitem[]{2753} Hines, D. C., Backmanm D, E, Bouwman, J., et al. 2006, ApJ, 638, 1070
\bibitem[]{2837} Hines, D. C., Schneider, G., Hollenbach, D. et al. 2007, ApJ (Letters), in press
\bibitem[]{2754} Hollenbach, D., Gorti, U., Meyer, M., et al. 2005, ApJ, 631, 1180
\bibitem[]{2755} Jura, M. 2004, ApJ, 603, 729 
\bibitem[]{2756} Jura, M., Ghez, A.M., White, R.J., McCarthy, D.W., Smith, R.C., Martin, P.G. 2005, ApJ, 445, 451 
\bibitem[]{2757} Kalas, P., Graham, J.R., Clampin, M.C., Fitzgerald, M.P. 2006, ApJL, 637, 57  
\bibitem[]{2758}  Kenyon, S. J., \& Bromley, B. C. 2002, ApJ, 577, L35  
\bibitem[]{2759} Kenyon, S. J. \& Bromley, B. C. 2005, AJ, 130, 269
\bibitem[]{2760} Kim, J.S., Hines, D., Backman, D., Hillenbrand, L., Meyer, M., et al. 2005, ApJ, 632, 659 
\bibitem[]{2761} Kim et al. 2008, in prep.
\bibitem[]{2762} Krivov, A.V., Mann, I., \& Krivova, N.A. 2000, AA, 362, 1127 
\bibitem[]{2763} Landgraf, M., Liou, J.-C., Zook, H., \& Grun, E. 2002, AJ, 123, 2857 
\bibitem[]{2766} Levison, H. F. \& Morbidelli, A. 2003, Nature, 426, 419  
\bibitem[]{2767} L{\"o}hne, Torsten; Krivov, Alexander V.; Rodmann, Jens, ApJ, in press (arXiv0710.4294)
\bibitem[]{2769} Low, F.J., Smith, P.S., Werner, M., Chen, C., Krause, V. et al.  2005, ApJ, 631, 1170
\bibitem[]{2770} Mamajek, E.E., Meyer, M.R.; Hinz, P.M.; Hoffmann, W.F.; Cohen, M., Hora, J.L., 2004, ApJ, 612, 496
\bibitem[]{2771} Mannings, V. \& Barlow, M. J. 1998, ApJ, 497, 330
\bibitem[]{2774} Matt, S. \& Pudritz, R. E. 2007, (arXiv0707.0306) To appear in proceedings of IAU Symposium No. 243: Star-Disk Interaction in Young Stars
\bibitem[]{2777} Metchev, S. A., Hillenbrand, L. A., \& Meyer, M. R. 2004, ApJ, 600, 435
\bibitem[]{2778} Metchev, S. et al. 2008, in preparation
\bibitem[]{2779} Meyer, M.R., Backman, D.E., Weinberger, A.J., \& Wyatt, M.C. 2007, in Protostars and Planets V, B. Reipurth, D. Jewitt, and K. Keil (eds.), University of Arizona Press, Tucson
\bibitem[]{2780} Meyer, M. R., Carpenter, J. C., Mamajek, E. E., et al. 2008, ApJL, submitted
\bibitem[]{2781} Meyer, M. R., Hillenbrand, L. A., Backman, D. E., et al. 2004, ApJS, 154, 422
\bibitem[]{2782} Meyer, M.R. et al. 2006, PASP, December issue
\bibitem[]{2783} Mo{\'o}r, A., {\'A}brah{\'a}m, P., Derekas, A., Kiss, Cs., Kiss, L. L., Apai, D., Grady, C., \&Henning, Th.  2006, ApJ, 644, 525
\bibitem[]{2784} Morbidelli, A., Tsiganis, K., Crida, A., Levison, H.F., Gomes, R., 2007, AJ (astro-ph/0706.1713)
\bibitem[]{2785} Moro-Mart{\'i}n, A. \& Malhotra, R. 2002, AJ, 124, 2305 
\bibitem[]{2786} Moro-Mart{\'i}n, A.\& Malhotra, R. 2003, AJ, 125, 2255
\bibitem[]{2787} Moro-Mart\'{\i}n, A., Carpenter, J.M., Meyer, M.R., Hillenbrand, L.A. et al. 2007a, ApJ, 658, 1312
\bibitem[]{2788} Moro-Mart\'{\i}n, A. Malhotra, R., Carpenter, J.M., Hillenbrand, L.A., Wolf, S., Meyer, M.R., Hollenbach, D., Najita, J., Henning, Th., 2007b, ApJ, 668, 1165
\bibitem[]{2789} Najita, J. \& Williams, J. P. 2005, ApJ, 635, 625
\bibitem[]{2790} Najita, J. et al. 2008, in preparation 
\bibitem[]{2791} Neuh{\"a}user, R., Walter, F. M., Covino, E., Alcal{\'a}, J. M., Wolk, S. J., et al.  2000. A\&AS, 146, 323
\bibitem[]{2792} Ozernoy, L.M., Gorkavyi, N.N., Mather, J.C. \& Taidakova, T.A. 2000, ApJL, 537, 147 
\bibitem[]{2793} Padgett, D. L., Cieza, L., Stapelfeldt, K. R., Evans, N. J., II, Koerner, D., et al. 2006, ApJ, 645, 1283
\bibitem[]{2794} Pascucci, I., Gorti, U., Hollenbach, D., et al. 2006, ApJ, 651, 1177
\bibitem[]{2795} Pascucci, I., Hollenbach, D., Najita, J., et al. 2007, ApJ, 663, 383
\bibitem[]{2796} Papovich, C., et al.\ 2004, ApJS, 154, 70
\bibitem[]{2797} Patten, B. M. \& Wilson, L. A. 1991, AJ, 102, 323
\bibitem[]{2798} Rhee, J.H., Song, I., Zuckerman, B., \& McElwain, M., 2007, ApJ, 660, 1556
\bibitem[]{2799} Rieke, G.~H., et al.\ 2004, ApJS, 154, 25
\bibitem[]{2800} Rieke, G. H., Su, K. Y. L., Stansberry, J. A., Trilling, D., Bryden, G.  et al. 2005, ApJ, 620, 1010
\bibitem[]{2805} Silverstone, M.D. 2000, Ph.\ D. Thesis, University of California, Los Angeles 
\bibitem[]{2806} Silverstone, M., Meyer, M., Mamajek, E., Hines, D., Hillenbrand, L., et al. 2006, ApJ, 639, 1138
\bibitem[]{2807} Smith, P. S., Hines, D. C., Low, F. J., Gehrz, R. D., Polomski, E. F., \&  Woodward, C. E. 2006, ApJL, 644, 125
\bibitem[]{2810} Spangler, C., Sargent, A. I., Silverstone, M. D., Becklin, E. E., \& Zuckerman, B. 2001, ApJ, 555, 923
\bibitem[]{2892} Stapelfeldt, K.R., Holmes, E.K., Chen, C. et al. 2004, ApJS, 154, 458  
\bibitem[]{2812} Stauffer, J. R., Rebull, L. M., Carpenter, J. M., et al. 2005, 130, 1834
\bibitem[]{2813} Stern, S.A., 1996a, AA, 310, 999  
\bibitem[]{2814} Stern, S.A., 1996b, AJ, 112, 1203 
\bibitem[]{2815} Stern, S.A. \& Colwell, J. 1997, ApJ, 490, 879  
\bibitem[]{2816} Su, K. Y. L., Rieke, G. H., Misselt, K. A., Stansberry, J. A., Moro-Mart\'{\i}n, A., et al. 2005, ApJ, 628, 487
\bibitem[]{2817} Su, K.~Y.~L., et al.\ 2006, \apj, 653, 675 
\bibitem[]{2819} Sylvester, R. J., Skinner, C. J., Barlow, M. J., \& Mannings, V. 1996, MNRAS, 279, 915
\bibitem[]{2820} Sylvester, R. J. \& Mannings, V. 2000, MNRAS, 313, 73
\bibitem[]{2821} Teplitz, V., Stern,S., Anderson, J., Rosenbaum, D., Scalise, R., \& Wentzler, P. 1999, ApJ, 516, 425
\bibitem[]{2822} Th\'ebault, P., Augereau, J. C., \& Beust, H. 2003, A\&A, 408, 775
\bibitem[]{2824} Th\'ebault, P. \& Augereau, J.-C. 2007, A\&A, submitted
\bibitem[]{2825} Wahhaj, Z., Koerner, D.W., \& Sargent, A.I. 2007, ApJ, 661, 368
\bibitem[]{2826} Weingartner, J. C. \& Draine, B. T. 2001, ApJ, 548, 296
\bibitem[]{2827} Williams, J. P., Najita, J., Liu, M. C., Bottinelli, S., Carpenter, J. M., Hillenbrand, L. A., Meyer, M. R., \& Soderblom, D. R. 2004, ApJ, 604, 414
\bibitem[]{2828} Wolf, S., \& Hillenbrand, L. A. 2003, ApJ, 596, 603
\bibitem[]{2829} Wood, B. E., Muller, H.-R., Zank, G. P., Linsky, J. L., \& Redfield, S.  2005, ApJL, 628, 143
\bibitem[]{2830} Wyatt, M. C. 2006, ApJ, 639, 1153
\bibitem[]{2831} Wyatt, M. C., Smith, R., Su, K. Y. L., et al. 2007, APJ, 663, 365
\end{thebibliography}
\end{document}